\documentclass[aps,epsf] {article}
\usepackage{graphicx}
\usepackage{epsfig}
\newcommand{\ltwid}
{\mathrel{\raise.3ex\hbox{$<$\kern-.75em\lower1ex\hbox{$\sim$}}}}
\newcommand{\beq}{\begin{equation}}
\newcommand{\eeq}{\end{equation}}
\newcommand{\beqs}{\begin{eqnarray}}
\newcommand{\eeqs}{\end{eqnarray}}

\newcommand{\ttt}{\theta_{13}}
\newcommand{\rms}{\rm\scriptstyle}
\newcommand{\nubar}[0]{\overline{\nu}}
\newcommand{\numu}[0]{\nu_{\mu}}
\newcommand{\nuebar}[0]{\overline{\nu}_{e}}

\newcommand{\nue}[0]{\nu_{e}}
\newcommand{\numubar}[0]{\overline{\nu}_{\mu}}

\newcommand{\dmsq}{\mbox{$\delta m^2$}}

\input defs.eoi_epsfig

\def\({ \left( }
\def\){ \right) }
\def\={\ =\ }

\def\+{\ +\ }
\def\-{\ -\ }

\def\r{\\ [2pt]} 
\def\hline{\noalign{\hrule \vskip2pt}}

\def\|{\ifmmode\Vert\else \char`\|\fi}
\ifx\oldzeta\undefined                          
  \let\oldzeta=\zeta                            
  \def\zzeta{{\raise 2pt\hbox{$\oldzeta$}}}     
  \let\zeta=\zzeta                              
\fi

\ifx\oldchi\undefined                           
  \let\oldchi=\chi                              
  \def\cchi{{\raise 2pt\hbox{$\oldchi$}}}       
  \let\chi=\cchi                                
\fi

\def\square{\hbox{{$\sqcup$}\llap{$\sqcap$}}}   

\def\frac#1#2{{#1 \over #2}}

\def\half{\ifinner {\scriptstyle {1 \over 2}}
   \else {1 \over 2} \fi}

\def\simge{\mathrel{%
   \rlap{\raise 0.511ex \hbox{$>$}}{\lower 0.511ex \hbox{$\sim$}}}}
\def\simle{\mathrel{
   \rlap{\raise 0.511ex \hbox{$<$}}{\lower 0.511ex \hbox{$\sim$}}}}

\def\buildchar#1#2#3{{\null\!                   
   \mathop#1\limits^{#2}_{#3}                   
   \!\null}}                                    
\def\overcirc#1{\buildchar{#1}{\circ}{}}

\def\slashchar#1{\setbox0=\hbox{$#1$}           
   \dimen0=\wd0                                 
   \setbox1=\hbox{/} \dimen1=\wd1               
   \ifdim\dimen0>\dimen1                        
      \rlap{\hbox to \dimen0{\hfil/\hfil}}      
      #1                                        
   \else                                        
      \rlap{\hbox to \dimen1{\hfil$#1$\hfil}}   
      /                                         
   \fi}                                         %

\def\subrightarrow#1{
  \setbox0=\hbox{
    $\displaystyle\mathop{}
    \limits_{#1}$}
  \dimen0=\wd0
  \advance \dimen0 by .5em
  \mathrel{
    \mathop{\hbox to \dimen0{\rightarrowfill}}
       \limits_{#1}}}                           

\def\overlay#1#2{\ifmmode%
\setbox0=\hbox{$#1$}%
\setbox1=\hbox to\wd0{\hss$#2$\hss}\else%
\setbox0=\hbox{#1}%
\setbox1=\hbox to\wd0{\hss#2\hss}\fi%
#1\hskip-\wd0\box1 }

\def\pmb#1{\leavevmode\setbox0=\hbox{#1}%
\kern-.02em\copy0\kern-\wd0
\kern.04em\copy0\kern-\wd0
\kern-.02em\raise.04em\box0 }

\def\vereq#1#2{\lower3pt\vbox{\baselineskip1.5pt \lineskip1.5pt
\ialign{$\m@th#1\hfill##\hfil$\crcr#2\crcr\sim\crcr}}}

\def\tensor#1{\protect\@ontopof{#1}{\leftrightarrow}{1.15}\mathord{\box2}}
\def\overstar#1{\protect\@ontopof{#1}{\ast}{1.15}\mathord{\box2}}
\def\overdots#1{\protect\@ontopof{#1}{\cdots}{1.0}\mathord{\box2}}
\def\overcirc#1{\protect\@ontopof{#1}{\circ}{1.2}\mathord{\box2}}
\def\loarrow#1{\protect\@ontopof{#1}{\leftarrow}{1.15}\mathord{\box2}}
\def\roarrow#1{\protect\@ontopof{#1}{\rightarrow}{1.15}\mathord{\box2}}

\def\@ontopof#1#2#3{%
{\mathchoice
{\@@ontopof{#1}{#2}{#3}\displaystyle\scriptstyle}%
{\@@ontopof{#1}{#2}{#3}\textstyle\scriptstyle}%
{\@@ontopof{#1}{#2}{#3}\scriptstyle\scriptscriptstyle}%
{\@@ontopof{#1}{#2}{#3}\scriptscriptstyle\scriptscriptstyle}%
}%
}

\def\@@ontopof#1#2#3#4#5{%
\setbox0=\hbox{$#4#1$}%
\setbox1=\hbox{$#5#2$}%
\setbox2=\hbox{}\ht2=\ht0 \dp2=\dp0 %
\ifdim\wd0>\wd1 %
\setbox1=\hbox to\wd0{\hss\box1\hss}%
\mathord{\rlap{\raise#3\ht0\box1}\box0}%
\else   %
\setbox1=\hbox to.9\wd1{\hss\box1\hss}%
\setbox0=\hbox to\wd1{\hss$#4\relax#1$\hss}%
\mathord{\rlap{\copy0}\raise#3\ht0\box1}%
\fi
}%

\def\lambdabar{\protect\@lambdabar}
\def\@lambdabar{%
\relax
\bgroup
\def\@tempa{\hbox{\raise.73\ht0
\hbox to0pt{\kern.25\wd0\vrule width.5\wd0
height.1pt depth.1pt\hss}\box0}}%
\mathchoice{\setbox0\hbox{$\displaystyle\lambda$}\@tempa}%
{\setbox0\hbox{$\textstyle\lambda$}\@tempa}%
{\setbox0\hbox{$\scriptstyle\lambda$}\@tempa}%
{\setbox0\hbox{$\scriptscriptstyle\lambda$}\@tempa}%
\egroup
}

\def\corresponds{{\lower.2ex\hbox{=}}{\rm\kern-.75em^\triangle}}
\def\succsim{\succ\kern-.9em_\sim\kern.3em}
\def\precsim{\prec\kern-1em_\sim\kern.3em}
\def\slantfrac#1#2{\kern1em^{#1}\kern-.3em/\kern-.1em_{#2}}

\begin{document}
\begin{center}

NuMI-PUB-GEN-0880
\\
{\Large \bf Detector R\&D for future Neutrino Experiments \\
with the NuMI Beamline} \\
October 21, 2002 {\it(revised November 11, 2002)}\\
{\it A report to the Fermilab Directorate from the Study Group
on Future Neutrino Experiments at Fermilab.}
\end{center}

\vspace*{0.6in}

\begin{center}
G.~Barenboim,$^1$
A.~Bodek,$^2$
C.~Bromberg,$^3$
A.~Bross,$^1$
L.~Buckley-Geer,$^1$
B.~Choudhary,$^1$
D.~Cline,$^4$
F.~DeJongh,$^1$
G.~Drake,$^5$
S.~Geer,$^1$
M.~Goodman$^*,^5$
A.~deGouvea,$^1$
D.~A.~Harris$^*,^1$
K.~Heller,$^6$
J.~Huston,$^3$
J.~Johnstone,$^1$
M.~Kostin,$^1$
J.~Learned,$^7$, 
P.~Litchfield,$^6$, 
M.~Marshak,$^6$, 
K.~McDonald,$^8$
K.~S.~McFarland,$^2$
S.~Menary,$^9$
M.~Messier,$^{10}$
D.~Michael,$^{11}$
R.~Miller,$^7$
N.~Mokhov,$^1$
J.~K.~Nelson,$^1$
E.~Peterson,$^6$, 
R.~Richards,$^7$, 
K.~Ruddick,$^6$, 
F.~Sergiamepietri,$^4$
P.~Shanahan,$^1$
R.~Shrock,$^{12}$
Y.~Seo,$^4$
R.~Stefanski,$^1$
M.~Szleper,$^{13}$
K.~Tollefson,$^3$
J.~Urheim$^6$ 
\end{center}
\vspace*{0.6in}
\begin{itemize}
\item[*] Editors
\end{itemize}
\begin{enumerate}
\item Fermilab National Laboratory
\item University of Rochester
\item Michigan State University
\item University of California at Los Angeles
\item Argonne National Laboratory
\item University of Minnesota
\item University of Hawaii 
\item Princeton University
\item York University
\item University of Indiana
\item California Institute of Technology
\item State University of New York at Stonybrook
\item Northwestern University 
\end{enumerate}
\section{Executive Summary}
\label{sec:summary}
\par This document is the result of a request 
from the Fermilab directorate to (i) investigate the detector technology 
issues relevant for future long baseline experiments and (ii) 
consider the associated detector R\&D that would be needed to prepare 
the way for future neutrino oscillation experiments using the NuMI beamline.  
Because of the narrow energy spread provided by an off-axis beam and 
the resulting low intrinsic
electron neutrino background, as well as the very favorable 
duty cycle of the NuMI 
beamline, a well-placed neutrino detector at the surface of the earth
could take the next important steps in neutrino oscillation 
physics.  The biggest outstanding issue in this field is whether or not 
the last unmeasured element of the leptonic mixing matrix, parameterized
by the mixing angle $\theta_{13}$, is nonzero.  If it is in fact non-zero, 
this opens the door to measurements of the neutrino mass hierarchy and,
if the solar neutrino oscillations are described by the LMA solution, 
searches for CP violation in the lepton sector.  In order to get to any 
of these measurements, an off-axis detector must be 
capable of measuring the $\nu_\mu(\bar\nu_\mu) \to \nu_e (\bar\nu_e)$ 
transition probabilities as well as the 
$\nu_\mu(\bar\nu_\mu)$ survival probabilities, 
at the energies present in these off-axis beams, which could lie anywhere 
from 0.6 to 3 GeV.  
Optimal baselines and energies will
depend on the physics goal of the experiment.  For example, an optimization
of the sensitivity for $\nu_e$ appearance from a $\nu_\mu$ beam 
assuming $\Delta m_{32}^2=3\times 10^{-3}$ eV$^2$ would lead to 
a baseline of $\sim$ 700-900 km and an energy of $\sim 2.2$ GeV. 

\par Consideration of future neutrino experiments can be separated into
three phases (starting now):  
\begin{enumerate}
\item[I.] \it{0-5 years}: the beginning of the MINOS
project.
\item[II.] \it{5-10 years}: off-axis experiments using the current 
NuMI beam.
\item[III.] \it{$>$ 10 years}: future superbeam/neutrino 
factory program with larger detectors.
\end{enumerate}
It is already clear that  a future program is desirable to search for 
CP violation if LMA is confirmed, or if a non-zero value for $\theta_{13}$ 
is found in I or II.  Even if LMA is not confirmed, a second phase is 
important to push the sensitivity for $\theta_{13}$, and possibly to measure
the mass hierarchy.  What is also clear from examining different detector concepts 
is that the detector one would chose to perform a phase II experiment 
may not be the one chosen for phase III.  If $\theta_{13}$ is still not seen 
in phase III, there is still a possibility of measuring it in a neutrino factory, 
but the detector issues associated with that experiment are far less 
challenging, and are documented in a previous report \cite{fnal_nufact}.  
Because a neutrino factory produces beams of $\nu_\mu$ and 
$\bar\nu_e$ or $\bar\nu_\mu$ and $\nu_e$, a detector simply has to 
identify the presence and charge of an outgoing muon to address both 
the atmospheric oscillation parameters and $\theta_{13}$. 

We outline in this document several detector possibilities for 
an experiment at an off-axis site along the NuMI beamline:  water Cerenkov, 
several versions of fine-grained calorimetry, liquid argon TPC, and mention in 
passing the AQUARICH concept.  In this executive summary we wish to mention 
the salient features of each technology, and the key issues to address 
before choosing the technology for either a phase II or phase III detector.

Conclusions about Water Cerenkov:  
\begin{itemize} 
\item Much expertise in the field with large detector performance 
\item 20 kton fiducial mass proof of principle exists 
\item Operation at the surface not obvious but perhaps possible (K2K) 
\item Could be promising for high angle lowest energy (sub-GeV) beams, but 
\item Monte Carlo studies show 
$\nu_e$ identification above 2 GeV compromised due to inability of detector 
to discriminate between high energy neutral current $\pi^0$ production, and 
charged current $\nu_e$ interactions.  
\item R\&D efforts being pursued elsewhere already for JHF to HK, which 
include developing cheaper and more robust photodetectors.  This won't 
change the background rejection capabilities, however.  
\item Since individual particle energy resolution is not a limiting factor, the 
AQUARICH technology is not likely to have very different conclusions 
than regular water Cerenkov devices.  
\end{itemize} 

Conclusions about Liquid Argon TPC's:  
\begin{itemize} 
\item Monte Carlo Studies show this to be the most efficient detector for 
keeping signal and rejecting background 
\item Cosmic ray studies in Pavia show that backgrounds at the ground level 
are manageable assuming acceptable data handling capabilities. 
\item Economies of scale and experience of Liquid Natural Gas industry promising for a large
(phase III) single-volume detector. 
\item Need to verify that particle identification works as well as predicted
in simulations--this could be a promising phase III detector, but we strongly 
recommend placing a prototype detector in a neutrino beam which could prove the 
performance in the first few radiation lengths of a neutrino interaction.  
\end{itemize} 

Conclusions about Fine-Grained Calorimetry: 
\begin{itemize} 
\item Monte Carlo studies show that for a $\sim 2$ GeV off-axis neutrino
beam, this detector has adequate background 
discrimination and energy resolution, and the processes that generate 
the signals are well-understood (thresholds well below those for 
water Cerenkov, for example, and there's a long history in the field of 
sampling calorimetry).  

\item Low Z absorber would provide the maximum amount of mass per 
readout plane, but low density induces large separations between consecutive
readout planes.  Backgrounds induced by operation at the surface must be verified.  

\item Different readout technologies have different risks associated with them: 
\begin{itemize} 
\item RPC's:  possibly the cheapest readout per $m^2$, but operational difficulties
have been encountered in the past. 
\item Streamer Tubes:  are likely to be the next cheapest readout.
\item Liquid or Solid Scintillator is the easiest to operate, no tricky gas or 
high voltage systems to build.  
\begin{itemize} 
\item Depending on light collection technique, the integration time could 
be quite long, implying bigger cosmic ray problems. 
\item Minimum R\&D, can use much of what was learned while designing MINOS. 
\item Gains in recent past to reduce fabrication costs for solid scintillator 
\item Liquid scintillator would be easy to install in situ.  
\item R\&D on solid scintillator currently being performed by the K2K 
collaboration for a new near detector.
\end{itemize} 
\end{itemize} 

\item Different absorber ideas have different risks associated with them: 
\begin{itemize} 
\item Is the cost of containing the water for a water-absorber detector prohibitively high? 
\item Would particle board warp too much to be acceptable for housing detector elements? 
\item Can any solid low z material provide enough mechanical support for readout?  
\end{itemize} 

\item Finally, before one embarks on a full-scale construction of a fine-grained  
calorimeter, one should certainly produce a prototype, where 
at least one dimension of the prototype would be the size of a single 
module.  
\end{itemize} 

There are a few issues which must be addressed regardless of detector 
technology:  for example, what is the the optimal segmentation that is required
to get an acceptable neutral current rejection factor? Also, does the detector 
technology respond as predicted to charged particle beams?  \\


\newpage

{\bf Recommendations} \\
\par For phase II, we specifically recommend focused R\&D on fine-grained 
calorimetry:  this technique appears to have the smallest amount of risk 
associated with it, and although there are several options for absorber and 
readout technology, the outstanding issues are largely engineering ones, 
and can be addressed relatively quickly.  

\par For both phases, we will need to improve our understanding of 
neutrino interactions in the NuMI Off-axis energy regime.  
In phase II this is critical to get to the 
best precision on measuring the $\nu_\mu$ 
disappearance probability, and in phase III this will be essential to 
optimize the design of what is likely to be a $>100$M\$ detector.  We 
therefore recommend that as early as phase I that there be a program 
established to study neutrino interactions in a location 
underground at the NuMI beamline facility.


\par For phase III, large water Cerenkov detectors or liquid
argon offer scaling advantages.  In addition to sensitivity for
$\theta_{13}$, if placed underground,
such detectors would be sensitive to proton decay
and other topics of underground physics.  Since the 
time scale for phase III R\&D will take longer, it is important that 
this effort start now.  We recommend building a small prototype to test
in (but slightly off the axis of) the NuMI beamline, somewhere in the 
near detector hall.  

\par Finally, the most sensible path to the physics
is not simply to 
improve the far detector's size and/or performance.  
Investments in both the proton source (as early as phase I) and the 
beamline itself (phase III) will improve the experiment's sensitivity dramatically, 
and in a more economical way than by simply increasing the detector size.  


\par The writers of this report look forward to joining
the R\&D programs and collaborations which are forming to pursue
future neutrino initiatives.  
\tableofcontents
\setcounter{page}{0}
\section{Introduction}

Although we as a field have been trying to detect neutrinos for a 
very long time, the different techniques we use are highly constrained 
by the fact that neutrinos are so very weakly interacting.  As an example, 
consider the historic first and the most recent reactor neutrino detector: 
Reines and Cowan used 17 tons of liquid scintillator instrumented with 
phototubes to see the neutrinos from the Savannah River reactor, and KamLAND 
is using 1000 tons of liqiud scintillator instrumented with phototubes 
to see neutrinos from reactors located all over Japan.  Of course the 
light collectors have improved and expanded, the timing, electronics, 
and data acquisition has improved, but the fundamental technique for 
studying reactor neutrinos has not changed.  

Just as the topology of neutrino interactions changes dramatically 
as the neutrino energy increases, so too does the detector technology. 
At the high energy limit (above a few GeV), 
neutrinos simply break apart the nucleus 
and one can measure the incoming neutrino energy from a charged current
interaction by calorimetric sampling 
measurements of the final state particles, 
which are classified as either the outgoing lepton, or the ``hadronic
shower''.  Again, the MINOS detector today is not fundamentally different
from the E1A experiment at Fermilab, which measured neutrino interactions
calorimetrically with mineral oil-based scintillator 
and spark chambers, followed by an iron spectrometer.  MINOS combines the 
spectrometer and calorimeter functions, 
but the detector concept has not changed.   

In this document we will discuss detectors which are being considered 
for use in the 1-3 GeV regime, in other words, the regime between the 
two extremes described above.  This regime is suitable for 
experiments which are off the axis of the NuMI beamline, which are 
very well-positioned to take the next steps in neutrino oscillation 
physics.  

The recent compelling evidence for neutrino oscillation,
and hence neutrino masses and lepton mixing, in solar
and atmospheric neutrino experiments, has
 led to
an intensive program of experimentation to explore further this physics.
Fermilab is strongly positioned in this program with the MiniBooNE experiment
now beginning data-taking \cite{miniboone} and the MINOS experiment under
construction \cite{minos}.  MINOS will use a neutrino beam with an energy of
order a few GeV, traversing a pathlength of $L = 735$ km, from Fermilab to
Soudan, MN and will perform several very important measurements, including (i)
$\nu_\mu$ charged-current event rate and energy spectrum, which will check the
results of the atmospheric neutrino experiments and is expected to measure the
values of $\sin^2 2\theta_{23}$ and $|\Delta m^2_{32}|$ to about 10 \%, and
(ii) the $\nu_\mu$ neutral-current event rate, which will provide a cross-check
on the oscillation fit and put constraints on the involvement of light
electroweak-singlet neutrinos in the oscillations.  The proton beam intensity
is anticipated to be between about 0.25 MW and 0.4 MW.  The above sensitivities
are based on 10 kton-yrs of data, i.e. about 2 yrs with the 5 kton MINOS
detector.  The MINOS program is nicely complementary to the CERN-Gran Sasso
neutrino program, in which the OPERA experiment is designed to explicitly
detect $\tau$ appearance \cite{cngs}.  In Japan, the rebuilding of
Super-Kamiokande is almost complete, 
and the K2K experiment is expected to begin
running again in late 2002.  Beyond this, there is an ambitious long-baseline
neutrino oscillation experiment that plans to use of an intense 
$\nu_\mu$ beam
from the Japan Hadron Facility JHF which will traverse a distance of 295 km to
the fully rebuilt Super-Kamiokande detector \cite{jhf} (JHF-SK).  This beam will
have an energy $\ltwid 1$ GeV and be produced by an intense 0.77 MW proton beam
at JHF.  The JHF-SK experimental program envisions measurements of $\nu_\mu$
disappearance to get high-precision determinations of $|\Delta m^2_{32}|$ and
$\sin^2 2\theta_{23}$ and a search for $\nu_\mu \to \nu_e$ oscillations down to
a sensitivity of $\sin^2 \theta_{13}$ of 0.0015 \cite{jhf} (see further below).
Assuming requisite funding for beamlines, etc. the JHF-SK experiment expects to
start commissioning in 2007-2008.  Later running with a $\bar\nu_\mu$ beam is
also planned, and consideration has been given to a second phase of the
JHF-Kamioka neutrino program involving the construction of a very large 1 Mton
water Cherenkov detector, Hyper-Kamiokande.

Given the large investment that the U.S. high-energy physics community has made
in MINOS, and the fact that, together with MiniBooNE, it will be the center of
the domestic U.S. accelerator neutrino program during the next 10 years, there
is strong motivation for planning upgrades and extensions of this experiment.
Several studies have discussed the physics that could be accessed with a
higher-intensity conventional neutrino beam, also involving upgrades to the
proton intensity, at Fermilab \cite{fnalsuperbeam}-\cite{pdphysics}.  Earlier
related efforts studied the physics reach of a neutrino factory, including the
U.S. studies \cite{fnal_nufact,bnl_nufact} and related studies in Europe and
Japan.  In the U.S. an effort is also underway to study the physics potential
of long-baseline neutrino oscillation experiments using the BNL AGS, upgraded
from 0.14 MW to 0.5 MW, with a very large, multi-hundred kton water Cherenkov
far detector \cite{bnl_nufact,bnl_superbeam}.

Here we address research and development for 
a second far detector using the
NuMI beam and optimized for the 
study of $\nu_\mu \to \nu_e$ oscillations.  This is known as the search
for $\theta_{13}$ (or $U_{e3}^2$).  As described in Section~2, this is
the only element of the MNS matrix which has not been measured (only
upper limits are presently available).  Measurement of a non-zero
$\theta_{13}$ is required to measure the neutrino mass 
hierarchy through matter effects, and ultimately to search 
for CP violation in future neutrino long-baseline experiments.  

\par
The MINOS far detector is comprised of magnetized iron slabs with plastic
scintillator and is a coarse-grained sampling calorimeter with muon ID and
momentum measurement.  It can provide some modest sensitivity to $\nu_\mu \to
\nu_e$ oscillations but is not optimized for 
this.  A major background is
neutral current $\pi^0$ production, since it 
is difficult to distinguish the
shower produced by the $\pi^0$ from the 
shower produced by an electron.  The
search for $\nu_\mu \to \nu_e$ oscillations motivates 
planning and constructing
a second, fine-grained, far detector which 
would take advantage of the NuMI
beam.  To reduce the high-energy part 
of the neutrino flux and thereby reduce
backgrounds to this search due 
to neutral current $\pi^0$ production, 
the use of an off-axis position for this second far detector becomes
crucial.

The physics goals of this program would thus include the following: 

\begin{itemize}

\item 

Further measurement of $\nu_\mu$ disappearance and determination of $|\Delta
m^2_{32}|$ and $\sin^2 2\theta_{23}$ to higher accuracy. 

\item 

Improving the reach for a non-zero probability for 
$\nu_\mu \to \nu_e$ of about a factor of 10 past the 
CHOOZ limit, or a probability sensitivity of about 0.3\%.  
(comparable to the JHF-SK program\cite{jhf}).  By careful choice
of baseline and energy, this measurement, when combined with 
a possible JHF measurement, could start to take the next important
steps of addressing the mass hierarchy and CP violation.  

\item

Given that some interesting baselines are quite long, of order $10^3$ km
(assuming that the typical neutrino energies explored will be above 1~GeV), the
study of matter effects (which is possible if one runs both neutrino and
antineutrino beams) should allow the determination of the neutrino mass
hierarchy. Such a measurement cannot be performed by the JHF-SK experiment in
Japan, as currently envisioned \cite{jhf}. It should be noted that this is
obtainable even if the solution to the solar neutrino puzzle is not in the LMA
region \cite{pd}.

\item 

Ultimately, perhaps, a measurement or limit on 
the CP-violating phase $\delta$.  This
will require analysis of certain parameter ambiguities \cite{ambiguities} and
of matter effects \cite{msw}-\cite{cp}, and the sensitivity will depend on the
pathlength(s) of the experiment.

\end{itemize}

Given this possible physics potential, there is
more than enough motivation to
carry out an R\&D study of various promising
types of detectors and simulations
of their response.  A study group has been
meeting at Fermilab for the past
year to discuss this.  Recently one
report was submitted \cite{para}.  We
believe that it is clear that this
work should be pursued with further R\&D.

\par The organization of this report is as follows:  
Theoretical Motivations for future neutrino experiments are outlined
in Section~\ref{sec:theory}.  Comments about the beam spectrum and the 
features of an off-axis beam are
given in Section~\ref{sec:offaxis}.  Section~\ref{sec:detector}
discusses issues such as energy resolution which are common to
any detector technology.  Sections~\ref{sec:absorption} 
and \ref{sec:sampling}
includes
subsections on a variety of possible detector choices.  Where
appropriate, a discussion of future R\&D for each detector choice
is made in a separate subsection.  Total absorption detectors 
considered include Liquid Argon, Water Cerenkov Detectors, and Water
RICH detectors.  Sampling detectors discussed readouts which include
solid scintillator, liquid scintillator, limited stream tubes and
resistive plate chambers.  In Section~\ref{sec:other}, we discuss 
a number of other R\&D issues which cannot be neglected, but do not
concern a specific detector technology, such as the importance of 
understanding both cosmic ray backgrounds and neutrino cross sections,
as well as target survival issues, the NuMI lattice,
and possible proton intensity upgrades.  Section~\ref{sec:summary}
summarizes the salient features and outstanding issues with each 
detector technology.  
\section{Theoretical Motivations} 
\label{sec:theory}
In a modern theoretical context, one generally expects nonzero neutrino masses
and associated lepton mixing.  There is currently strong experimental evidence
for neutrino masses and mixing.  One source of this evidence is from solar
neutrino experiments, most recently, Super-Kamiokande and SNO \cite{sol}.  This
data can be fit by oscillations of the solar $\nu_e$'s into $\nu_\mu$ and
$\nu_\tau$, with the relevant $\Delta m^2_{21} = m(\nu_2)^2 - m(\nu_1)^2 \simeq
3 to 20 \times 10^{-5}$ eV$^2$ (at 90\% confidence level) and mixing angle 
$\tan^2\theta_{12}\simeq 0.4$ \cite{sol_res}.  Strong evidence also comes from
atmospheric neutrino experiments, especially Super-Kamiokande \cite{atm}, with
confirming results from Soudan-2 and MACRO.  The atmospheric neutrino data can
be fit by $\nu_{\mu} \rightarrow \nu_\tau$ oscillations with $|\Delta m^2_{32}|
\simeq 1.6 - 3.9 \times 10^{-3}$ eV$ ^2$ (also at 90\% confidence level) 
and maximal mixing, $\sin^2 2 \theta_{23} =
1$ \cite{atm}.  A pioneering long-baseline accelerator neutrino oscillation
experiment, K2K \cite{k2k}, has obtained data consistent with Super-Kamiokande
results.  The sum of data from these experiments can be explained in terms of
oscillations involving the three neutrinos $\nu_e$, $\nu_\mu$, and $\nu_\tau$,
members of electroweak doublets.  This data excludes light electroweak-singlet
(``sterile'') neutrinos as playing a large role in the oscillations.  There is
also a reported observation of $\bar\nu_\mu \to \bar\nu_e$ oscillations by the
LSND experiment \cite{lsnd}, which is not confirmed, but also not completely
excluded, by the similar KARMEN experiment \cite{karmen}. The LSND claim will
be tested by the MiniBooNE experiment at Fermilab \cite{miniboone}, which is
currently in an early commissioning phase, and expects to present definitive
results in about two years.

In the standard model generalized to include neutrino masses and without mixing
with light sterile neutrinos, the weak leptonic charged current has the form
$J_\lambda = \bar\nu_L U \gamma_\lambda \ell_L$, where the vector of neutrino
mass eigenstates is $\nu =(\nu_1, \nu_2, \nu_3)$ and the analogous vector of
charged lepton mass eigenstates is $\ell = (e, \mu, \tau)$.  The $3 \times 3$
unitary lepton mixing matrix $U$ depends on three Euler rotation angles
$\theta_{12}$, $\theta_{13}$, and $\theta_{23}$, and a phase $\delta$, all of
which can be probed in neutrino oscillation experiments, and (potentially) two
other Majorana phases which cannot be directly tested.  For $|\Delta
m^2_{23}|\gg\Delta m^2_{12}$ and ignoring matter effects, the probability for
$\nu_\mu \to \nu_\tau$ is
\beq
P(\nu_\mu \to \nu_\tau) = 
\sin^2(2\theta_{23})\cos^4 \theta_{13} \sin^2 \phi_{32}
\label{pnumunutau}
\eeq
where
\beq
\phi_{ij} = \frac{\Delta m^2_{ij}L}{4E}
\label{phi_ij}
\eeq
and $\Delta m^2_{ij} = m(\nu_i)^2 - m(\nu_j)^2$.  Since the best fit to the
Super-Kamiokande data has maximal mixing, it follows that $\theta_{23} \simeq \pi/4$ and
$\theta_{13} \ll 1$.  The latter constraint also arises from the CHOOZ reactor
antineutrino experiment \cite{chooz}. The CHOOZ limit is dependent on the input
value used for $|\Delta m^2_{32}|$; for the current central value $2.5 \times
10^{-3}$ eV$^2$, this is $\sin^2(2 \theta_{13}) < 0.11$, while for $|\Delta
m^2_{32}|=2.0 \times 10^{-3}$ eV$^2$, it is $\sin^2(2 \theta_{13}) < 0.18$
\cite{chooz}.

In vacuum (e.g. \cite{bargersuperbeam,tmo}) 
\beqs
P(\nu_\mu\to \nu_e)&=&2 \sin(2\theta_{13}) s_{23}c_{13}s_{12}
(s_{12}s_{23}s_{13}-c_{12}c_{23} c_\delta)\sin^2\phi_{32} \nonumber \\
&+&2 \sin(2\theta_{13}) s_{23}c_{13}c_{12}
(c_{12}s_{23}s_{13} +s_{12}c_{23} c_\delta)\sin^2\phi_{31} \nonumber \\
&-& 2 \sin(2\theta_{12}) c_{13}^2 \biggl [
s_{12} c_{12}(s_{13}^2 s_{23}^2-c_{23}^2)+s_{13} s_{23} c_{23}
(s_{12}^2-c_{12}^2)c_\delta \biggr ] \sin^2\phi_{21}
\nonumber \\
&+& \frac{1}{2} \sin(2\theta_{12}) \sin(2\theta_{13})\sin(2\theta_{23})c_{13}
s_\delta \bigg[ \sin \phi_{32}\cos \phi_{32} \nonumber \\
&-&\sin\phi_{31}\cos\phi_{31} +\sin\phi_{21}\cos\phi_{21} \bigg],
\label{numunuefull}
\eeqs
while the same expression applies for $P(\bar{\nu}_\mu\to \bar{\nu}_e)$ 
after performing $\delta\rightarrow -\delta$. 

In the limit $\Delta m^2_{12}\rightarrow 0$ (this applies if the solution to
the solar neutrino puzzle is in the LOW region), 
\beq
P(\nu_\mu \to \nu_e) = \sin^2(2\theta_{13})\sin^2 \theta_{23} \sin^2 \phi_{32},
\label{pnumunue}
\eeq
in vacuum. 
Hence, since one knows $\sin^2 \theta_{23}$ and $|\Delta m^2_{32}|$ moderately
well from the atmospheric neutrino data (and these determinations will become
more accurate with MINOS), it follows that a measurement of, or search for,
$\nu_\mu \to \nu_e$ will immediately yield the value of, or a limit on, 
$\sin^2(2\theta_{13})$.  

In general, however, (this applies if the solution to the solar neutrino puzzle
is in the LMA region, which is favored by the current solar neutrino data)
several of the terms in Eq.~(\ref{numunuefull}) can be of the same order of
magnitude. In this case, a signal for $\nu_{\mu}\to\nu_e$ (or lack thereof)
does not easily translate into a clean measurement of (or upper bound on)
$\theta_{13}$, unless all other parameters are precisely known. It should also
be noted that, if matter effects \cite{msw} are important, the survival
probabilities will also depend on another observable, namely the neutrino mass
hierarchy, which we choose to parametrize by the sign of $\Delta m^2_{23}$.
(Whether matter effects are visible depends on the energy of the neutrino beam
and on the baseline length; they tend to be more important for higher energies
and longer pathlengths.)  This ultimately implies that, in order to determine
or constrain $\theta_{13}$, it is necessary also to determine several other
neutrino oscillation parameters, including the neutrino mass hierarchy and the
CP-odd phase $\delta$. This can only be achieved by comparing different
channels (neutrino and antineutrino oscillations) and/or different neutrino
beams and different baselines.

Turning the picture around, the study of the subleading $\nu_{\mu}\to \nu_e$
transition ultimately enables (indeed, requires) one to explore leptonic CP
violation.  This leptonic CP violation involves the phase $\delta$ (and two
Majorana phases which cannot be directly probed with neutrino oscillations) and
is measured by the rephasing-invariant quantity \cite{j} determined via the
product $Im(U_{ij}U_{kn}U^*_{in}U^*_{kj})$,
\beq
J = \frac{1}{8}\sin(2\theta_{12})
\sin(2\theta_{23})\sin(2\theta_{13})\cos \theta_{13} \sin\delta
\label{j}
\eeq
Because one knows that $\sin^2 2\theta_{23} \simeq 1$ from the atmospheric data
and $\sin^2 2\theta_{12} \simeq 0.8$ in the LMA fit to the solar data, it
follows that $J$ in the leptonic sector may be much larger than its value of
few $\times 10^{-5}$ in the quark sector.  Furthermore, the values of $|\Delta
m^2_{32}|$ and $\Delta m^2_{21}$ are such that the corresponding three neutrino
mass eigenstates are sufficiently large and nondegenerate so as to allow
possible experimental exploration of leptonic CP violation in long-baseline
neutrino oscillation experiments.  It should be noted that comparison of
neutrino and antineutrino disappearance also allows one to probe violation of
CPT \cite{cptv}.

\section{NuMI Off Axis Neutrino Spectra} 
\label{sec:offaxis}

The NuMI beamline is designed to produce relatively wide-band neutrino
beams of peak energies ranging between 3.5 and 14GeV.  Its
focusing system consists of two horns, which, depending on their
relative  spacing and their spacing from the NuMI target, can focus
pions of   varying momentum ranges.  Because muon neutrinos come
predominantly  from 2-body meson decays, the neutrino
energy at a far detector is determined simply by geometry and
kinematics:  
$$ E_\nu = \gamma m_h *
\frac{m_h^2-m_\mu^2}{m_h^2}*\frac{1}{1+\theta^2\gamma^2} 
$$ where
$\gamma$ is simply the parent meson's relativistic boost, $m_h$ is 
the parent meson's mass, and $\theta$ (assumed to be $<<1$) is the angle
in radians between the detector and the parent meson's direction of flight. 

\begin{figure}[htp]  
\begin{center}
\includegraphics*[width=4in]{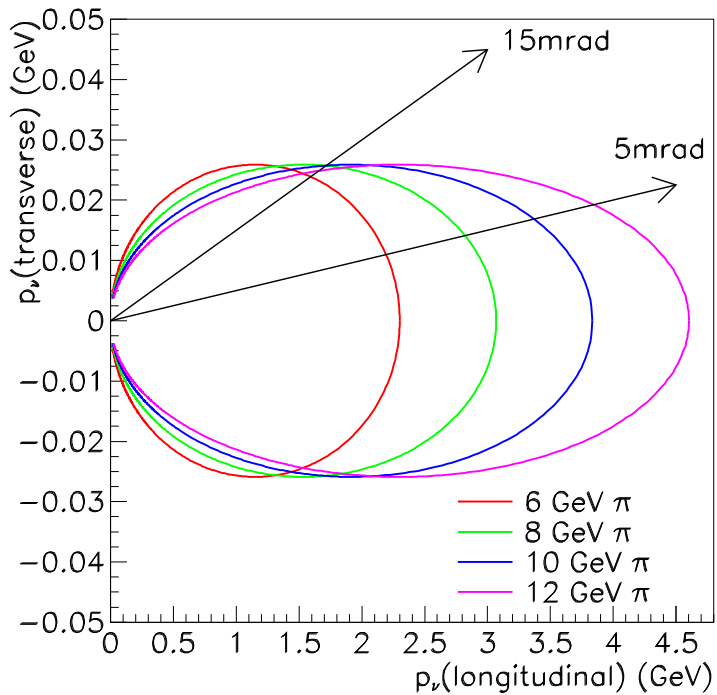}
\parbox{5.5in} 
{\caption[ Short caption for table of contents ]
{\label{fig:shooff} Neutrino Energy for $\pi\to \mu\nu_\mu$  decays 
as a function of angle from the pion momentum.  
}}
\end{center}
\end{figure}
\begin{figure}[htp]  
\begin{center}
\includegraphics*[width=.9\textwidth]{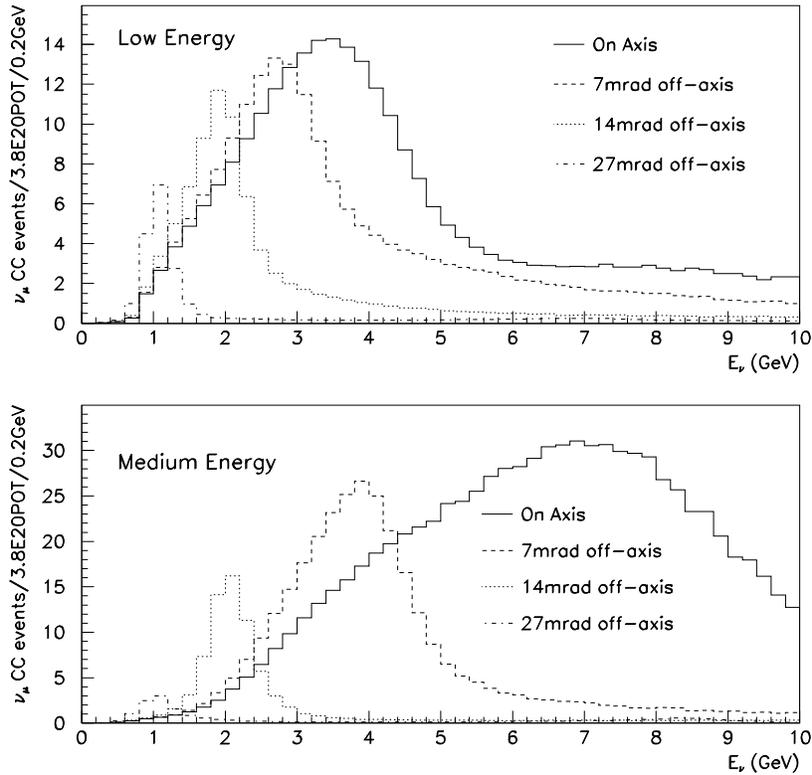} 
\parbox{5.5in} 
{\caption[ Neutrino Event Spectra off the NuMI beamline axis]
{\label{fig:oaspect} Neutrino Energy Spectra for the low (top) and 
medium (bottom) energy configurations at NuMI, for a detector 
at a distance 735km from Fermilab but at various angles away 
from the NuMI beamline axis.  
}}
\end{center}
\end{figure}

For a perfectly-focused meson beam, and a detector at $\theta=0$, 
the neutrino energy is simply proportional to the 
parent meson's energy.  However for a detector at a non-zero angle, 
the energy is considerably less, and no longer proportional.  Figure 
\ref{fig:shooff} shows the transverse versus longitudinal neutrino momentum 
for perfectly focused pions of different energies.  The neutrino energy is 
simply the length of the line from the origin to any point on the circle.  
Note that at a particular
angle, the pions of many different energies contribute neutrinos of the 
same energy--at this angle the flux from two-body decays has the narrowest
energy spread.  The intrinsic $\nu_e$ background, since it arises 
from three-body decays, is not peaked in this way, so the relative signal
to background at this angle is maximized.  For 
detectors which are 0 and 15mrad from the beamline axis, the neutrino spectra 
which can be produced by the NuMI beamline is shown in figure 
\ref{fig:oaspect}.  Notice also that the actual flux of neutrinos at 
2GeV is higher for
an off-axis beam, which means the average oscillation probability 
($\sin^2 (\Delta m_{32}^2L/4E)$ for that beam is in fact higher as well.  

\section{Detector Issues and Comparison}
\label{sec:detector}
\subsection{Introduction}
As was discussed in the previous sections, the next big step for 
our field to take in neutrino oscillations, is to see if there are 
transitions between muon and electron neutrinos at the atmospheric
neutrino mass splitting.  For conventional neutrino beams, this means
at the minimum building a detector that can distinguish between 
electrons and muons.  However, given that the oscillation probability 
we are trying to measure has already been limited to less than 5\% at 
90\% confidence level by CHOOZ, and that the intrinsic electron neutrino 
contamination in a conventional beam can be a few percent, detectors will 
have to do significantly more.  Furthermore, not only are 
there likely to be other final state particles
present in the neutrino charged current interaction, confusing the 
signal, but neutral current interactions, in which there is no final 
state lepton, can also provide a background through
the production of neutral pions.  

\subsection{Detector Requirements for $\theta_{13}$}

\begin{figure}[h]  
\begin{center}
\includegraphics*[width=4in]{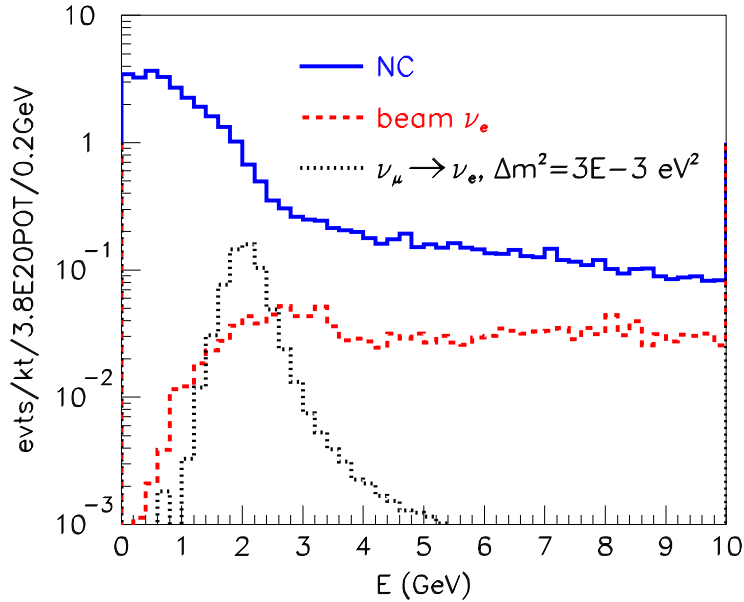}
\parbox{5.5in} 
{\caption[ Visible energy distribution for a 15mrad off-axis
NuMI beam ]
{\label{fig:backgds} Visible energy distribution for a 15mrad off-axis
beam:  for a signal at $\delta m_{32}^2=3 \times 10^{-3}eV^2$, the 
intrinsic $\nu_e$, and the neutral current background, assuming 
no particle identification, and perfect energy resolution.  
}}
\end{center}
\end{figure}

A summary of the detector challenge in the NuMI Off-axis beam 
can be found in Figure \ref{fig:backgds}.  This plot shows the 
true visible energy distribution for different possible events 
at the detector:  a signal at $\Delta m_{32}^2=3 \times 10^{-3}eV^2$, the 
intrinsic $\nu_e$, and the neutral current background, assuming 
no particle identification.  Note that in order to reduce the 
neutral current background to the level where it is comparable 
to the signal at the CHOOZ limit, in the limit of perfect energy resolution, 
one must have a background rejection factor of about 4 or 5.  As you 
add in energy resolution you need even better background rejection, since the 
backgrounds are flat or steeply falling in energy, while the signal is very 
peaked.  

\subsubsection{Neutral Current Rejection in a superbeam}

     The dominant background process that any detector must be 
ready to cope with is that of neutral currents.  Neutral pions 
are often produced in neutral current interactions, and the two 
photons to which they decay can easily be mistaken for an electron, 
in certain detectors.  In this section we will discuss how different
detectors might see these neutral pions.  

\begin{figure}[htb]  
\begin{center}
\includegraphics*[width=4in]{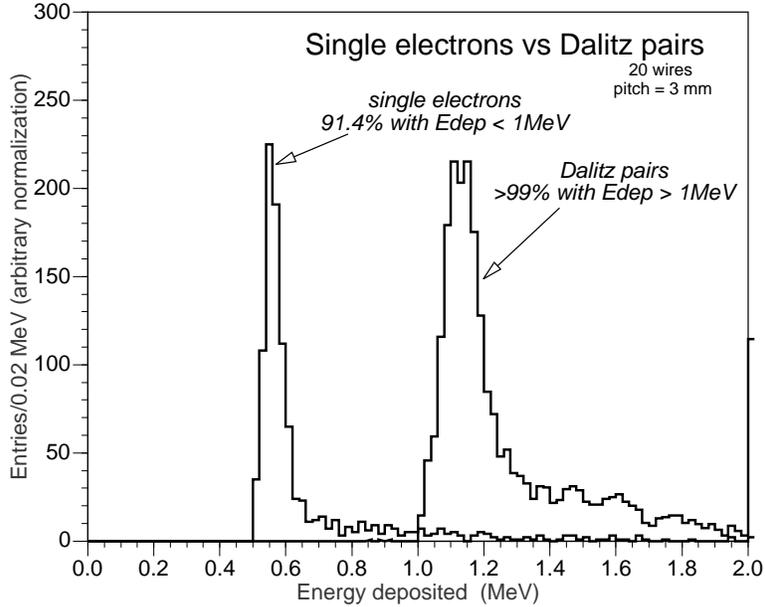}
\parbox{5.5in} 
{\caption[ Energy loss distribution for converted photons and 
electrons in the ICARUS detector ]
{\label{fig:eldal} Energy loss distribution in the first two 
radiation lengths of a neutrino interaction for converted photons and 
electrons in the ICARUS detector.  }}
\end{center}
\end{figure}

     If one had an extremely fine-grained detector, then discriminating 
between electrons and neutral pions would be straightforward:  electrons
have only one charged particle with an electromagnetic shower, while
pions will decay to photons, which convert to two electrons.  So 
in a liquid argon detector, for example, one can simply look at the 
energy lost by a track 
in the first few radiation lengths after the event vertex, and
converted photons will have twice the energy loss as single
electrons.  This is 
shown in Figure \ref{fig:eldal}, which is from the ICARUS proposal.  
Presumably, if the electrons have enough energy to travel a few 
radiation lengths, then above that minimum energy this cut would be 
extremely efficient at removing the neutral current events, without
significant loss of signal.  

\begin{figure}[htp]  
\begin{center}
\includegraphics*[width=4in]{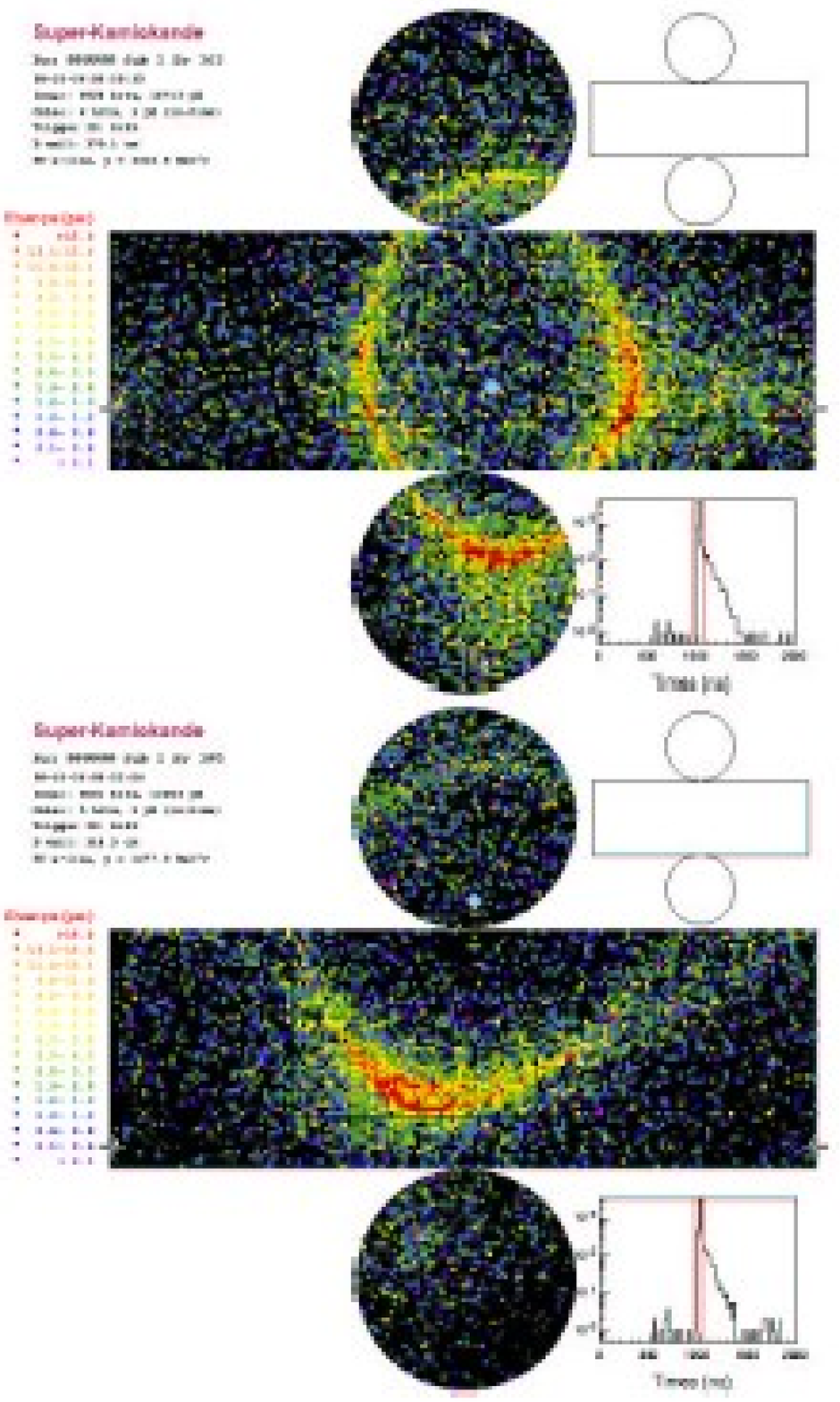}
\parbox{5.5in} 
{\caption[ Signal and Background events in a water cerenkov detector ]
{\label{fig:skevts} (top) Signal and (bottom) neutral current background
event displays for the Super-Kamiokande detector simulation.  The 
visible neutrino energy for both events is in the signal region.  }}
\end{center}
\end{figure}

     Still another way to discriminate between electrons and neutral pions
is to see the two photons separately, as can be done in the water cerenkov
or the aquarich technology.  For low energy $\pi^0$'s, the two photons 
are very well separated, and the only significant background occurs when 
there is a very asymmetric $\pi^0$ decay, producing only one 
electromagnetic shower.  As the pion energy gets larger, however, 
the two cerenkov rings from the photons get closer together, and then 
resolving two rings becomes too difficult given the intrinsic widths of the 
rings themselves.  Figure \ref{fig:skevts} shows the event displays in 
the Super-Kamiokande monte carlo for an electron neutrino charged current
interaction, and one for a neutral pion with two rings that are very 
close to overlapping.  

Even if a detector cannot discriminate two photons
separately, the electromagnetic shower from a $\pi^0$ may be wider 
than that from a signal electron.  In order to do this, the segmentation
of the detector must be finer than one Moliere radius $R_M$, which 
characterizes the width of an electromagnetic shower (on average, 
90\% of the shower's energy is contained in a cylinder of radius $R_M$).  
According to the PDG, the Moliere Radius for solids is 
well-approximated by the formulae 
$$
R_M = X_0 \frac{E_s}{E_c} , E_s = 21.2MeV, E_c=\frac{610MeV}{Z+1.24} 
$$ 
.  In table \ref{materials} we give the salient features of many 
of the materials considered for detectors in this document.  

\begin{table}[htb] 
\begin{center} 
\begin{tabular}{lccccc} 
          &     &            & \multicolumn{2}{c}{Radiation} & Moliere \\ 
          &     & Density    &  \multicolumn{2}{c}{Length}  & Radius \\ 
 Material &  $<Z>$  & $(g/cm^3)$ & $(g/cm^2)$ & $(cm)$ & $(cm)$ \\\hline\hline 
Argon     &    18    & 1.4         &  19.55  &   14   & 9.4    \\ 
Water     &    3ish  & 1           &   36.1  &   36   & 5.4      \\
Carbon    &     6    &  2.3        &   42.7  &   19   & 4.7      \\ 
Steel     &     26   & 7.9         &  13.84  &   1.8  & 1.7       \\
Plastic   &    3ish  &  0.7        &  43.7   &   62   & 9.4      \\ 
(Polystyrene) &        &           &         &        &    \\
\hline
\end{tabular} 
\end{center} 
{\caption[ Vital Characteristics of various materials ]
{\label{materials} Defining characteristics of various materials.  
The radiation length, given in $g/cm^2$, represents how much mass
one would get for a single detector plane.  The Moliere radius, given
in $cm$, indicates what the transverse segmentation would have to be 
better than for a fine-grained calorimeter.  For any of these materials, 
the effective longitudinal segmentation would have to be significantly
better than one radiation length. Reference:  K. Hagiwara {\em et al}, 
Physical Review {\bf D66}, 010001-1 (2002).   }}
\end{table} 

\begin{figure}[hbt]
\centerline{\psfig{figure=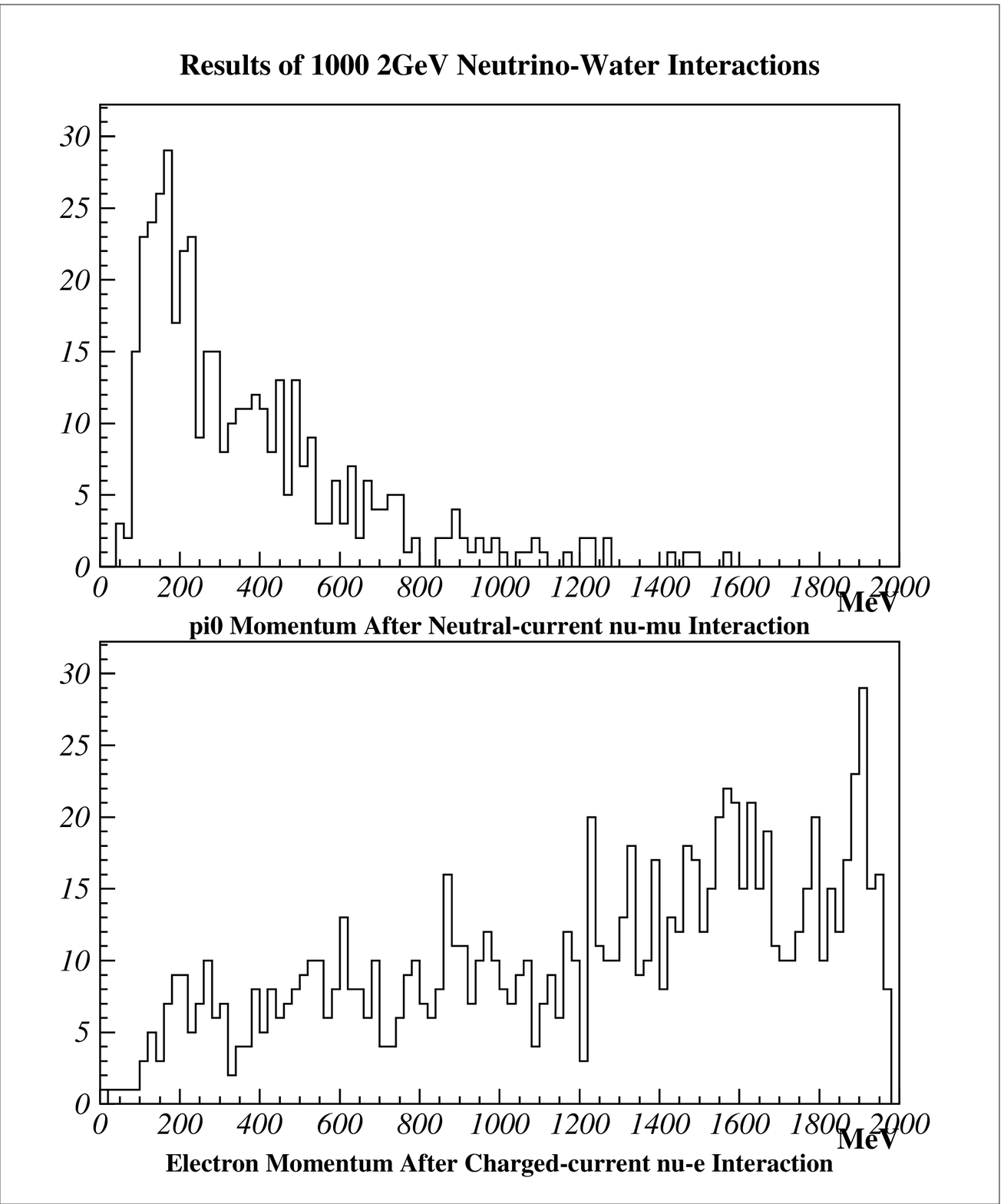,bbllx=0pt,bblly=33pt,bburx=536pt,bbury=629pt,width=.7\textwidth,clip=1}}
\caption[]{The momentum distributions of $\pi^0$'s and electrons produced
in neutrino-water interactions.}
\label{fig:momenta}
\end{figure}

     Finally, the one remaining difference between electron neutrino 
charged current events and neutral current events is their ``electron 
candidate'' energy distribution.  For real charged current interactions, 
the ratio of lepton energy to total energy is roughly flat (and peaked
towards one for antineutrino interactions!), while for neutral current
interactions, the distribution of $\pi^0$ energy is peaked at very low
energies.  Figure \ref{fig:momenta} shows the momenta for electrons
in charged current events in 2GeV neutrinos, as well as the neutral 
pion momentum distribution for neutral current events of 2GeV neutrinos, 
as generated by the NUANCE neutrino event monte carlo.  Although a cut on the
electron energy will not provide a large rejection factor compared to the
signal acceptance, it would still provide some discrimination.  

%

\subsubsection{$\nu_e$ Signal Efficiency, Background Rejection, and Mass} 

     For a given $\nu_\mu$ beam with a particular intrinsic electron 
neutrino contamination, there are various approaches one might take:  
one could try to build a very fine-grained detector with a high efficiency
which removes all the backgrounds, or one could try for something more 
coarse-grained, with the assumption that even if the background level
is higher, because the coarse-grained detector could presumably built 
to be more massive for the same amount of money.  

     Clearly the goal for any detector is to remove as much of the 
background as possible while keeping the signal efficiency high.  
However, because of the intrinsic electron neutrino background in 
the beam, it is not worthwhile to reduce the neutral current background
well below the intrinsic $\nu_e$ background at the expense of signal 
efficiency.  We therefore define the neutral current backgrounds for
different detectors by how large they are compared to the intrinsic
beam background under the oscillation peak.  Table \ref{tab:effbkg} 
shows the results for both signal efficiency and background rejection 
from different geant-based analyses which have used the NuMI off-axis 
beam at 15mrad as input.  

\begin{table}[tbh] 
\begin{tabular}{lcccc} 
 &  Signal  & NC & 
&  \\
Detector & Efficiency & fake rate & NC/$\nu_e$ 
& Reference \\ \hline\hline 

Liquid Argon TPC & 0.90 &  0.001     & $<0.1$ &  ICARUS TDR \\ 
Steel/Scintillator &  0.40  & 0.2\%  &  $\sim$1  &  \verb+hep-ph/0204208+ \\
Plastic/RPC        &  0.35  & 0.2\%  &  $\sim$1  & \verb+hep-ex/0210005+    \\
Water Cerenkov    &   0.24  & 1\%    &  2      &  this document         \\
\hline 
\end{tabular} 
\caption[]{Signal efficiency and NC backgrounds for different detectors--
note these numbers are approximate and represent current status at the
writing of this document.}
\label{tab:effbkg}
\end{table} 

     In order to understand how to get the largest reach on measuring 
$\nu_\mu \to \nu_e$ for a given 
investment in money, it is useful to first see how much mass one would 
need of these different kinds of detectors for comparable sensitivities, 
and then see what the cost is for these different detectors of varying 
sizes.  Furthermore, 
not all detectors' costs scale as their mass:  for example, 
for a water cerenkov device,
if one simply scaled up the Super-Kamiokande detector 
in all dimensions, a large part of the
cost would grow as the area of the vessel, since that determines the number
of phototubes required.  Finally, the real mass that counts here is the 
fiducial mass, which again is not a linear function of the detector mass
(see the following section).  

\clearpage
\begin{figure}[htb]  
\begin{center}
\includegraphics*[width=4in]{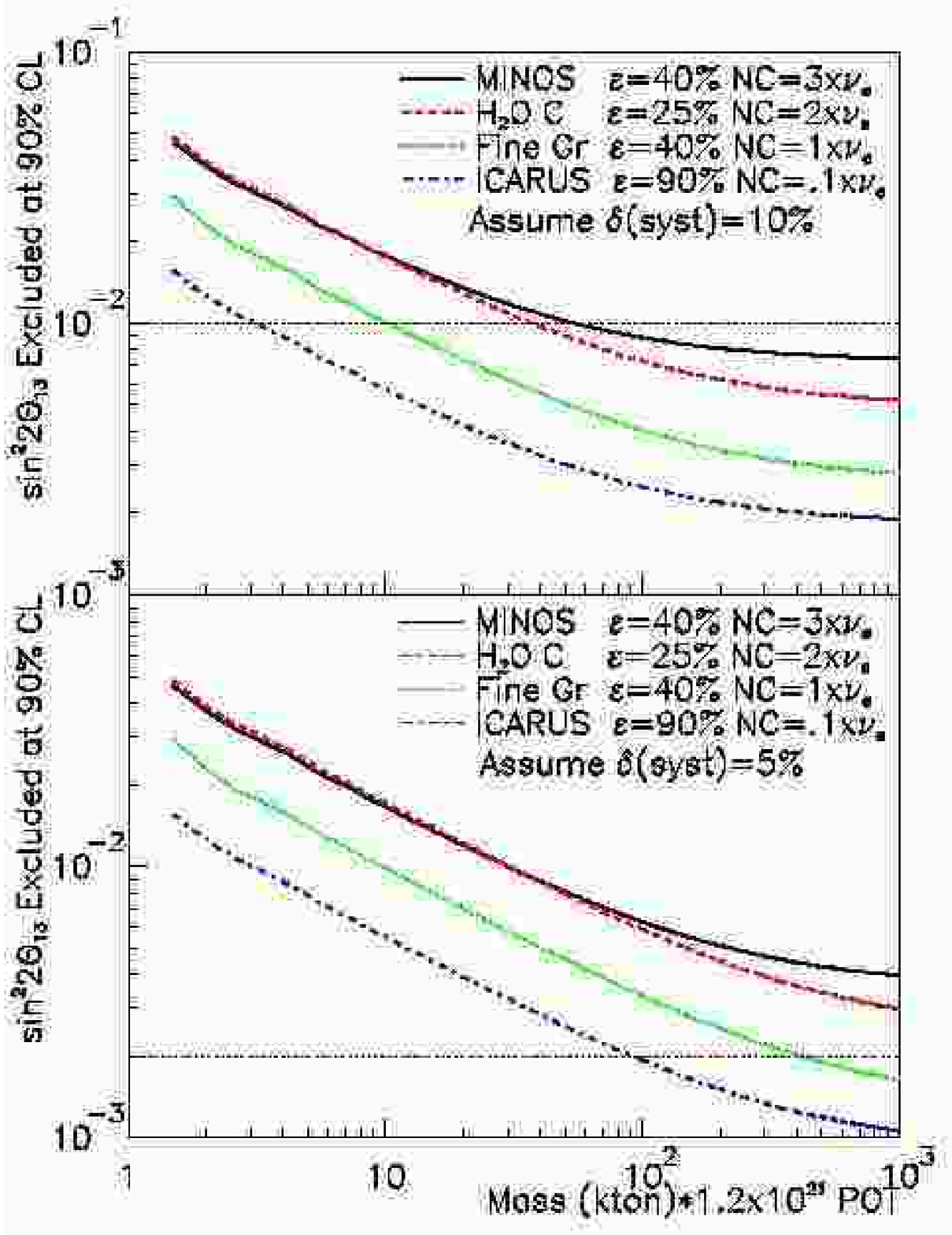}
\parbox{5.5in} 
{\caption[ Reach in $\sin^2 2\theta_{13}$ versus detector mass ]
{\label{fig:reachvmass} 90\% confidence level limit one could achieve
in the absence of a signal as a function of detector mass, for 
different detector assumptions.  The top plot assumes a 
background uncertainty of 10\% and the bottom plot assumes a 
background uncertainty of 5\%, where the bottom plot is relevant in 
the event of a proton driver upgrade, where the proton intensity would 
be increased by a factor of 5 above nominal.  
}}
\end{center}
\end{figure}

     Figure \ref{fig:reachvmass} shows the 90\% confidence level limit
that one could achieve in the NuMI off-axis beamline, for a detector
at 735km, 15mrad off axis, for different detectors as a function of 
detector mass.  This analysis assumes no solar mass term effects, 
i.e. the probability for $\nu_\mu \to \nu_e$ oscillations is simply 
$P=0.5\sin^2 2\theta_{13}\sin^2(\Delta m_{32}^2 L/2E)$.  Note that for a 
sensitivity which is a factor of 
10 past the CHOOZ limit, one would need
approximately 5 ktons fiducial of Liquid Argon, 20kton of a 
fine-grained calorimeter, and 80kton of a water cerenkov device.  
Note also that for a sensitivity which would be significantly better, 
one would start having significant systematic errors, and so one would 
need to plan on reducing those below 10\%, if one were to embark on a 
much larger detector or more powerful proton source.  

\subsection{Detector Requirements for $\Delta m_{23}^2, \theta_{23}$}

Although the primary motivation for an off-axis experiment is the 
search for a non-zero probability of $\nu_\mu$ to $\nu_e$ and its 
CP conjugate, it is important to remember that this experiment also 
has the potential to drastically improve the precision on the 
atmospheric neutrino parameters $\Delta m_{32}^2$ and $\theta_{23}$, 
through the disappearance measurements in $\nu_\mu$ and $\bar\nu_\mu$ 
beams.  The extent to which $\theta_{23}$ is different from $\pi/4$ has 
important constraints on understanding the underlying symmetry breaking 
which gives rise to neutrino oscillations in the first place.  Furthermore,
when 
we ultimately want to determine whether or not CP violation is present
in the lepton sector, degeneracies and
correlations between the measured probabilities and the mixing angles 
themselves~\cite{degenerate1,degenerate2} will require very precise 
disappearance as well as appearance measurements.  

\begin{figure}[tbp]
\epsfxsize=\textwidth\epsfbox{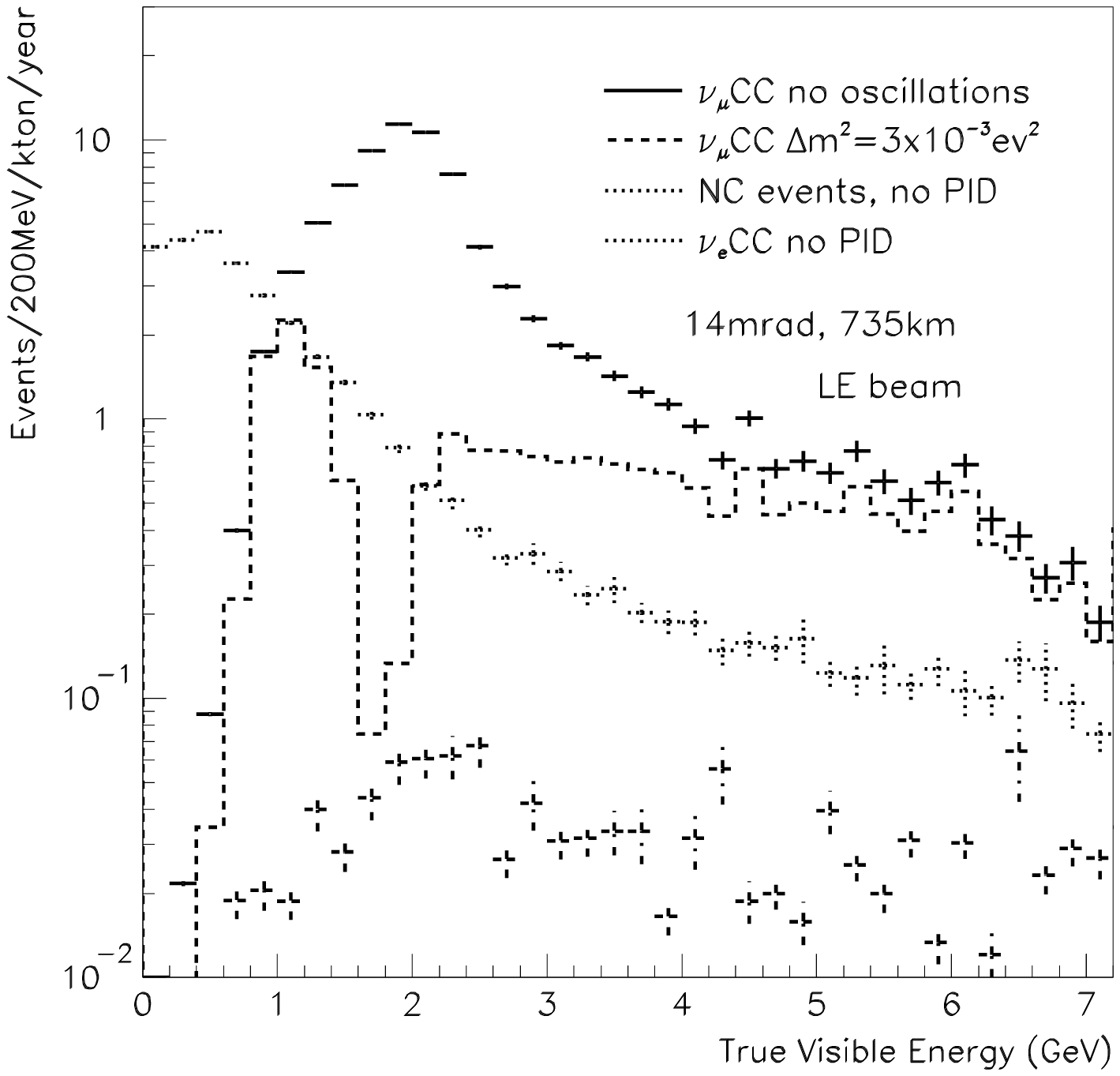}
\caption{$\nu_\mu$ signal candidates, $\dmsq=3\times10^{-3}$eV$^2$,
$\sin^22\theta_{23}=1$, shown with possible sources of background.
Note that backgrounds which cannot be eliminated must be subtracted 
in order to measure 
$\dmsq_{23}$ from the location of the dip and $\theta_{23}$ from the
depth of the dip.}
\label{fig:numu-disapp}
\end{figure}

So although (a) the disappearance probability is expected to be large based on 
atmospheric results, (b) the neutrino beam produced is predominantly
muon neutrinos, and (c) the fact that muons are easier 
to identify than electrons, there is still the same nagging issue of 
neutral current events.  Figure~\ref{fig:numu-disapp} shows the 
$\nu_\mu$ event rate for an off axis beam (in the low energy 
configuration), with and without oscillations, and also shows what 
the visible energy distribution is for neutral current events, assuming 
no particle identification but perfect energy resolution.
It is clear from this picture that the level
at which a detector can distinguish an outgoing muon from, for example, the 
most energetic 
outgoing charged pion from a neutral current event is very important. 
This will determine whether
or not the events ``in the dip'' will be charged current events which 
entered through detector energy resolution, or neutral current events, 
which of course are not affected by (active) neutrino oscillations.  
What is quite possible is that the most important uncertainty in  
this measurement is the neutral current background prediction, which 
is likely to be dominated by uncertainties in the cross sections, as will 
be discussed in Section \ref{cross}.  
Once particle identification cuts are made (based on a detailed understanding
of the detector response), this promises to be a very powerful constraint on 
the atmospheric parameters $\Delta m^2_{32}$ and $\sin^2 2\theta_{23}$.  

\subsection{Fiducial Versus Total Mass} 
While the physics return on any detector is proportional to the
useful fiducial mass, the total cost is usually proportional
to the total mass.  That ratio depends on the geometry of the neutrino
interaction, and the size and shape of the detector.  In general,
the fiducial efficiency is larger for a large detector.
We will consider
this ratio here for a liquid argon detector  constructed in the form 
of a right circular
cylinder of radius 
$r$ and height $h = 2 r$.
The volume is $V = 2 \pi r^3$, 
and the total mass of argon (density 1.4 metric tons/m$^3$) is
\begin{equation}
M_{\rm total}({\rm tons}) = 2.8 \pi r^3,
\label{lar2}
\end{equation}
for radius $r$ in meters.

Because the time projection chamber has 
electrodes at 250 kV, the instrumented
volume must be set back from the 
cryostat wall by some distance $ \approx
0.5$ m.  Therefore, the instrumented mass of argon is
\begin{equation}
M_{\rm instrumented}({\rm tons}) = 2.8 \pi (r - 0.5)^3
\approx \left( 1 - {1.5 \over r} \right) M_{\rm total}.
\label{lar3}
\end{equation}

\begin{figure}[htp]  
\begin{center}
\includegraphics*[width=4in]{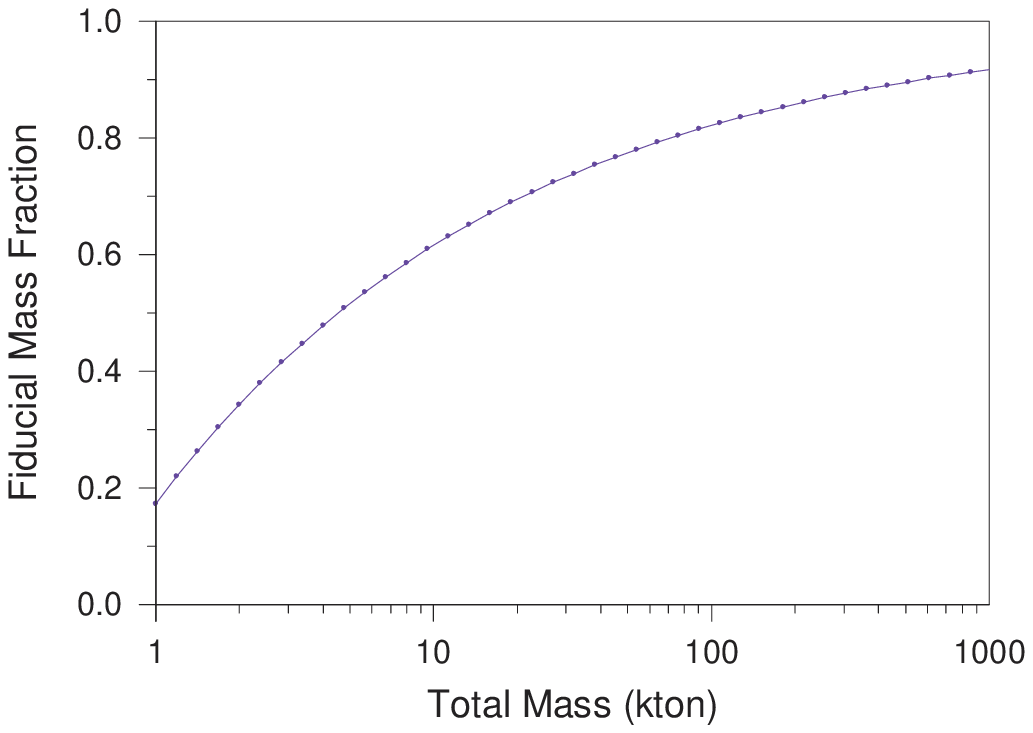}
\parbox{5.5in} 
{\caption[ Short caption for table of contents ]
{\label{lar_fid} The fraction of the total mass of liquid argon detector
that is in the fiducial mass for detection of electron neutrino appearance
events, according to expression (\ref{lar4}).
}}
\end{center}
\end{figure}

The fiducial volume is smaller than the instrumented volume because a neutrino
interaction must be well contained within the detector to be useful in the
physics analysis.  Taking the physics emphasis to be electron neutrino
appearance events, assume the characteristic volume of an interaction is a
cylinder of about 5 Moliere radii ($\sim 0.5 m$) and about 18
radiations lengths deep (2.5 m). The events should not start closer than,
say, 0.5 m from the edge of the instrumented volume, to insure that
they originate from a neutral particle.  Hence the radial depth of
the fiducial volume is less than that of the instrumented volume by
(2.5 + 0.5)/2 = 1.5 m, while the radial height and width are smaller
by 0.5 m.  These offsets must be combined with the high-voltage
offset of 0.5 m, leading to the expression
\begin{equation}
M_{\rm fiducial}({\rm tons}) = 2.8 \pi (r - 2) (r - 1)^2
\approx \left( 1 - {4 \over r} \right) M_{\rm total},
\label{lar4}
\end{equation}
which is illustrated in Fig.~\ref{lar_fid}.

For a sampling calorimeter, which is expected to be roughly square in 
cross section, one could proceed with a simpler argument.  
For a sampling detector, 
one generally assumes that 1 meter is required between the edge of 
the detector and the interaction vertex.  Consider first the loss from 
a cut in the transverse position of the vertex--for 
a 20kton detector which has a 
$10m \times 10m$ cross section, one is already losing (36\%) of the events 
from a transverse vertex position cut, while for a $20m\times 20m$ 
cross-section detector, the same 1m cut incurs a loss of only 10\% 
loss.  Now consider the cut in the longitudinal position of the vertex:  
for that same $20m\times 20m$ detector,
to accrue 22 total ktons, the detector would have to be about $55$m 
long (assuming a density of 1), or about 150 radiation lengths (assuming 
water).  To contain most of an electromagnetic shower, all accepted 
events would 
have to start some 18-20 radiation lengths upstream of the downstream end of 
the detector, which would again incur about a 10-13\% loss 
in fiducial acceptance.
Clearly an optimization for width and height for each 
absorber material is required, and for any detector technology 
considered it should 
be kept in mind that to contain all the hadronic as well as electromagnetic
showers, one would have to make considerably larger fiducial cuts.   
\section{Total Absorption Detectors}
\label{sec:absorption}
\subsection{Liquid Argon TPC}

Among the options for a large neutrino detector, a liquid argon time
projection chamber (TPC) \cite{Rubbia77} provides the greatest amount of
information, in the form of fine-grain tracking as well as 
total-absorption calorimetry, via a very simple mechanical structure which 
is therefore very cost-effective when implemented on a large scale.

The power of a liquid argon detector is especially noteworthy for
detection of charged-current electron-neutrino interactions of 0-2 GeV, 
where detailed tracking provides excellent rejection against
neutral-current muon-neutrino interactions with a final-state $\pi^0$.
Hence, it is the most effective detector per unit mass for $\nu_\mu \to \nu_e$ 
appearance measurements (of 
$\sin^2 2 \ttt$, the sign of $\Delta m^2_{23}$, CP violation, ...) 

\par
A liquid argon detector is a total absorption calorimeter with
time-projection readout via the signal of drifting electrons collected in crossed
planes of wires.
The effective pixel size is about $5 \times 5 \times 1$ mm$^3$,
compared to the radiation length of 14 cm and nuclear interaction length of 55 cm.
At a drift field strength of 500 V/cm, the drift velocity is about 1 mm/$\mu$s, so the
drift time over, say, 5 m would be 5 ms.  
Even if operated at the Earth's surface with
no shielding, a liquid argon TPC has only 
about 1 (localized) cosmic-ray track per
m$^2$ of horizontal surface per drift time, 
so events appear very clean.

\par
A liquid argon detector of 100 ktons or more also has competitive
capability for nucleon decay searches \cite{tm-01-03},
particularly because it has high efficiency for the decay $p \to K^+ \bar\nu_\mu$
that is favored in many SO(10) SUSY models \cite{Pati},
as well as for atmospheric, solar and supernova neutrino physics.  However,
pursuit of these additional physics topics will likely require the
detector to be sited underground, at considerable additional
expense.  Here, we emphasize a detector on the surface for use in a 
pulsed neutrino beam.

The reliability and stability of their electronic readout has led to the
use of liquid argon calorimeters in numerous electron and hadron beam
experiments over the past 20 years.  A liquid argon TPC differs from
these sampling calorimeters in having a long (2-5 m $\Leftrightarrow$ 
2-5 msec) drift length.  For stable operation over a long
drift path the oxygen content of the liquid argon must be less than 
0.1 ppb \cite{Bettini91}, which can be maintained by continuous 
filtration of the argon
(both in liquid phase and in the boiloff/recondensation phase) using
commercial Oxisorb cartridges \cite{oxisorb}.

\begin{figure}[htp]  
\begin{center}
\parbox{3in}{\includegraphics*[width=3in]{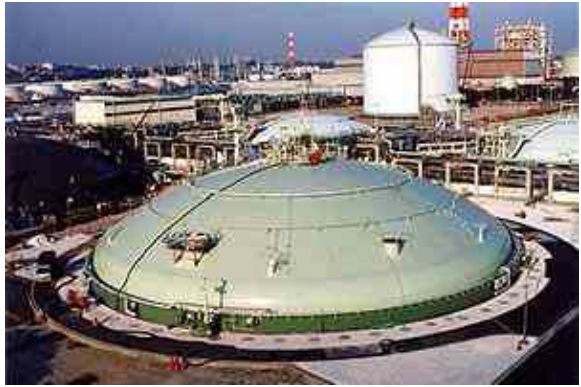}}
\parbox{5.5in} 
{\caption[ Short caption for table of contents ]
{\label{larfig1} Photograph of cryogenic storage tank of volumes $\approx 200,000$
m$^3$.
}}
\end{center}
\end{figure}

To obtain economies of scale, a large liquid argon detector should be implemented in a
single cryostat, such as those commonly used in the liquefied natural gas industry.
Cryogenic volumes of up to 200,000 m$^3$ (= 280 kton if liquid argon) are now in
use, as sketched in Figure~\ref{larfig1}.

\begin{figure}[htp]  
\begin{center}
\includegraphics*[width=5in]{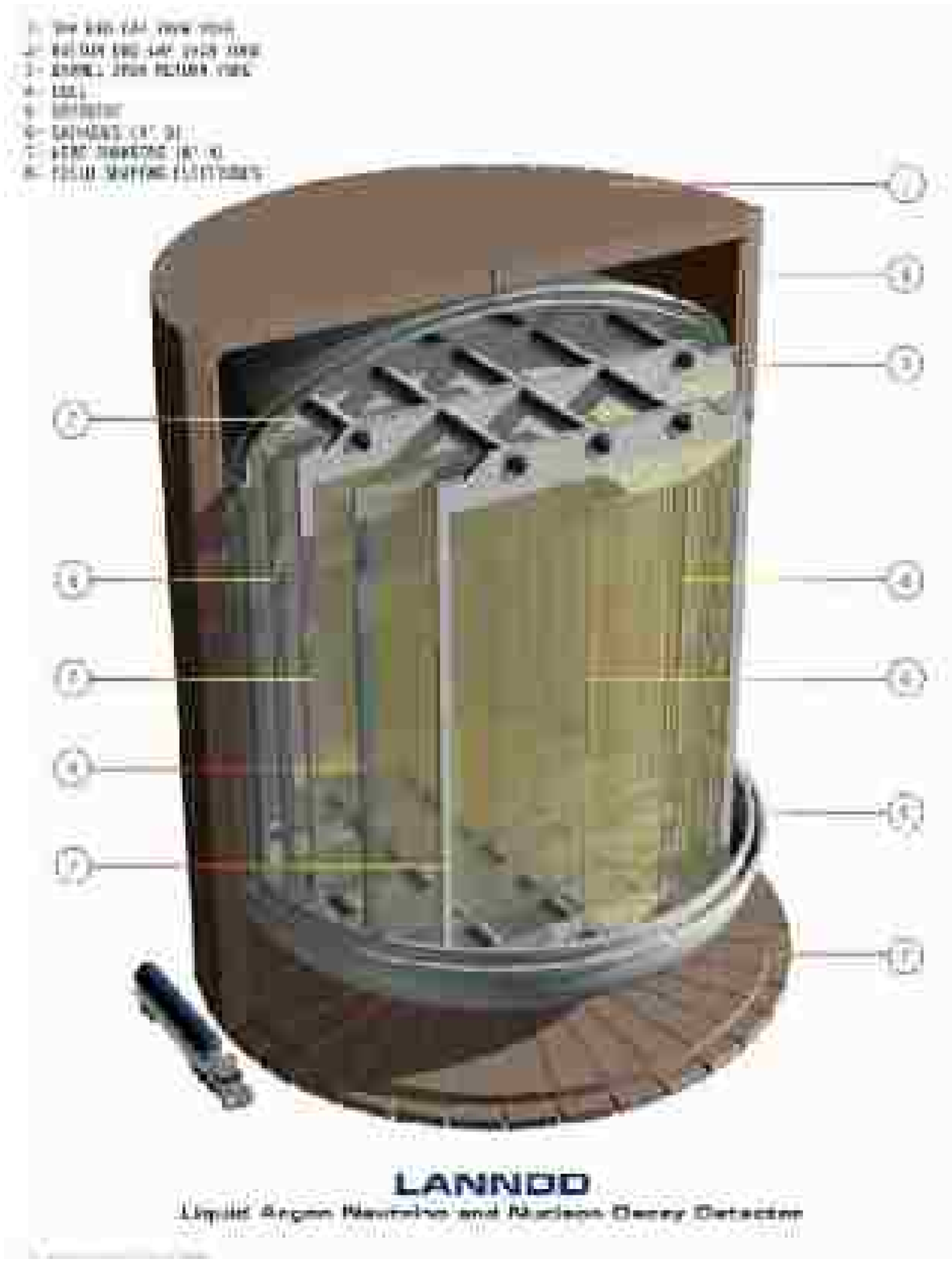}
\parbox{5.5in} 
{\caption[ Short caption for table of contents ]
{\label{lanndd1} Concept of a 70-kton Liquid Argon Neutrino and Nucleon Decay
Detector (LANNDD) \cite{lanndd,franco1}.
}}
\end{center}
\end{figure}

\begin{figure}[htp]  
\begin{center}
\includegraphics*[width=5in]{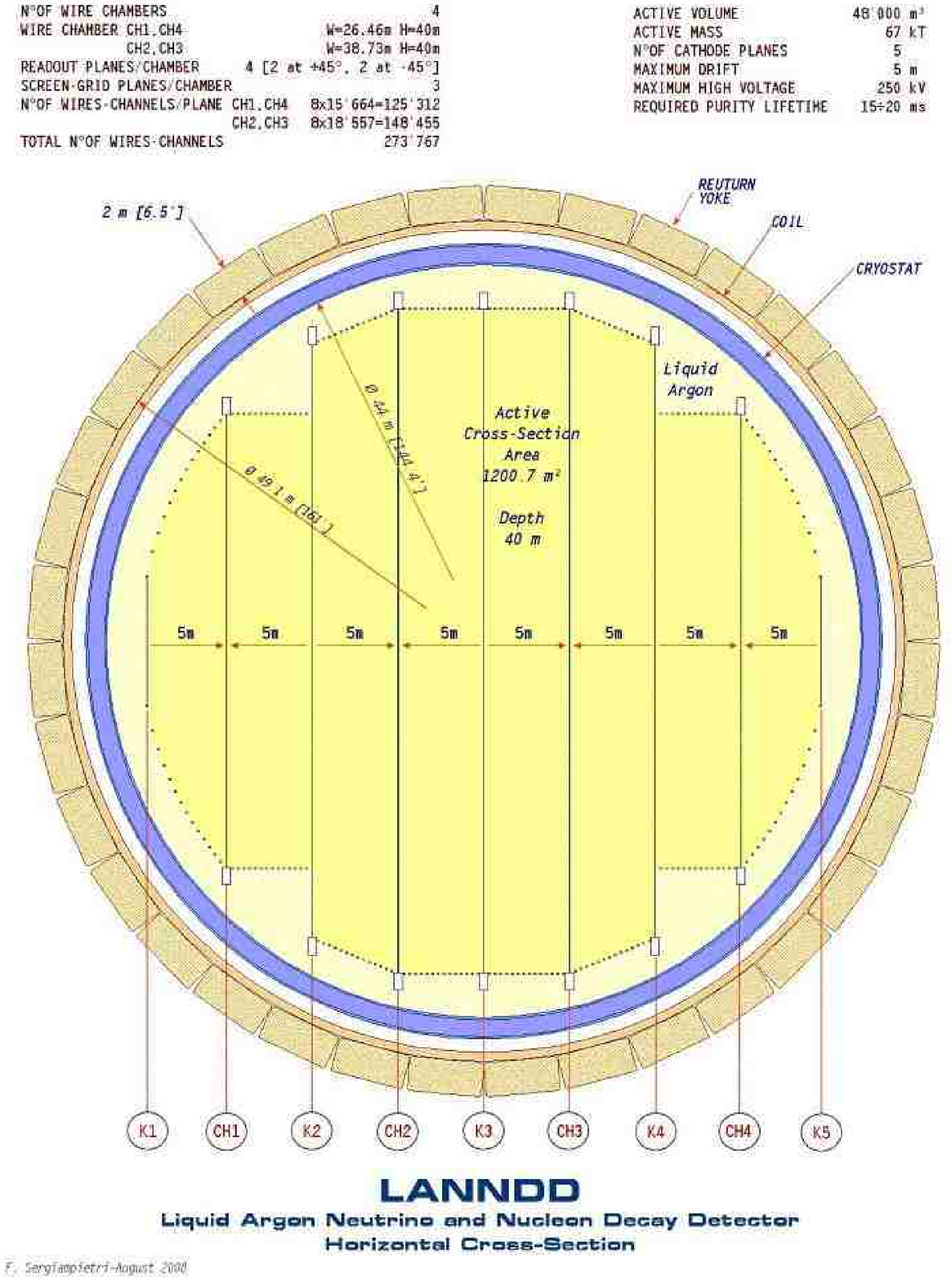}
\parbox{5.5in} 
{\caption[ Short caption for table of contents ]
{\label{lanndd_top} Top view of the electrode arrangement of a 70-kton 
Liquid Argon Neutrino and Nucleon Decay
Detector (LANNDD) \cite{franco1}.
}}
\end{center}
\end{figure}

An overall concept of a large liquid argon detector is shown in 
Figure~\ref{lanndd1}.  
The diameter of a large liquid argon detector will be greater than (twice) the 
maximum drift distance ($\approx 5$ m as limited by oxygen impurities), 
so the readout must consist of a set of parallel anode and
cathode planes that subdivide the detector, as shown in Figs.~\ref{lanndd1} and
\ref{lanndd_top}.  Then, the number of readout channels scales as the surface
area of the detector.

The data acquisition for a liquid-argon TPC can operate in 
a pipelined, deadtimeless mode,
with zero-suppression \cite{icarus_daq}.
This could permit fully live
operation at the Earth's surface for nucleon decay 
studies, in addition to triggered
data collection of neutrino interactions from a 
pulsed accelerator beam.  The data rate
would, of course, be quite high in this case.  
It may nonetheless be less costly to
implement a high rate data-acquisition system than 
to reduce the untriggered data rate by
siting the detector deep underground.

\underline{Readout Channel Count}

The instrumented volume of a liquid argon time projection chamber is organized
into a set of cells of depth 3-5 m along the direction of the electric field
($\approx$ 500 V/cm), as illustrated in Figure~\ref{lanndd_top} for a 70-kton 
detector.  

In a typical gas-phase TPC, the ionization electrons drift to an
anode-wire plane at which Townsend amplification occurs, and the readout
is based on time digitization of the induced signal size on pads on a 
nearby cathode plane.

In a liquid argon TPC no amplification of the ionization electrons is required,
as a minimum ionizing particle creates about 50,000 electron-ion pairs per cm.
However, the use of a pad readout plane would lead to a prohibitively large
channel count.  Instead, the signals are detected on two (or three) crossed
wire planes (per cell).  The ionization electrons pass by the first (and
second if a total of three) of these planes, inducing signals on the wires,
and are then collected on the second (or third) plane.  The use of three
readout planes, $x$-$u$-$v$, allows superior rejection of ``ghost''
images in case of multiple ``hits'' within a given time slice.

For operation of a large detector at the Earth's surface, it may be
preferable to have three readout planes to provide greater separation
of neutrino events from ``accidental'' cosmic-ray events.

The ICARUS detector uses three readout planes, with a wire spacing of 3 mm
on each plane.  The time-sampling frequency of 2.5 MHz corresponds to
sampling over 0.6 mm along the drift direction.  Thus, the effective
pixel size of the ICARUS readout is $3 \times 3 \times 0.6$ mm$^3$.

The maximum wire length in the present ICARUS detector is about 9 m.  It
is proposed that longer wires be used in a larger detector, so that all
wire connections can be made near the outer surface of the detector.
Longer wires have larger capacitance (proportional to their length), 
and hence a given charge leads to
a smaller voltage signal.  As the present ICARUS detector operates near
the limit of acceptable signal/noise, some change will have to be made
for successful operation with longer wires.

A simple solution is to increase the wire spacing from 3 mm by the factor 
$L / 9$, where $L$ is the wire length.  This increases the signal in
the same ratio, while the capacitance increases by $L/(9 \ln(L/9))$,
and the signal/noise ratio actually improves by $\ln(L/9)$.

Of course, this solution increases the effective pixel size in the two
transverse coordinates (but not along the drift coordinate where the
time sampling frequency determines the pixel size).  Additional study
is required to determine whether such larger pixels would have a
detrimental effect on the identification of electron neutrino
interactions against the background of neutral current interactions.

For the present study, we assume that use of a larger wire spacing and
an $x$-$u$-$v$ readout is appropriate for a large detector.  For the
particular example of a 100 kton detector ($r = 22.5$ m), 
where the wire length would be about 45 m, 
we considered use of a 1-cm wire spacing.  The wire
capacitance increases by $5/ \ln 5 = 3.1$ compared to 9-m-long wires,
and the signal size is 3.3 times larger than that for 3-mm wire
spacing.  Then a cell arrangement similar to that shown in
Figure~\ref{lanndd_top} (but with 10 cells of depth 4.4 m rather 
than 8 cells of 5-m depth) leads to a channel count slightly less than
300,000.

The channel count for detectors of other mass is scaled from this number
according to the surface area, \ie, as $M^{2/3}$.

\underline{Preliminary Cost Estimate}

As well as being the highest-performance large detector for neutrinos, 
a liquid argon TPC is also one of the least costly.  
We have made a preliminary cost estimate for
a 100 kton detector, assigning costs in two categories:
\begin{enumerate}
\item Costs proportional to detector mass.
\begin{enumerate}
\item
Liquid argon @ \$700k/kton (delivered to a site in Minnesota)
based on a preliminary budget estimate from the largest USA vendor 
of argon \cite{Praxair}.  

The annual production of liquid argon in the
USA is about 1Mton, so filling of a large liquid argon detector in
a timely manner is a significant perturbation on the entire USA market.
The largest argon production facilities are located in Chicago and the
Gulf Coast.  To fill a 100 kton detector in one calendar year would require
a tank truck every 2 hours, 24 hours a day.
\item
The on-site cryogenic system for purification and recondensation of liquid argon,
sketched in Figure~\ref{lanndd_cryo}, is estimated to cost \$10M \cite{Mulholland}.

\begin{figure}[htp]  
\begin{center}
\includegraphics*[width=5in]{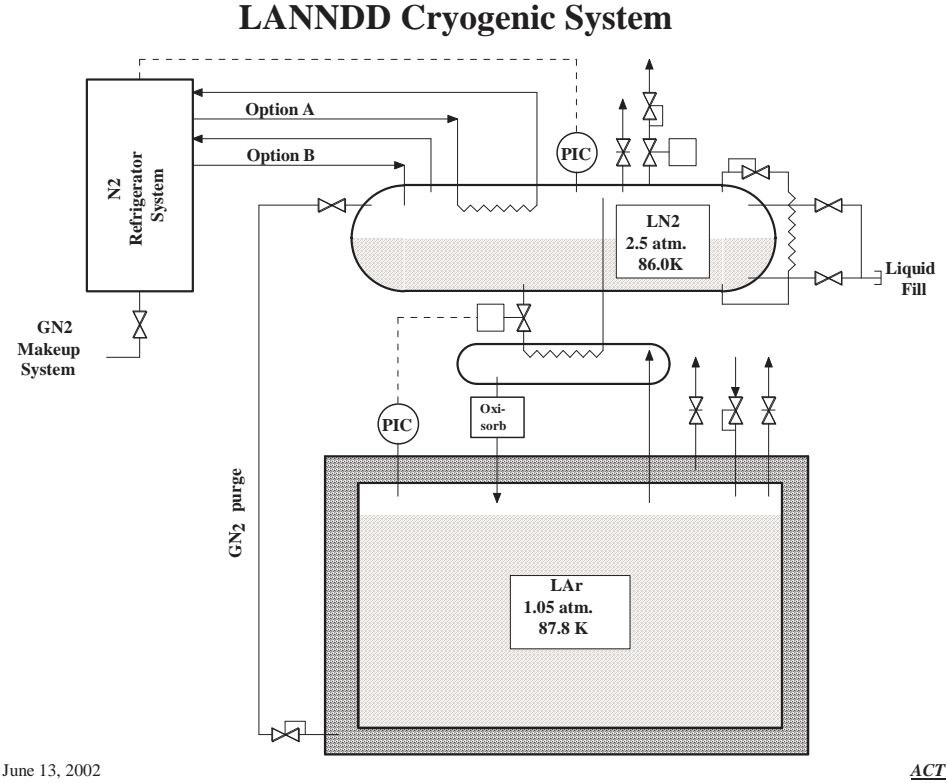}
\parbox{5.5in} 
{\caption[ Short caption for table of contents ]
{\label{lanndd_cryo} Cryogenic system for a 100-kton liquid argon detector
\cite{Mulholland}.
}}
\end{center}
\end{figure}

\end{enumerate}

\item
Costs proportional to detector surface area.
\begin{enumerate}
\item
Site preparation, estimated at \$10M.
\item
Cryogenic storage tank, estimated at \$20M based on a quotation
from the leading USA vendor of liquefied natural gas storage
tanks \cite{Mulholland}.  

The inner vessel of a large cryogenic
storage tank is welded together from nickel-steel plates.  Present
welding technology limits the plate thickness to 60 mm, and
consequently the present maximum height of a tank is 30-40 m.  That
is, tanks for more than 100 kton of liquid argon will require an
advance in plate welding technology.  Such advances are, of course,
of interest to the tank manufacturers independent of our application. 

\item
Readout electronics, \$30M for 300k channels of commercial electronics
designed for ICARUS, based on a discussion with CAEN \cite{CAEN}.  

If equivalent electronics were to be produced ``in house'' at Fermilab,
the cost might be 30-50\% less.

\item
Computer systems, estimated at \$10M.

\end{enumerate}
\end{enumerate}

In view of the very preliminary nature of these estimates, we add a 33\%
contingency.  Table~\ref{100kt_cost} summarizes this cost estimate.

\begin{table}[htbp] 
\begin{center}
\parbox{5.5in}  
{\caption[ Short caption for the List of Tables. ]
{\label{100kt_cost} Preliminary cost estimate for a liquid
argon detector of 100 kton total mass.
}}
\vskip6pt
\begin{tabular}{lr}
\hline\hline
Component & Cost \r
\hline
Liquid argon (industrial grade) & \$70M \\
Cryo plant, including Oxisorb purifiers & \$10M \\
Surface site preparation & \$10M \\
Cryogenic storage tank & \$20M \\
Electronics (300k channels) & \$30M \\
Computer systems & \$10M \\
\hline
Subtotal & \$150M \\
\hline
Contingency & \$50M \\
\hline
Total & \$200M \\
\hline\hline
\end{tabular}
\end{center}
\end{table}

\begin{figure}[htp]  
\begin{center}
\includegraphics*[width=4in]{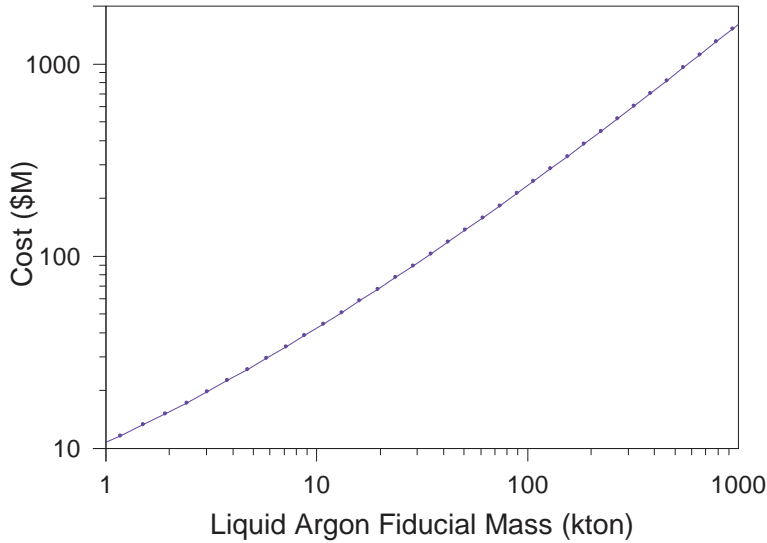}
\parbox{5.5in} 
{\caption[ Short caption for table of contents ]
{\label{lar_cost_fid} Preliminary cost estimate of a liquid argon detector
as a function of fiducial mass, according to expressions (\ref{lar4})
and (\ref{lar1}).
}}
\end{center}
\end{figure}

The cost estimate scales with total detector mass $M$ in ktons according to
\begin{equation}
\mbox{Cost\ in\ \$M} = 1.333 \left[ 80 {M \over 100} + 70 \left( {M \over 100}
\right)^{2/3} \right].
\label{lar1}
\end{equation}
The cost
estimate as a function of fiducial mass can now be obtained using
eq.~(\ref{lar4}), with results shown in Figure~\ref{lar_cost_fid}.

\subsection{R\&D Program for Liquid Argon}

The ICARUS project continues to be the major source of R\&D into 
hardware, software, and simulation of liquid argon time projection chambers.  
However, the ICARUS concept is presently limited to modules of about 1 kton.
Before a larger module could be constructed, several issues should
be addressed by additional R\&D activity.

\medskip 

\noindent{\bf Hardware R\&D Topics}

\medskip

\begin{enumerate}
\item
Liquid-phase purification of industrial grade argon via Oxisorb.
\item
Mechanics and electronics of wires up to 60-m long.
\item
Cryogenic feedthroughs, possibly including buffer volume at 150K for
low-noise FET's.
\item
Verification of operation of a liquid argon TPC at 10 atmospheres
(as at bottom of a 100-kton tank).
\end{enumerate}

R\&D proposals to study item 4 have been submitted
\cite{microlanndd_cern,argonprop,small_lar_rnd}, but not yet
funded.  New initiatives are needed to address items 1-3, and the following:

\medskip

\noindent{\bf Simulation Studies}

\medskip

\begin{enumerate}

\item
What is maximum wire spacing consistent with 
good background rejection of neutral current events, 
\ie, good $\pi^0$ identification?
\end{enumerate}
\underline{Near Detector in the NuMI Beam}

Associated with the opportunity for use of a large liquid argon
TPC in the NuMI off-axis beam $\approx 1000$ km from Fermilab
is the need for a near detector to characterize the neutrino flux
and to measure the low energy neutrino-argon cross section for
energies up to 3-4 GeV.  A near
detector of fiducial mass of 1.5 tons at 1 km from the NuMI target
is suitable for this, as it would detector about $10^5$ charged-current
$\nu_\mu$ interactions per year \cite{Bodek}.

Outside the fiducial volume for the neutrino interaction  vertex, 
a near detector must contain electromagnetic
and hadronic showers associated with the neutrino interaction.  It will
not be possible to range out the final-state muons in a detector of
modest size, since the $dE/dx$ loss for muons is about 200 MeV/m.
Therefore it will be necessary to immerse some or all of the liquid
argon TPC in a magnetic field, or to follow the liquid argon
TPC with a magnetic spectrometer.

\begin{figure}[htp]  
\begin{center}
\includegraphics*[width=5in]{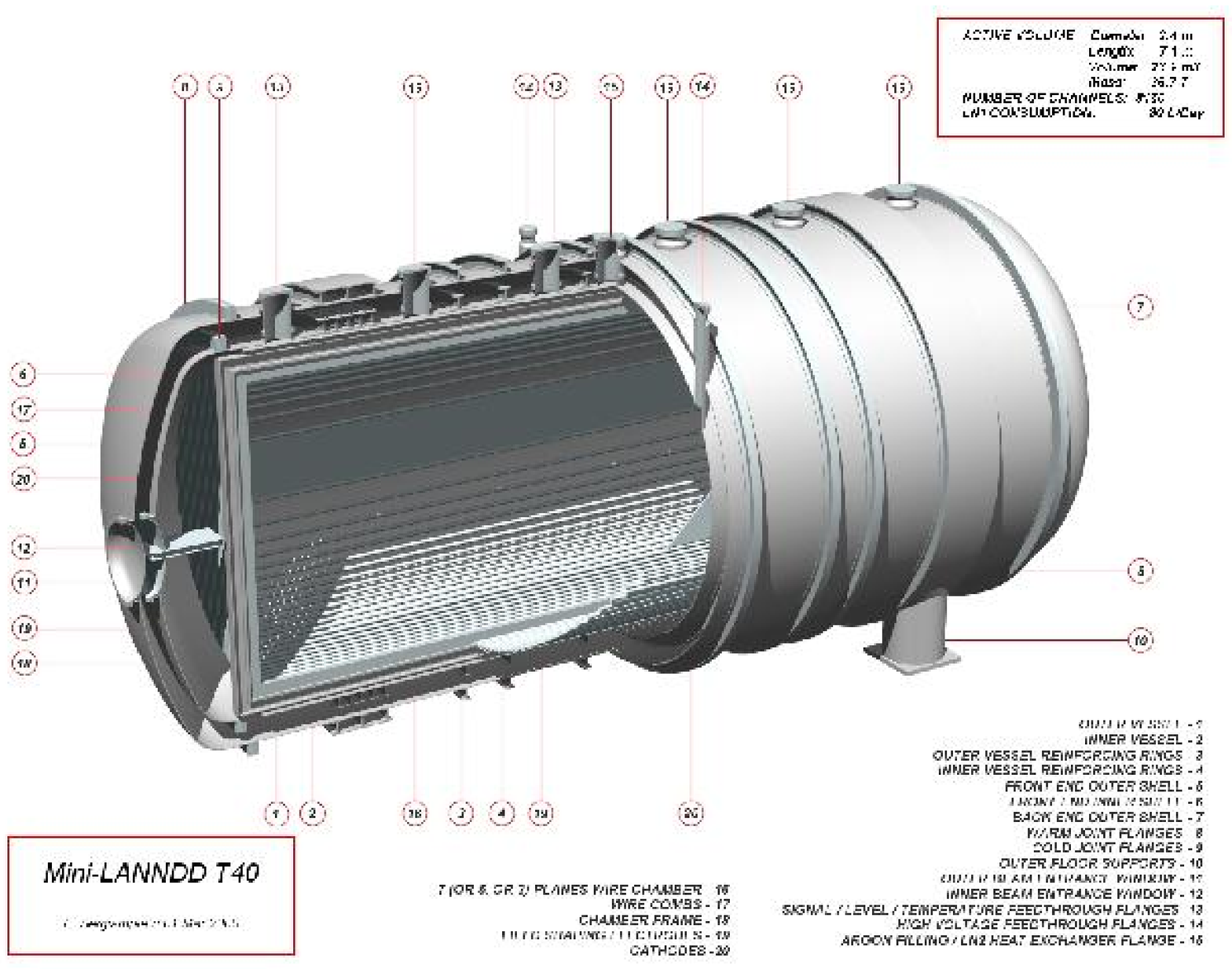}
\parbox{5.5in} 
{\caption[ Short caption for table of contents ]
{\label{t40} Concept of a 40-ton liquid argon detector for use
as an off-axis near detector in the NuMI beam. 
}}
\end{center}
\end{figure}

Electromagnetic showers in liquid argon, whose radiation length is
0.14 m, are well contained within a
cylinder of about 0.5 m radius and 2.5 m length.  Low-energy hadronic
showers are well contained within a cylinder of about 0.8 m radius and
5 m length, since the hadronic interaction length is about 0.8 m.
A simple geometry for a near detector would be a cylinder of 2.4 m
diameter and 7 m length, as shown in Figure~\ref{t40} which is
based on an earlier concept for a hadronic beam test of a liquid
argon TPC \cite{calo93}.   The fiducial
volume of this detector would be a cylinder 0.8 m in diameter and
2 m long, with volume of 1.0 m$^3$ and fiducial mass of 1.4 tons.
The total mass of liquid argon would be 37 tons.  The readout
channel count would be about 8,000.

Such a detector is intermediate in scope between the ICARUS 10-ton
prototype \cite{tm-00-06} (whose fiducial mass is zero for neutrino
interactions) and the ICARUS T-600 modules \cite{icarus_tm_01_09}.
The principal costs of a 40-ton detector would be for the cryostat,
the cryogenic system, and the electronic readout, very roughly \$1M 
each.

\begin{figure}[htp]  
\begin{center}
\includegraphics*[width=3in]{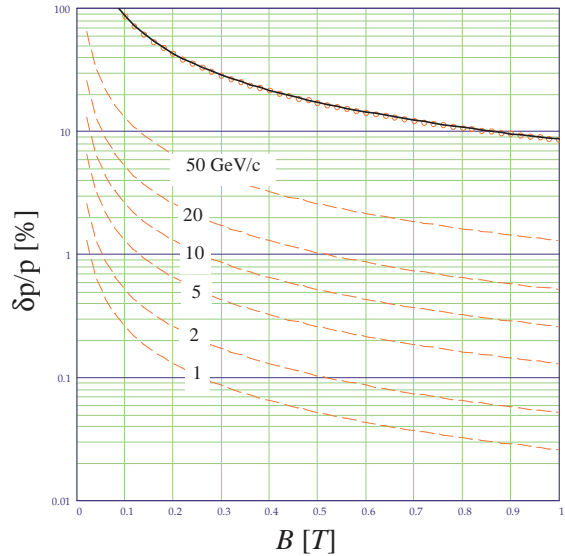}
\parbox{5.5in} 
{\caption[ Short caption for table of contents ]
{\label{res} 
Momentum resolution {\it vs}.\ magnetic field for muons crossing 
20 $X_0$ in liquid Argon. Dashed curves: contribution of the detector 
resolution at momenta 1, 2, 5, 10, 20 and 50 GeV/$c$. Circles: 
contribution of the multiple scattering independent of momentum. 
Solid thick curve: combined contribution of detector resolution and 
multiple scattering in the range 1-50 GeV/$c$.
}}
\end{center}
\end{figure}

As noted above, it is necessary to measure the final-state muon 
momenta, which could be accomplished by superimposing a magnetic
field over only part of the liquid argon detector.  As indicated
in Figure~\ref{res}, a magnetic field of 0.5 T on the downstream
3 m of the detector would provide 15\% momentum resolution up to
5 GeV muon energy.  This fields could be provided, for example, by
reconfiguring the superconducting coils originally used on the
Fermilab 15$'$ bubble chamber.

\subsection{Water Cerenkov Detectors}

Water Cherenkov detectors have been closely associated with neutrino
physics since the early success of the Kamiokande and IMB detectors in
the 1980's. These detectors consisted of a large volume of water
surrounded by planes of photomultipliers.  Neutrino interactions with
the water produce charged particles which emit Cherenkov light. The
pattern of Cherenkov light is recorded by PMT's on the walls of the
detector. The neutrino event vertex is reconstructed based on the PMT
hit times. Particle types are reconstructed by the pattern of
Cherenkov light with muons being characterized by collapsed rings
($<42^\circ$ in radius) at the lowest energies (100's of MeV), sharp
ring patterns at medium energies (a few GeV), and long tracks at the
highest energies (several GeV). PMT hit patterns from electrons are
typically much more diffuse when compared to the patterns resulting
from muons. For single-ring events, the Super--Kamiokande detector has
achieved a particle ID efficiency of 98\% for interactions in the GeV
range.

Currently, the Super--Kamiokande and SNO detectors are the state of
the art in water Cherenkov detectors. These detectors have had great
success with neutrino measurements from solar neutrino energies (a few
MeV up through atmospheric neutrino energies (100's of GeV) and
Super--Kamiokande has been the target of the K2K long-baseline
experiment.

Several properties of water Cherenkov detectors make the technology an
excellent candidate for doing neutrino physics at neutrino energies
at 1~GeV and below. A great deal of experience building and operating
these detectors exists in the high energy physics community. Also, it
is possible to achieve very large mass using this detector technology
while keeping channel counts (and hence costs) under control. Further,
the detectors have good energy resolution, can be very accurately
calibrated, and have excellent particle ID performance.

As the energies of the neutrino interactions rises above 1~GeV, the
analysis of neutrino interactions in water Cherenkov detectors becomes
more difficult. As the neutrino energies increase, so too do the
multiplicities of the final state interactions. This increase has two
effects. The first is to degrade particle identification as the
additional Cherenkov rings are likely to overlap one another at least
partially. The second is a degradation of the neutrino energy
resolution as more of the neutrino energy tends to go into making
pions below Cherenkov threshold.

\begin{figure}[htb]
  \begin{center}
  \includegraphics[width=.8\textwidth]{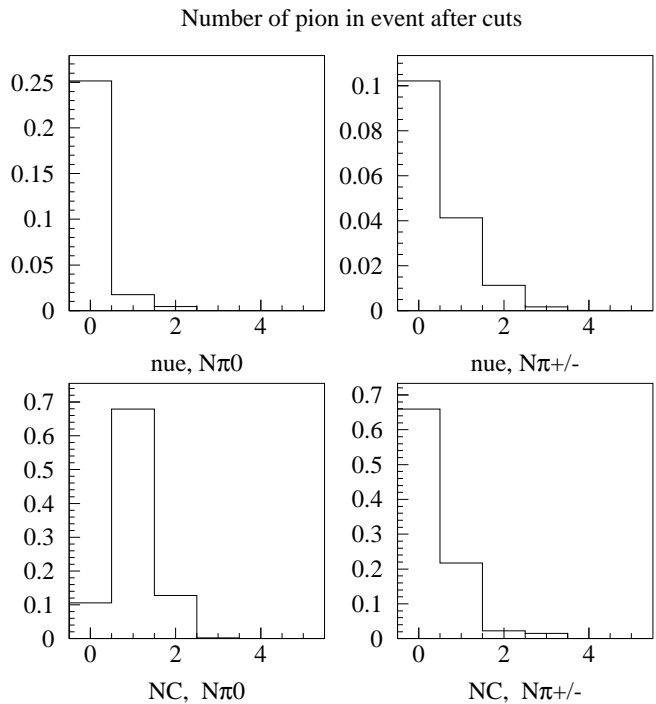}

  \caption{ The number of neutral (left panels) and charged (right
  panels) pions in single ring $e$-like signal events (top panels) and
  neutral current-background events (bottom panels). The
  neutral-current backgrounds are dominated by events with a single
  energetic $\pi^0$. }

  \label{fig:wc-npions}
  \end{center}
\end{figure}

For use in the NuMI beam, 
the most important effect of beam energies over
1~GeV is on the topology of single $\pi^0$'s produced in
the detector. At energies above 1~GeV the opening angle of the
$\gamma$'s produced by a $\pi^0$ is significantly smaller than the
Cherenkov angle of $42^\circ$. Further, the conversion distance of the
$\gamma$'s is comparable to the detector vertex resolution
(~10~cm). These two effects combine to make it difficult
to distinguish electron neutrino interactions at 2~GeV 
from 
$\pi^0$'s created via neutral-current interactions. As shown in
Figure~\ref{fig:wc-npions} the neutral-current component of the
single-ring $e$-like sample of a Super--Kamiokande-like detector
placed off-axis in the NuMI beam is dominated by events with a single
energetic $\pi^0$.

The combination of these effects suggest that the best strategy for
using a water Cherenkov detector as part of a long-baseline program is
to lower the neutrino beam energy to less than one GeV. For example,
the beam energy of the K2K experiment peaks at roughly 0.7~GeV, and a
similar beam energy is planned for the JHF-SK project. In the case of
the NuMI beam, the neutrino beam energy is fixed at roughly
2~GeV. Thus the main issue for the possible use of a water Cherenkov
detector in a NuMI off-axis experiment is the quality of the separation
between a possible electron neutrino signal and  neutral-current
backgrounds at neutrino energies of 2~GeV.

For this report, the selection efficiency of a Super-Kamiokande-like
detector has been estimated assuming neutrino fluxes at a location
14~mrad off the NuMI beam axis at a distance of 735~km. The analysis
combines several event variables:
\begin{itemize}
\item{\bf Single ring: }

The selection of single ring events retains roughly 40\% of the<CR>
potential electron neutrino signal at 2~GeV while reducing the number<C+
of neutral current interactions by 85\%.

\item{\bf Number of decay electrons: }
For electron neutrino QE events this number should be exactly zero,
and tends to be one or larger for muon neutrino interactions and
neutral-current interactions which produce sub-threshold pions.

\item{\bf Ratio of charge in ring to total event charge}
In the case of $\pi^0$, there is a tendency for the fitted Cherenkov
ring to have a great deal of light located outside the fitted cone.

\item{\bf Cherenkov angle }
Single ring fits to $\pi^0$ events tend to fit with slightly larger
Cherenkov angles resulting from the separation of the two $\gamma$'s.

\item{\bf Muon particle ID likelihood }
Removes un-oscillated muon-neutrino interactions

\item{\bf Shower particle ID likelihood }
The likelihood fit of a single electron shower to a $\pi^0$ decay tends
to produce less likely fits than it does when fitting true electron
neutrino interactions

\item{\bf Angle to beam direction }
Neutral current $\pi^0$ production has a strong coherent component
which is highly correlated with the neutrino beam direction. Hence,
one does better to select events slightly off the neutrino beam axis.

\end{itemize}

\begin{figure}[htb]
  \begin{center}
  \includegraphics[width=.8\textwidth]{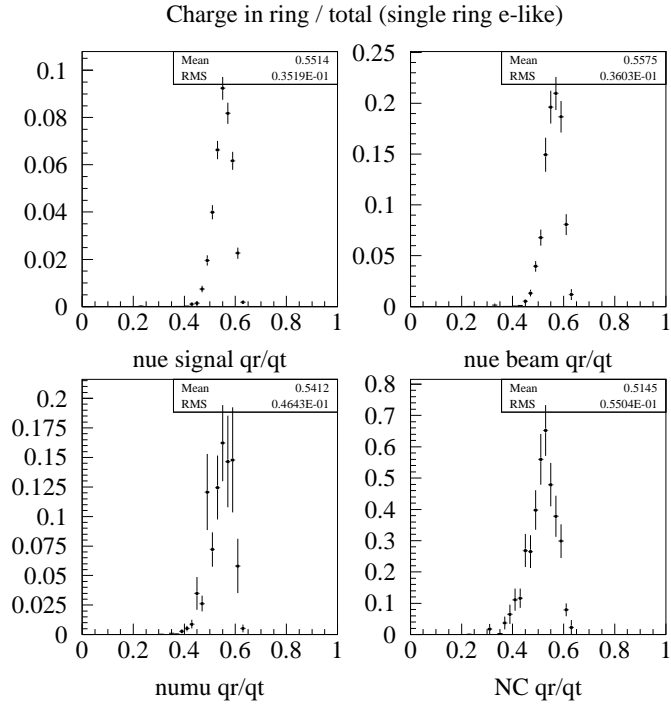}  

  \caption{ Ratio of PMT charge in fitted ring to total event charge,
  shown for electron-neutrino signal events, beam electron-neutrino
  background events, beam muon-neutrino background events, and
  neutral-current background events. }

  \label{fig:wc-qrqt1}
  \end{center}
\end{figure}

\begin{figure}[htb]
  \begin{center}
  \includegraphics[width=\textwidth]{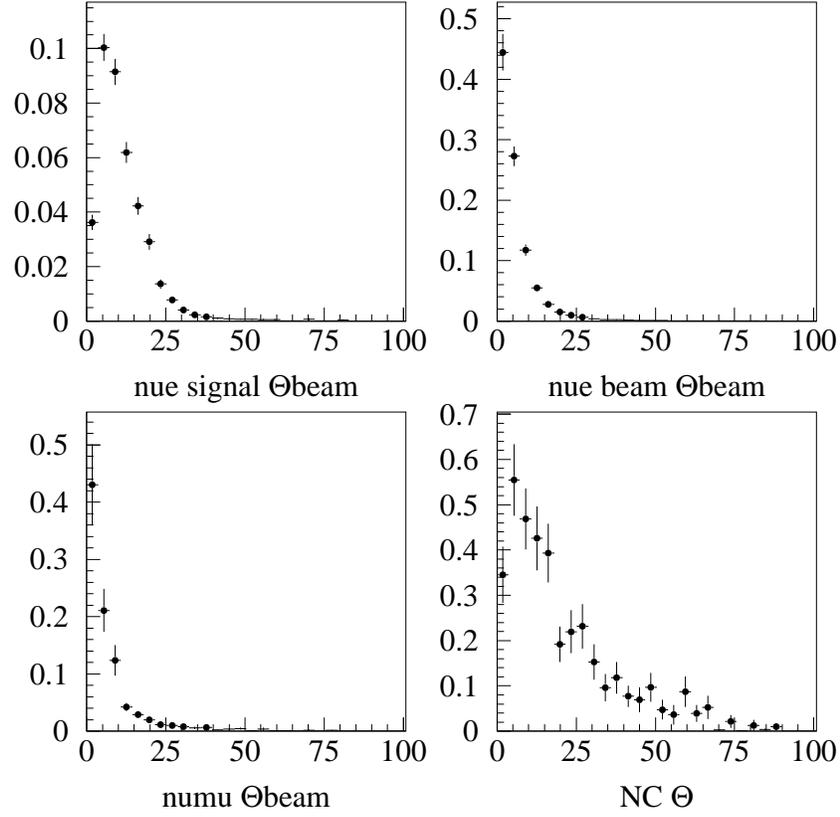}

  \caption{ The angle between event direction and 
beam direction in degrees, for electron-neutrino signal 
  events, beam electron-neutrino
  background events, beam muon-neutrino background events, and
  neutral-current background events. }

  \label{fig:wc-qrqt2}
  \end{center}
\end{figure}

\begin{figure}[htb]
  \begin{center}
  \includegraphics[width=\textwidth]{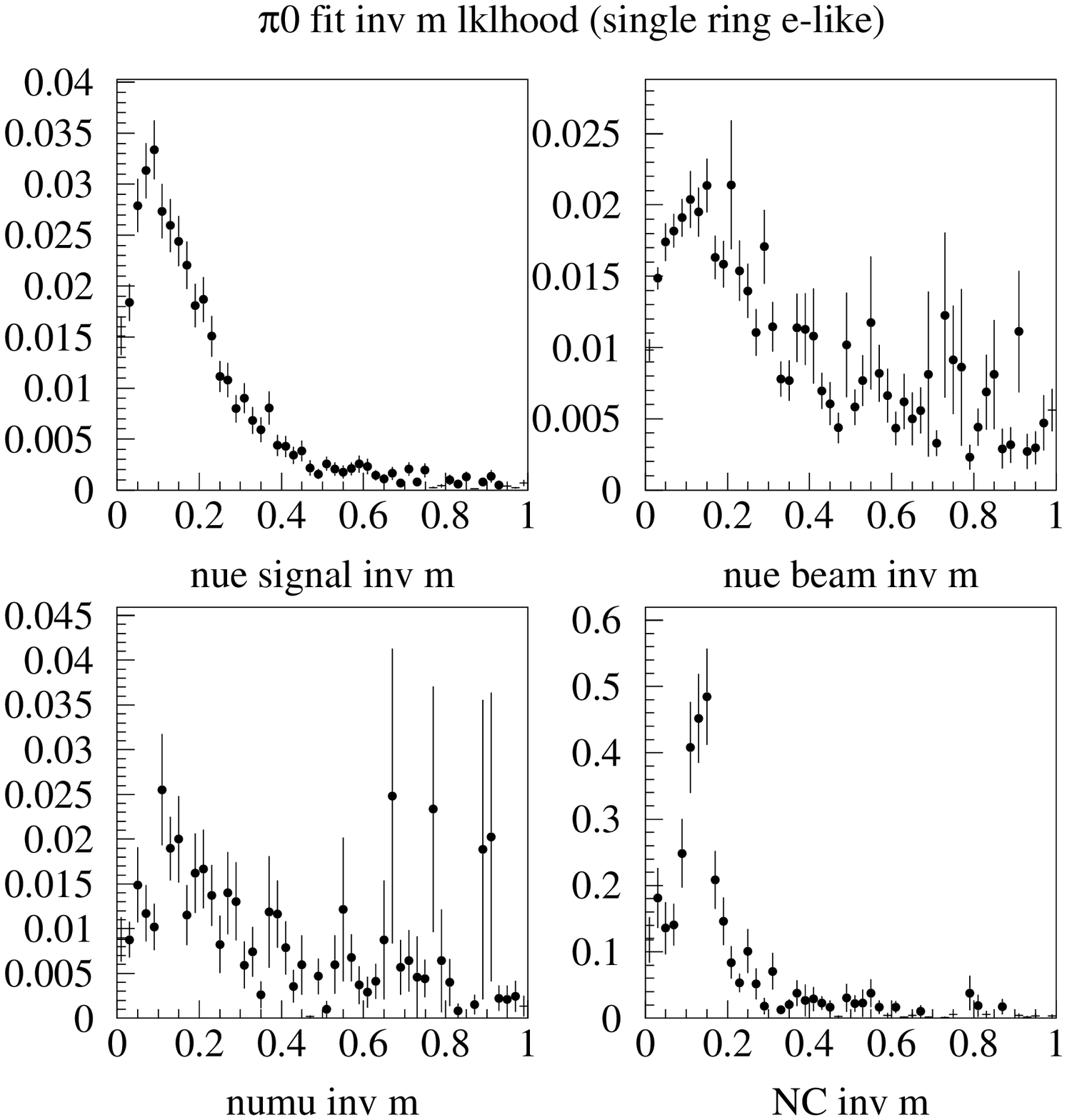}

  \caption{ The invariant mass of $\pi^0$ fit, 
 for electron-neutrino signal events, beam electron-neutrino
  background events, beam muon-neutrino background events, and
  neutral-current background events. }

  \label{fig:wc-qrqt3}
  \end{center}
\end{figure}

In addition to these event variables, there are several variables
which result from a fit to the events assuming a $\pi^0$ is present:
\begin{itemize}

\item{\bf Likelihood of $\pi^0$ fit}
Tends to be larger for true $\pi^0$ events than for electron-neutrino
events.

\item{\bf Ratio of dimmer ring to brighter ring }
True $\pi^0$ events tend to have roughly equally bright
rings. Electron neutrino events fitted as $\pi^0$ tend to have dimmer
second rings.

\item{ \bf Invariant mass of fitted rings }
should yield something close to the $\pi^0$ mass for true $\pi^0$
events.

\end{itemize}
Samples of distributions of these event variables are shown in
Figures~\ref{fig:wc-qrqt1},~\ref{fig:wc-qrqt2}, and~\ref{fig:wc-qrqt3}.


\begin{figure}[htb]
  \begin{center}
  \includegraphics[width=.7\textwidth]{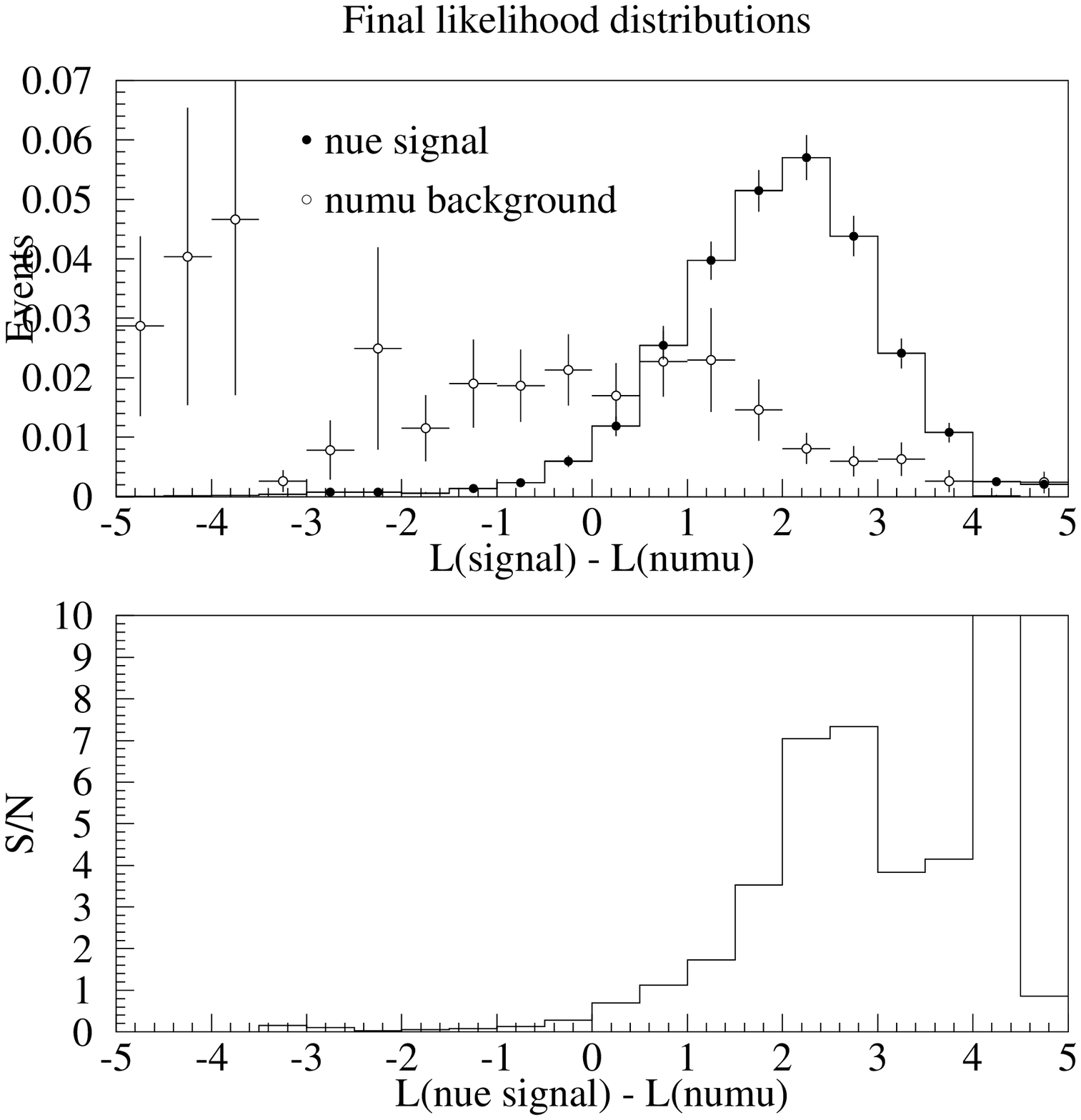}
  \includegraphics[width=.7\textwidth]{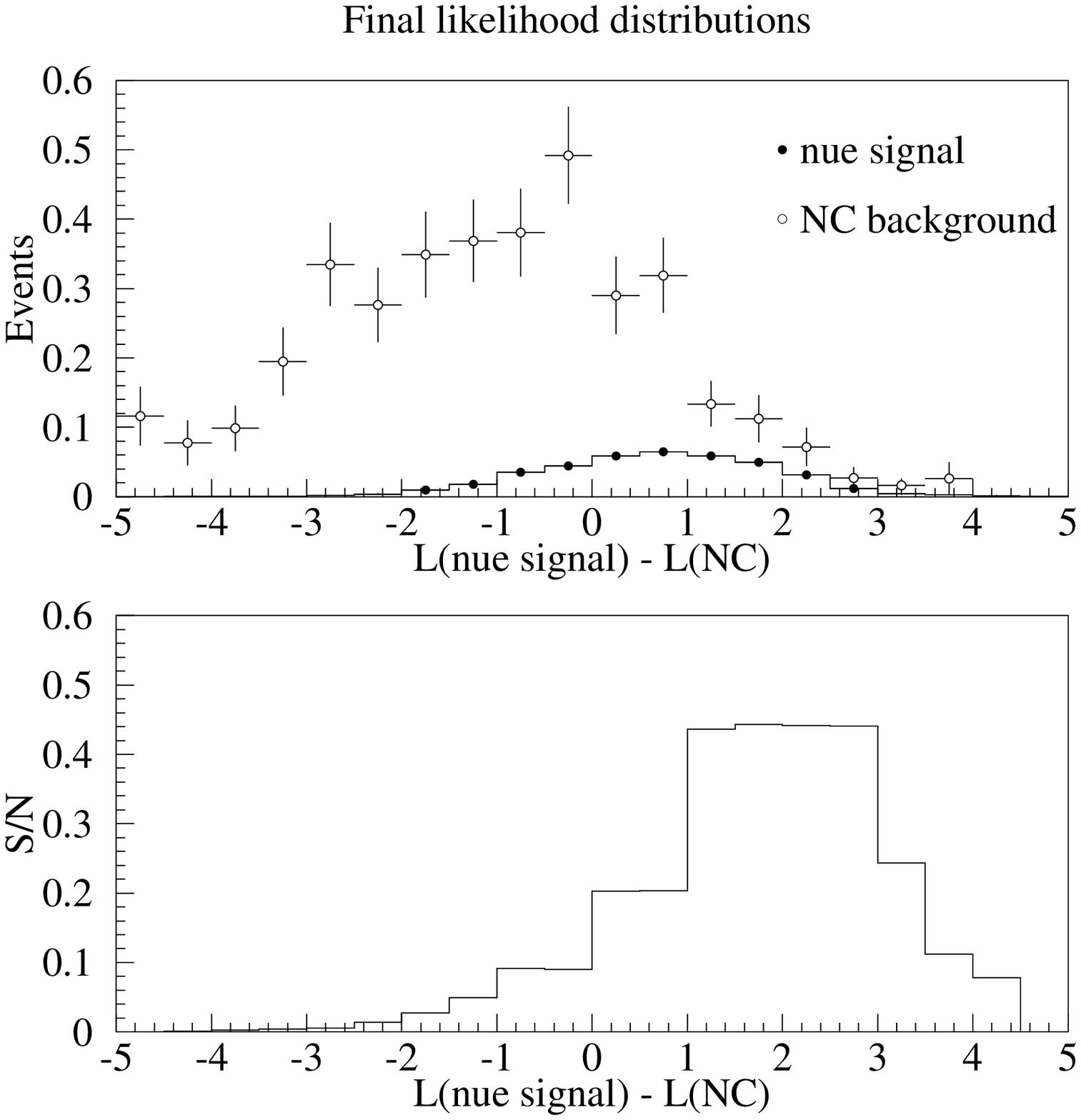}

  \caption{ Left: The final log likelihood difference for
  electron-neutrino signal events and muon-neutrino events. Top shows
  raw distributions, bottom shows ratio of the electron-neutrino
  signal to the muon-neutrino background. Events to the right of zero
  are accepted into the final sample.  Right: Final log likelihood
  difference for electron-neutrino signal events and neutral-current
  events. Events to the right of 1.0 are accepted into the final
  sample.  }

  \label{fig:wc-numulkh} 
  \end{center}
\end{figure}

Using these variables, a combined electron-neutrino, neutral-current,
and muon-neutrino likelihood is calculated for each event. The final
distributions of the final log likelihood distributions are shown in
Figure~\ref{fig:wc-numulkh}. The final signal sample contains events
which are significantly more likely to be electron neutrino signal
events than muon-neutrino backgrounds or neutral-current backgrounds
are selected.

\begin{figure}[htb]
  \begin{center}
  \includegraphics[width=.7\textwidth]{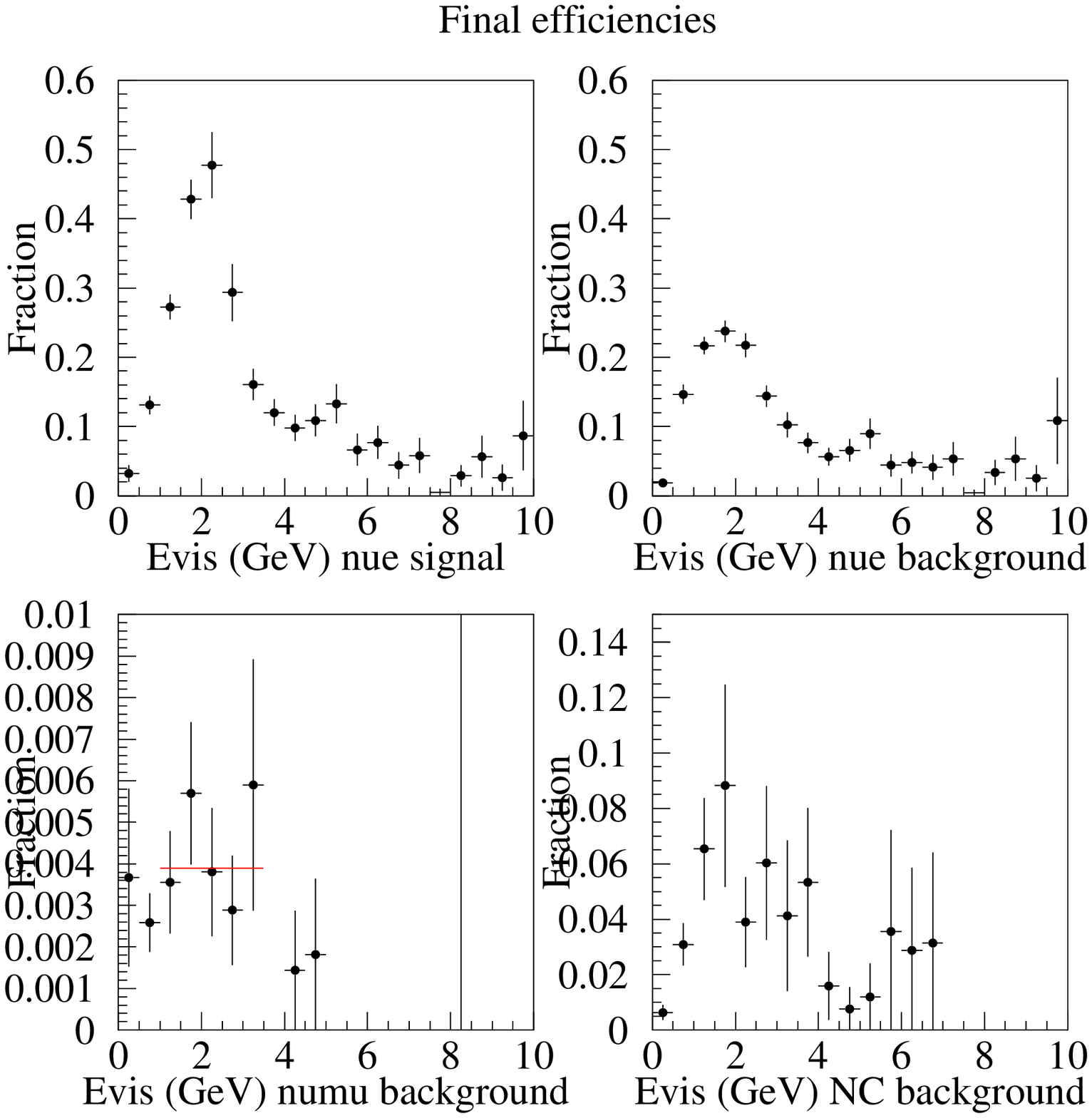}
  \includegraphics[width=.7\textwidth]{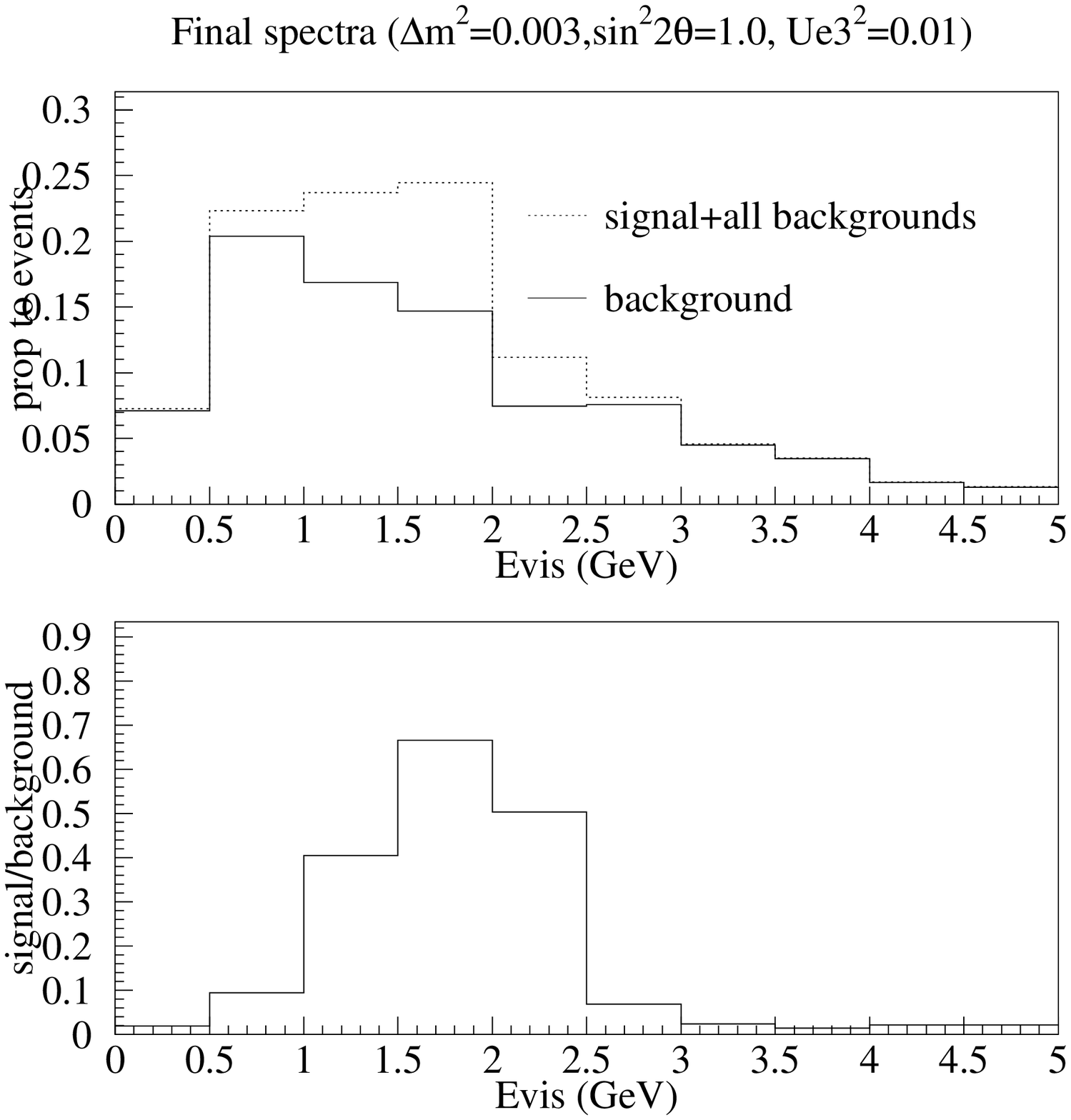}

  \caption{ Left: The final fractions of events accepted as signal
  events after cuts the log likelihood differences. Right: The final
  spectra of signal and background assuming $\nu_\mu \leftrightarrow
  \nu_e$ with $\Delta m^2 = 0.003~$eV$^2$, $\sin^2 2 \theta = 1.0$ and
  $U_{e3}^2 = 0.01$.  
  }

  \label{fig:wc-eff}
  \end{center}
\end{figure}

This selection retains roughly 20-50\% of the electron neutrino
signal, while rejecting roughly 19 out of every 20 neutral-current
background events. After event selection, muon-neutrino backgrounds are
extremely small. The final fractions of signal and background events
selected by this analysis are shown in Figure~\ref{fig:wc-eff}. The
figure also shows an example of the neutrino spectrum that results
from this analysis for the case of $U_{e3}^2 = 0.01$ and $\Delta m^2 =
0.003~$eV$^2$.

While the signal extraction for events in the energy range of 2~GeV
could be improved with more study, this analysis demonstrates the
challenges of operating a large water Cherenkov detector at energies
of 2~GeV. Continued research into optimizations of the water Cherenkov
technology for higher energy events may also lead to improvements. For
example, faster photo-detectors, or complete PMT waveform readout may
help resolve the difference in the conversion points for the two
$\gamma$'s produced in a $\pi^0$ decay helping to reduce this
background.

\subsection{Water RICH Detectors}

The use of a water RICH detector for neutrino physics was first proposed by
Tom Ypsilantis and collaborators in 1999\cite{AquaRICH_paper}. The 
AquaRICH (or AQUA-RICH) 
experiment as proposed was a 125 m diameter, spherical detector
containing 1 Mt of water. The detector
was to be sited outdoors in a natural pit under a 50 m water shield. 
Spherical reflecting mirrors were to focus the Cherenkov light which was
detected using hybrid photodiodes. AquaRICH was described by Ypsilantis 
as ``a Super-Kamiokande with spectacles.'' 

By using the RICH technique, particle velocities can be deduced from 
the ring radius while the ring center determines track direction with
$\sigma_{\theta x}$ and $\sigma_{\theta y}$ $\approx$ 6 mrad. Track
reconstruction is possible using the
time-evolution of photon detection coordinates which will require
time resolution on the order of 1 ns\cite{ypsag}. 
The track length is 
proportional to the number of detected photons. The new aspect of
AquaRICH is the ability to measure particle momentum using the measured
change in the Cherenkov angle from particle trajectory deflections 
due to multiple scattering\cite{gross}. 
A GEANT simulation was used to examine several 
algorithms for performing this measurement and, for pathlengths of
4~m, momentum
resolutions of 11\% (4 GeV/$c$) to 21\% (24 GeV/$c$) were found for 
muons\cite{gz}. Longer tracks were found to have improved accuracy, up to 
4\% resolution for 18 GeV/$c$ muons. 

\begin{figure}[hbt]
\centerline{\psfig{figure=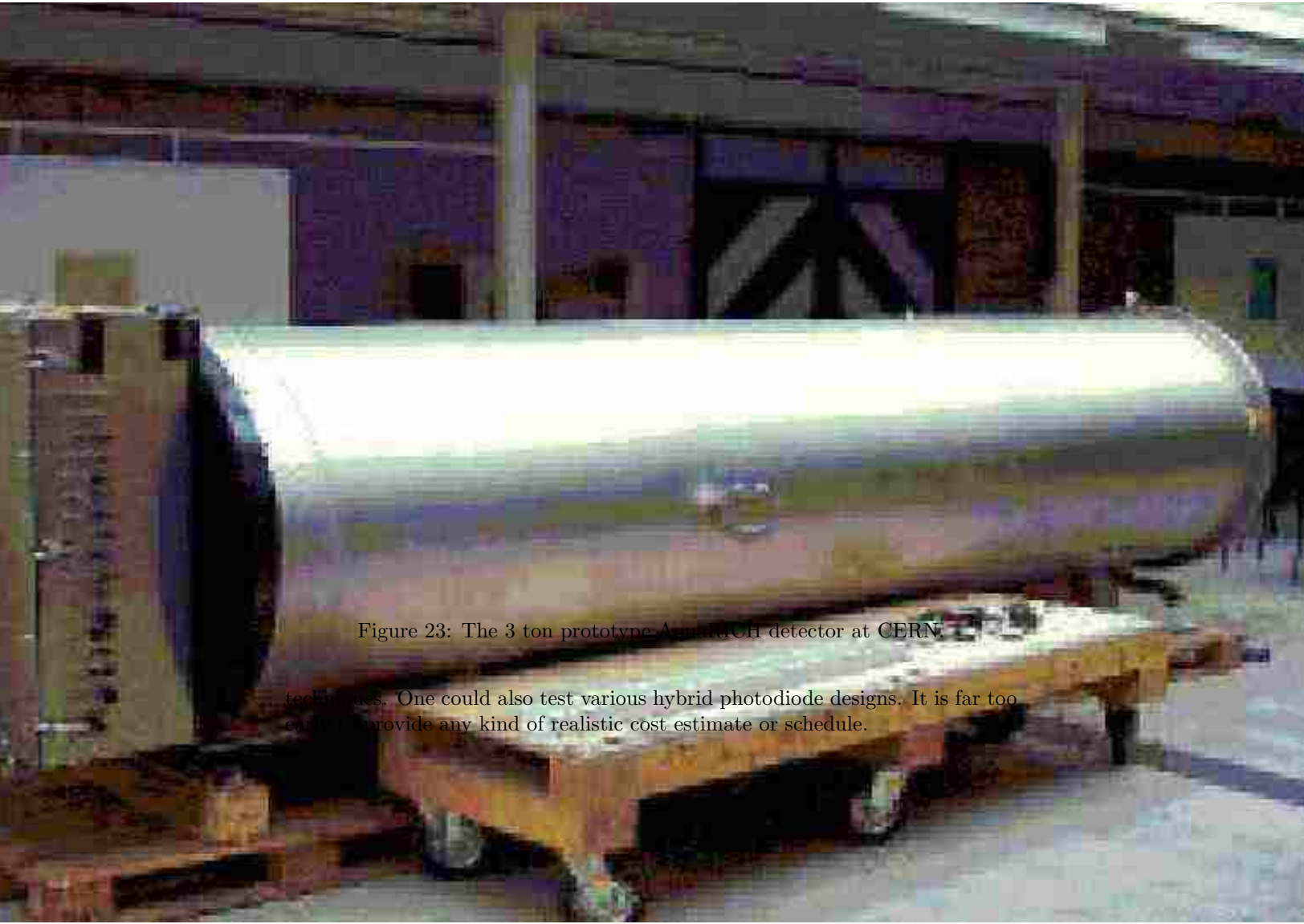,height=6in}}
\vskip -5.5cm
\label{fig:prototype}
\caption{The 3 ton prototype AquaRICH detector at CERN.}
\end{figure}

The spherical geometry of the AquaRICH
proposal allows for the detection of atmospheric neutrinos as well as
supernova detection and proton decay measurements.
The original AquaRICH was designed to go in Gran Sasso as a detector 
for a long-baseline neutrino experiment. The water was housed in a
a large rectangular box with the curved mirror at one end. A prototype
AquaRICH detector with 3 tonnes of water was built at CERN and is shown in
Figure \ref{fig:prototype}. 

\begin{figure}[hbt]
\centerline{\psfig{figure=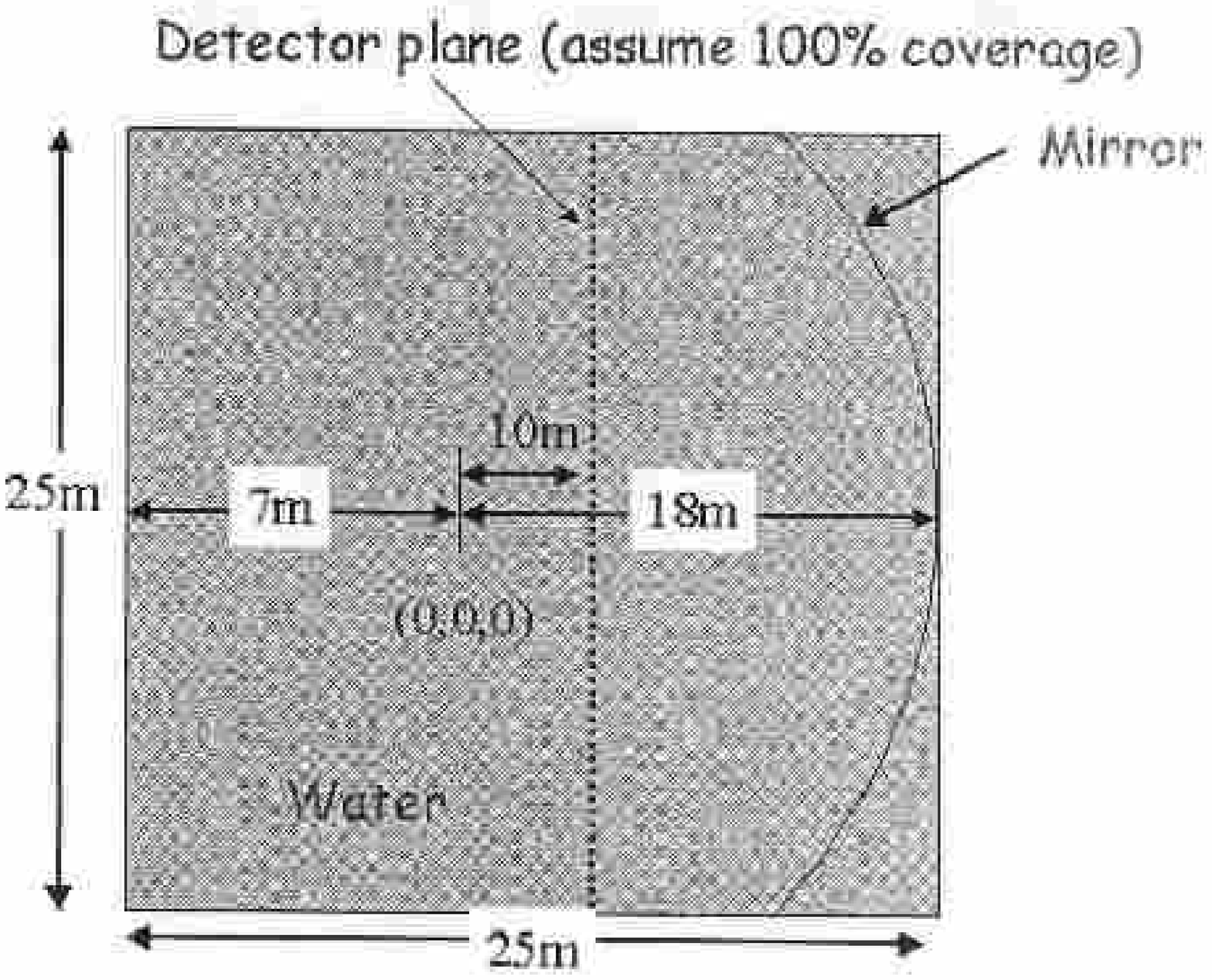,height=5in}}
\vskip -4cm

\caption[]{The geometry of the AquaRICH used in GEANT for Off-axis studies.}
\label{fig:GEANT_AquaRICH}
\end{figure}

It is this geometry which seems more relevant for the NuMI off-axis
detector. Just as a first look at the use of such a detector for NuMI off-axis,
the geometry as shown in Fig. \ref{fig:GEANT_AquaRICH} was used in 
a GEANT simulation\cite{richard}. No effort was made at this initial stage to 
simulate a realistic photodetection system. Of particular interest was how
slow $\pi^0$'s looked in the detector. To this end
{\tt NUANCE} was run to obtain the momentum distribution of $\pi^0$'s 
produced in neutrino interactions. Some results are shown in 
Figure \ref{fig:momenta}. The distribution of Cherenkov photons
from electrons and muons was also studied.

\begin{figure}[hbt]
\centerline{\hskip 1.5cm \psfig{figure=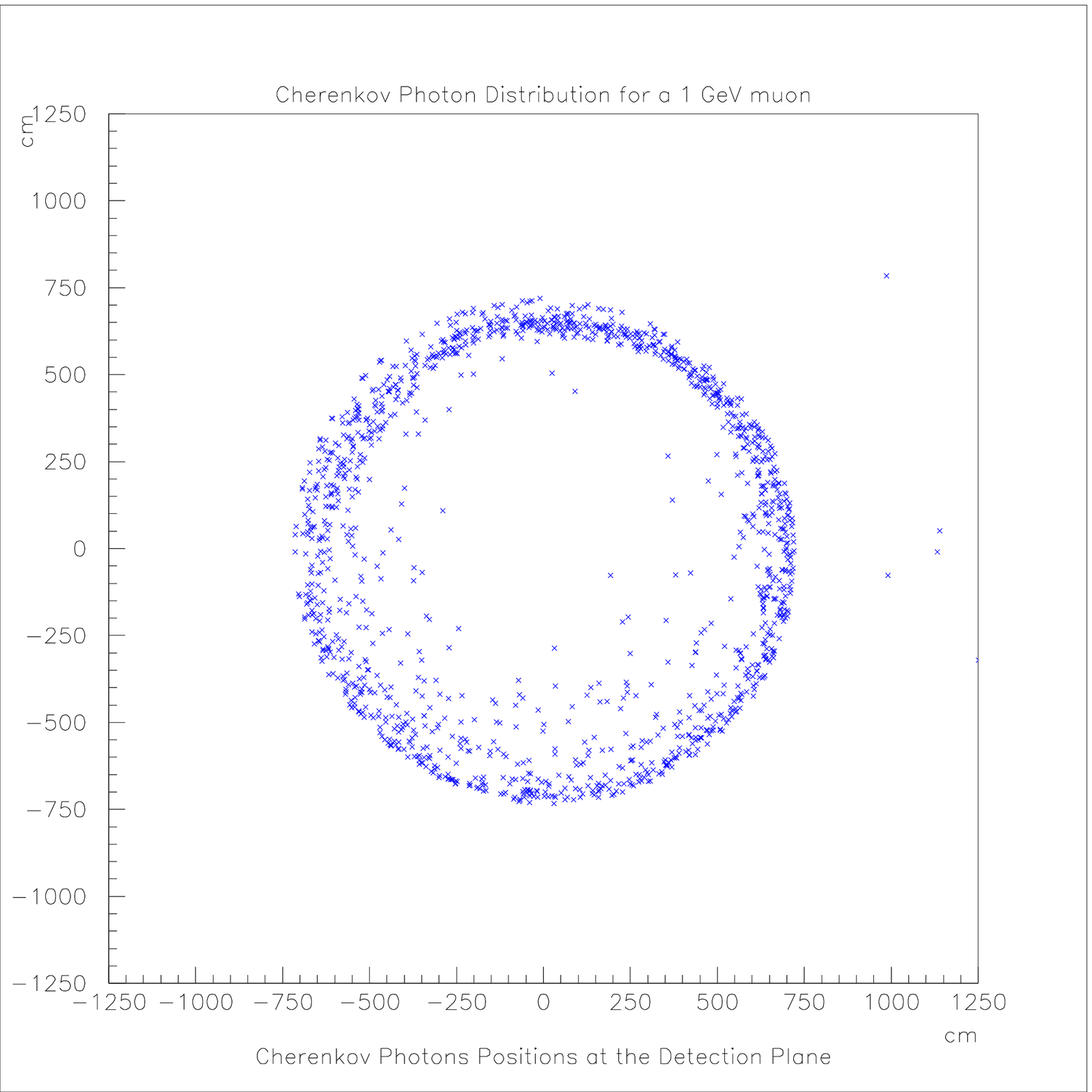,height=4in}}
\caption[]{Pattern of Cherenkov photons on the detector plane due to
a 1 GeV/$c$ muon.}
\label{fig:1GeVmu}
%
%
\centerline{\hskip 1.5cm \psfig{figure=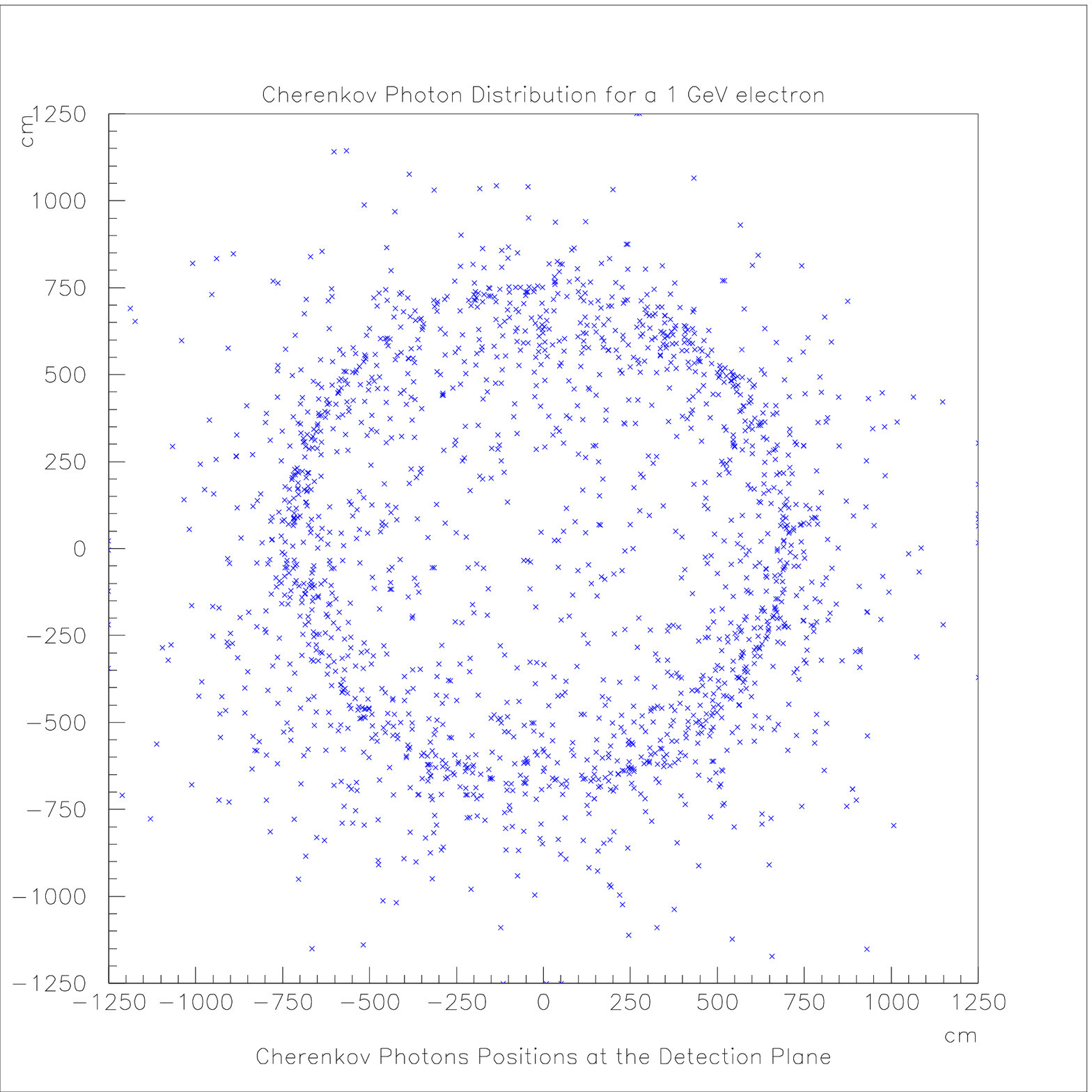,height=4in}}
\caption[]{Pattern of Cherenkov photons on the detector plane due to
a 1 GeV/$c$ electron.}
\label{fig:1GeVe}
\end{figure}

\begin{figure}[hbt]
\centerline{\hskip 1.5cm \psfig{figure=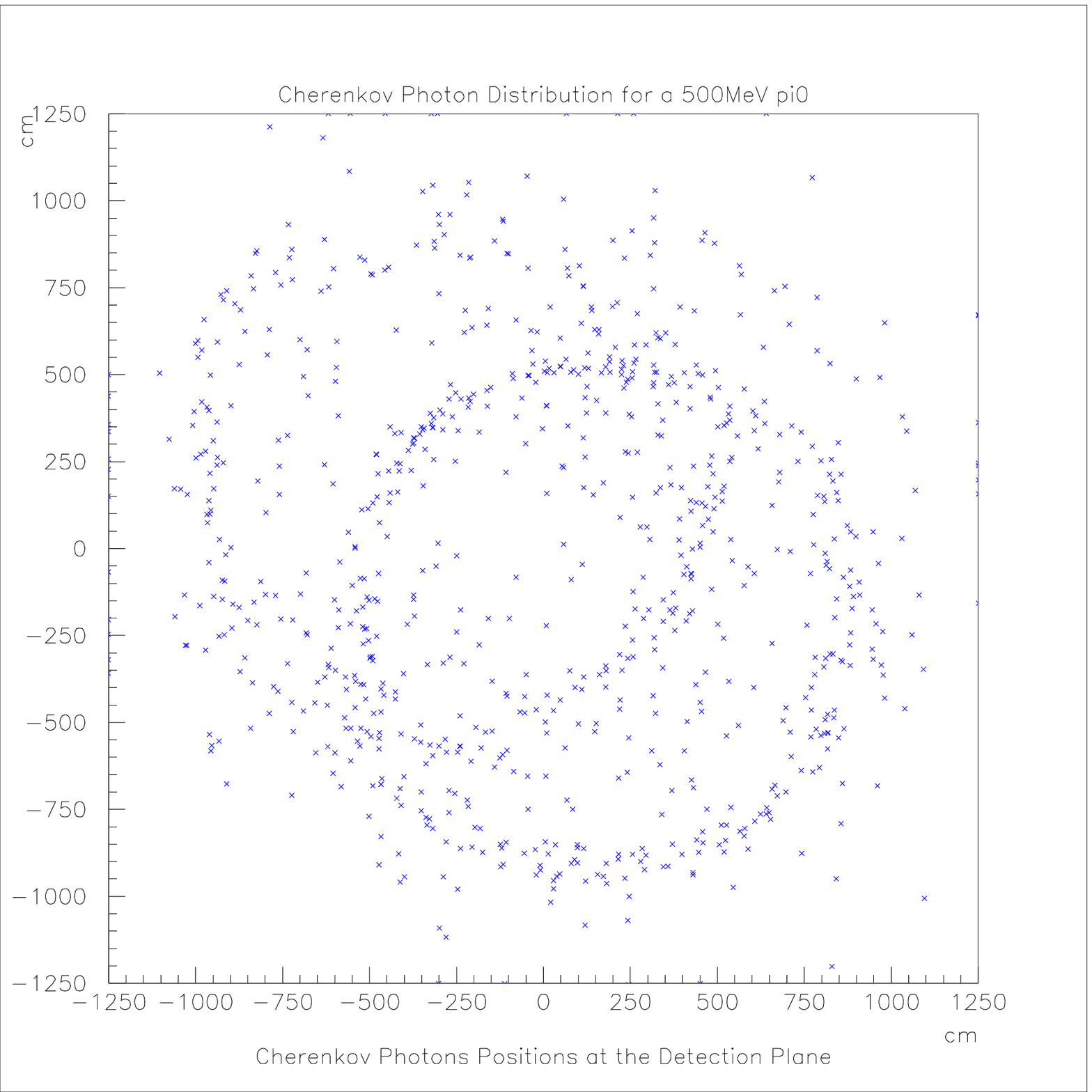,height=4in}}
\caption[]{Pattern of Cherenkov photons on the detector plane due to
a 500 MeV/$c$ $\pi^0$.}
\label{fig:500pi}
%
\centerline{\hskip 1.5cm \psfig{figure=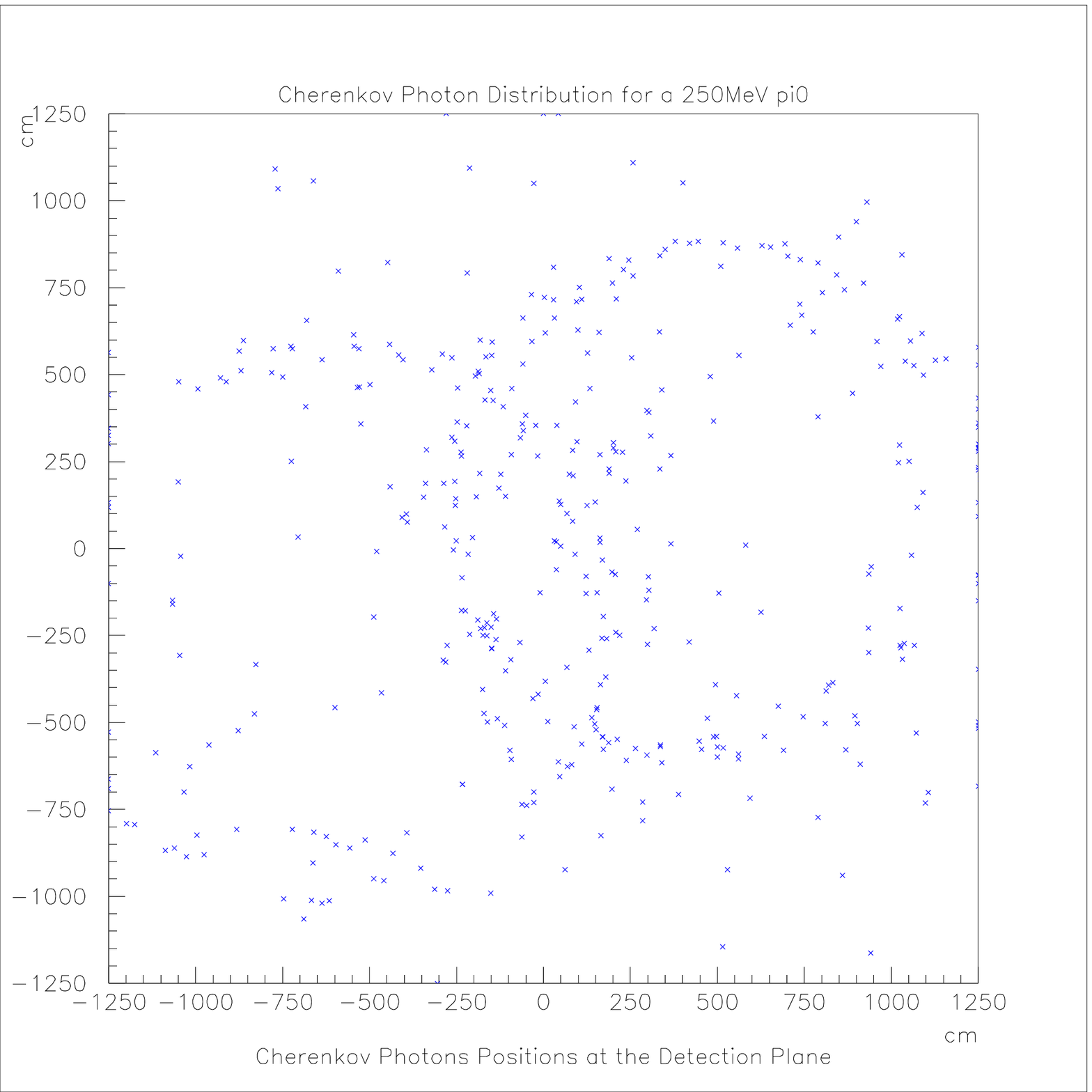,height=4in}}
\caption[]{Pattern of Cherenkov photons on the detector plane due to
a 250 MeV/$c$ $\pi^0$.}
\label{fig:250pi}
\end{figure}

The simulation was performed using 
single particles introduced along the central axis of the detector. 
The resultant rings for a 1 GeV/$c$ muon and electron as well as a 500 MeV/$c$ 
$\pi^0$ and a 250 MeV/$c$ $\pi^0$ are shown in Figures 
\ref{fig:1GeVmu},
\ref{fig:1GeVe},
\ref{fig:500pi}, and
\ref{fig:250pi}, respectively.

\subsection{R \& D Issues for RICH detectors}

The key issue for AquaRICH R \& D at the present is manpower. A first step
would clearly be to take the prototype and put it in a beam somewhere to
test the viability of the geometry and, in particular, the momentum 
reconstruction 
techniques. One could also test various hybrid photodiode
designs. It is far too early to provide any kind 
of realistic cost estimate
or schedule.

\section{Sampling Detectors}
\label{sec:sampling}

Sampling detectors have long been used in neutrino experiments to 
economically build detectors which nevertheless have adequate event
information to identify a neutrino interaction and calorimetrically 
measure the the total energy left by the neutrino.  To use a 
sampling detector for a $\nu_e$ appearance measurement, one takes
advantage of the fact that the radiation length of materials is usually 
a factor of 3 to 10 smaller than the interaction length, and therefore
the final state electron produces a very sharp burst of energy very 
close the event vertex.  A typical analysis using a fine-grained 
calorimeter can be found in \cite{para}, but the most important features
are cuts on the longitudinal energy profile of the event, and the 
ratio between the electromagnetic energy and the total visible energy
of the event.  Because the segmentation assumed is so good, the 
energy resolution of the proposed detectors is in fact slightly better 
than the width of the peak of the off-axis beam itself, and ensures that 
the gains realized in principle by going off axis can be achieved 
in practice.  In the following two sections we discuss 
absorber issues and readout options separately, for the most part.  
However, it is clear that ultimately there will be some amount 
of coupling between the choice one makes in each category.  

\subsection{Absorber Issues}
Various active detector technologies have been proposed for fine grain
calorimeter options for  phase 1 (i.e. 20 kton) detectors.  A
number of these ideas appear feasible for meeting the required
background rejection and energy resolution but with varying degrees of
affordability.

\par
The 5.4 kton MINOS Far Detector, which is over two-thirds complete
(October 2002), is
the largest sampling calorimeter even built and provides an interesting
model for understanding fine-grained absorbers and detector fabrication
issues.  The design of a new fine-grained detector starts
with a
MINOS-like device.  We increase the number of samples 
by roughly a factor
of four, while simultaneously reducing the readout pitch by a factor of
two and we increase the 
mass by roughly a factor of four.  While the cost
of active detector elements and their readout systems is a significant
issue, the increased sampling required for these detectors also places
severe requirements on the absorber's material costs, structural design,
fabrication, and element installation.

\par
The steel absorbers in MINOS are also the structural elements used for
support of the active detector technology. Roughly 35\% of the far
detector's budget was evenly divided between installation manpower
and absorber materials. MINOS was able to purchase low cost absorbers by
approaching the steel industry and accepting rejected grades of
mass-produced steel (e.g. steel too brittle to be used for stamping auto
body panels) and by using low-tolerance industrial processes (the edges
of the plates are raw submerged arc plasma cuts) to achieve significant
savings. Due to adoption of standard high-volume industrial processes,
the cost of the iron plates is dominated by the bulk cost of the
materials.

\par
About a third of the labor costs for MINOS installation are associated
with it being deep underground and magnetized. Neither of these is 
a requirement for a new NuMI off-axis detector
The remaining labor costs
are roughly split between:

\begin{itemize}
\item Installation and cabling of the detector components
\item Receiving, staging detector components, and rigging
\end{itemize}

An off-axis fine-grained calorimeter will have many more  active
detector, components and absorber layers (as much as a factor of twenty
more for some options) but only a factor of four more mass.   To achieve
an affordable detector careful attention needs to be paid to materials
costs.  However, an even greater level of vigilance needs to be applied
to the fabrication methods and integration to avoid having the absorber
and the detector installation dominate the detector's cost.

\par
The lessons from extrapolating MINOS include:

\begin{itemize}
\item Try to use mass-produced standard commercial materials,
\item Try to use standard automated fabrication processes developed for
bulk industrial applications,
\item Any operation done many time is expensive so efficiency,
creativity in design, and industrial automation are crucial for a cost
effective detector construction.
\end{itemize}

\par
As mentioned in the introduction to this section,
physics considerations influence the choice for preferred absorber
materials. Low-Z materials allow more mass for the same sampling pitch
and hence decrease the amount of instrumentation in the detector at
constant mass.  This drives one to consider them the front-runners for a
sampling design. On the other hand the relatively low cost,
availability, and structural properties of steel make it attractive to
keep as a considered option.

Issues and status of four broad classes of absorber 
materials will be
outlined in this section:

\begin{itemize}
\item Iron
\item Low Z solids
\item Low Z granular media
\item Low Z liquids
\end{itemize}

The first two have the advantage of combining the function
of absorber with structural support.
The granular media could
either be loose and held in a container or molded into a solid.
Finally, a liquid, such as water would need vessels to contain the
liquid and those vessels could also be part of the detector's structural
integration.

\underline{\bf Iron}
The availability and structural properties of steel make it attractive.
A reasonable sampling depth in a steel detector would be 
0.5\,cm. It would be fairly straightforward to design a calorimeter
using large, thin, suspended steel planes that also support the active
detector elements.  The steel plates would be rolled, plasma cut, and
plasma punched to the required shapes.
Detector elements would be tacked to
the sheets with welds, and the assembly hung from support trusses over
the detector.  Since this detector would not be magnetized, the steel
does not require a laminated design as was done for MINOS.  Hence, the
welding and rigging processes would be significantly more efficient than
those used in MINOS.

A fine-grained steel detector would be a fairly straightforward
extrapolation from MINOS and the design is not particularly aggressive
structurally. The bulk cost of iron is higher than some of the low Z
materials mentioned later but has the advantage of being well understood
structurally.

Since it is well understood how to develop a detailed design for a steel
calorimeter and it is not favored based on its implications for active
detector costs, no significant resources should be expended to further
an iron-based design unless comparably cost effective
 structural solutions
cannot be practically achieved in the low Z options.

\underline{\bf Low Z Solids and Granular Media}

Low-cost low-Z materials include recycled plastic, pulp products, and
agricultural products\cite{LOI}.  They could either be solids like
molded plastics, materials mixed with a curing agent like
particleboard, or loose materials in plastic containers.  This is an
area that requires further investigation on a number of fronts.

Custom molded plastics have been investigated and while feasible appears
to not be cost effective \cite{vic}.  The raw materials appear to be
cost effective but a packaging concept remains to be developed. Recycled
plastics have also been investigated. One major concern is availability
and uniformity of the materials in sufficient quantities
\cite{Richards}. Finally, use of agricultural materials such corn
byproducts (used in many packing materials) is cost effective but has a
number of issues \cite{Richards}.

Any of these substances could also be blown as loose material into a
vessel. The technical issues faced by this option are similar to those
described in the following section on liquid containers.

Many of the granular materials could also be glued into a structural
solid.  To be pursued, any forming process would require identification
of industrial scale facilities and, possibly, an R\&D program with an
industrial partner to be economical. One of the most interesting
existing products is particleboard.  A common construction material in
the US, it is low cost, and can be formed into a variety of planar
geometries.  As mentioned elsewhere in this document, the properties of
particleboard would allow one to make it the structural element in the
detector.

R\&D related to integration based on a particleboard absorber would be
required and should be pursued.

\underline{\bf Water}

Water is inexpensive
 and has both reasonable Z and low density.  A feasible
design has been proposed using extruded PVC containers with a matrix of
internal walls and end seals to contain the liquid.  These vessels would
be shipped empty, installed, and filled in situ. This has the attractive
feature that the components are light while being rigged and assembled.
Work related to this type of container was carried out in the design
phase of one of the proposed MINOS active detector technologies
\cite{border}.  The end seal technology was developed using standard PVC
sealing materials and an injection molded cap.  The costs of the
extrusions and caps are not prohibitive and the estimated labor costs
appear
to be reasonable \cite{heller}.  This is an area worth continued
investigation related to technical production and design features for
ease in integration and fabrication.

\subsection{R\&D on absorber issues}

An overarching theme for any fine-grained calorimeter design is reducing
the costs in materials, fabrication, and installation labor.  To produce
a very massive but fine grained detector will require that absorber
fabrication, detector integration, and installation concepts be given
significant attention early in the conceptual design process.  Issues
include a detector integration design with specific active detectors,
route for extrapolation to industrial scale production, integration with
a structural design, and effective concepts for reduction of the
detector installation labor requirements.

Specific recommendations:

\begin{itemize}
\item A very modest conceptual design program based on steel absorbers
could achieve a feasible design at relatively modest effort and well
understood costing.  Such a costing could be made using the vendors
currently making MINOS components.  It's main purpose, however, would be
as a cost benchmark for competing technologies.

\item There should be further investigation into 
an integrated particleboard design with
additional detector technologies as well as cost effective detector
integration/assembly schemes.

\item There should be further investigation of granular 
media in search of
alternative solutions to particleboard with increased flexibility and
cost effective containers for loose media.

\item Continued R\&D for water based absorbers should be pursued working
towards an integrated design and proven fabrication.

\item A final area for further attention
is integration of these 
detector concepts with the experimental facility's conceptual design and
integration, and outfitting.
\end{itemize}

\subsection{Solid Scintillator Detectors}

Solid scintillator technology has a long and successful tradition in
particle physics experiments, including several projects at FNAL.
These traditions involve past and current local R\&D and construction
efforts, ranging from MINOS, through CDF and D0 to CMS, as well as
designs of the neutrino oscillation project P860 and the STAR experiment.
All those projects triggered a lot of R\&D activity on the production
of high quality a lower cost extruded scintillator at Fermilab.
The detector element consists of a solid material with scintillating
properties, a WLS fiber, a light guide and signal amplifier.
A traversing charged particle loses energy in the
scintillator, a part of it characterized by the quantity called
scintillation efficiency, $\epsilon \sim$ 3\%, is deposited in form
of light.  Part of this light, defined by the fiber capture efficiency
$\chi \sim$ 5\%, is transported through the light guide and amplified
by a quantity $\Theta$ being a product of the Quantum Efficiency and
the gain, the latter being typically of the order of 5$\times 10^4$
for solid-state detectors and upwards of $10^7$ for PMTs.
The Quantum Efficiency of typical photomultipliers, used e.g. in
MINOS, is of the order of 13\%.

\subsection{R\&D Issues for Solid Scintillator Detectors}

A complete in-house facility at Fermilab includes a Scintillator Detector
Development
Laboratory, a Thin-Film facility, CNC routing and machine development.
The production of extruded scintillator may today proceed at a rate four
times higher than that for MINOS.
Better quality and some cost reduction
over MINOS are also expected, with costs falling possibly below \$5 per kg.

\begin{figure}[htb]
\hspace{3cm}\epsfig{file=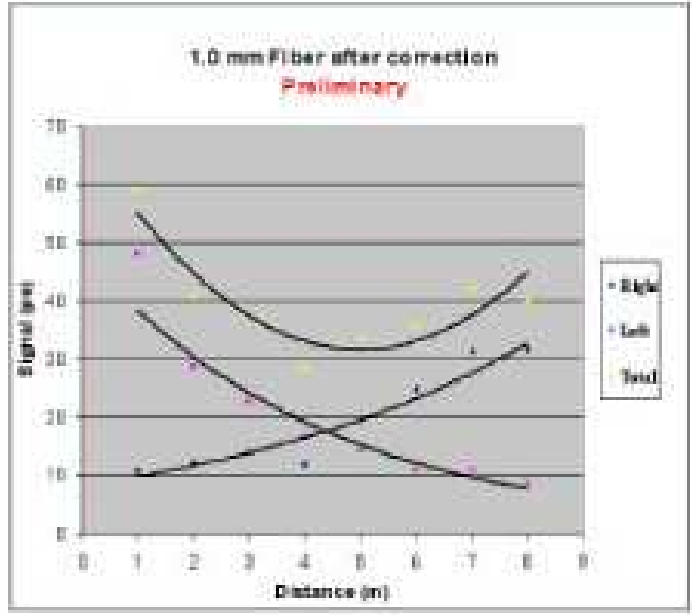,width=10cm}
\label{fig:bross1}
\caption{Photo-electron yields from a MINOS 4 cm wide extrusion obtained
by using a 1 mm thick WLS fiber and VLPC readout.  Data were corrected
by a factor $(1./0.965)^2$ to account for the fact that the VLPC cassette
uses 0.965 mm fiber.}
\end{figure}

\begin{figure}[htb]
\hspace{3cm}\epsfig{file=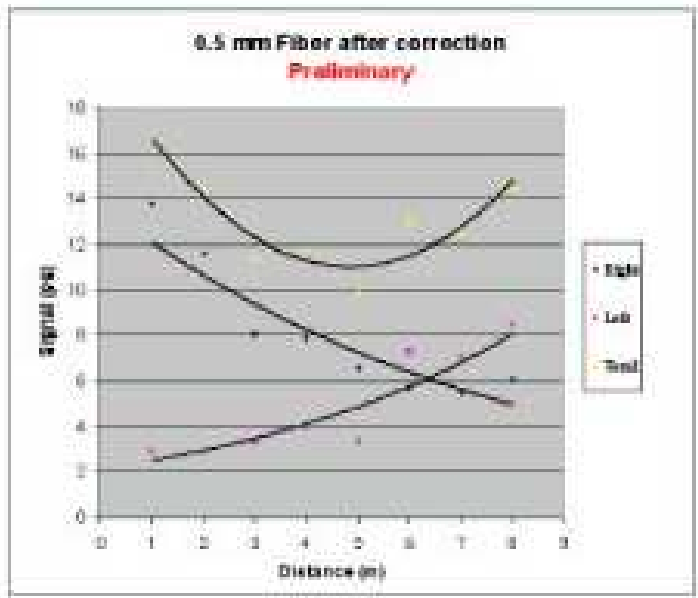,width=10cm}
\label{fig:bross2}
\caption{Photo-electron yields from a MINOS 4 cm wide extrusion obtained
by using a 0.5 mm thick WLS fiber and VLPC readout.  This as well as the
previous plots indicate that there is still something wrong (bad fiber,
connectors, etc.?), since the curves are not left-right symmetric.  A
better plot should become available from Alan Bross for the final document.}
\end{figure}

Other, perhaps more challenging, R\&D issues involve the fiber and
photodetector.  Readout optimization is a major component that largely
defines the overall cost of the detector.  Given a 5 times higher
longitudinal and 2 times higher transverse segmentation than for
the MINOS far detector, as well as a 4 times greater detector mass,
a linear extrapolation of the corresponding MINOS costs would lead to
unacceptably large numbers.  Essential cost reduction can be obtained
by applying currently available new technologies which have appeared since
the time of the MINOS far detector design.
Once physics simulations have specified fiducial volume, cell size, and
sampling fraction, optimization of a scintillator based detector is almost
entirely driven by the photo-electron yield of the basic cell to minimum 
ionizing particles.  A signal sufficient to do efficient tracking is normally
the sole criterion.  Optimization of the cell geometry, wavelength shifting
(WLS) fiber, coupling of the scintillator to the WLS fiber, and the selection of
an appropriate photodetector will determine the photo-electron yield and the 
final detector cost.  Since the cost
of the WLS fiber goes like the cross-sectional area of the fiber as does the
photodetector cost, reducing the readout fiber diameter has a tremendous impact
on the detector cost.
  Replacing the standard
MINOS PMT readout by a VLPC based one lead to
 an increase of the
Quantum Efficiency from 13\% to $\sim$80-85\%.  This should enable a reduction
of the original MINOS fiber diameter of 1.2 mm to 0.4 mm while keeping
a similar overall detector performance.
We have studied the yield from cosmic-ray muons traversing MINOS scintillator
extrusions that are readout with conventional WLS fiber of varying diameters
coupled to Visible Light Photon Counters (VLPCs).  Preliminary data for 1.0
and 0.5 mm fiber is shown in figures \ref{fig:bross1} and 
\ref{fig:bross2}.  At this point we see that the
photo-electron yield using 0.5 mm fiber and VLPCs is higher than the nominal 
yield in MINOS (which uses 1.2 mm WLS fiber).

Although at present relevant data exist only for the
MINOS 4 cm wide extrusions, it is expected that going to a 2~cm 
extrusion,
as proposed for NuMI Off-axis, may only somewhat improve the results.
This opens the possibility of an at least ten-fold cost reduction
compared to the numbers obtained from naive MINOS-based extrapolations.
Additional tests with different extrusions and the latest VLPC type
are due to become available shortly.  On the other hand, 0.4 mm is
believed to be near to the critical fiber diameter below which
attenuation effects become unacceptably large and this marks the
borderline of a possible cost reduction by going to thinner fiber.

The same data from the MINOS extrusion studies indicate a light
attenuation length along the fiber of $\sim$5 meters for a single
ended readout.  Accordingly, double sided readout was chosen for MINOS
to ensure a secure light yield at all positions.  The price is a
two-fold increase in the number of readout channels.  In a larger
detector (a detector 12 meters across is considered in the steel-scintillator
design), the issue is yet more important.  It has been proposed to overcome
this problem by applying the technique of mirroring of the far end light.
Tests of the D0 fiber tracker indicate that an effective attenuation
length of as much as 16 meters is obtained with a single ended readout
and far end mirroring.  This renders a single ended readout completely
sufficient for a 12 meters large detector.  With all the above
modifications, the obtained light yield can be expected to be equal
to or larger than that of the MINOS detector at any position.
New measurements are planned with both 4 cm and 2 cm wide extrusions
and with 1.2, 1.0 and 0.5 mm WLS fibers, both double ended and single
ended with a mirror at the far end.

A ten-fold cost reduction of the fiber is followed by a similar cost
reduction of the photodetector.  A 5$\times$10 element array with a
0.4 mm fiber would cost the same or less than a 2$\times$4 element
array with a 1 mm fiber used at D0 (about \$240).  Although a 10 times
higher density requires cold end electronics, this technology has been
successfully applied in the past by Boeing.  R\&D work on the next
generation VLPCs is currently under way in the Lawrence Semiconductor
Research Laboratory (LSRL), which has collaborated with Boeing, has
been previously awarded a phase I R\&D grant and recently a new phase II
grant for this purpose.  Their research aims at producing high density
VLPC arrays and demonstrating an order of magnitude cost reduction
compared to D0.

From all the presently known input, the total cost of a 20 kton
steel-scintillator detector has been recently estimated to about
\$100 M.  The ongoing extensive R\&D activity in the field can be
expected to reduce this number.
\subsection{Liquid Scintillator}
\underline{\bf Introduction}
Liquid scintillator is a proven technology that 
has been used in large quantities in detectors over long periods 
of time.  It gives good energy resolution and charged particle 
tracking efficiency.  Clearly liquid scintillator has properties 
similar to those of solid scintillator.  Its primary advantage is
 its low cost for a given photon yield.  Liquid scintillator allows 
for a flexible geometry and segmentation.
Because it is a liquid, it can be added to the detector 
after it is assembled to minimize assembly cost.  It can even be 
removed if repair is necessary or if the detector needs to be 
disassembled and moved.   
Modern off-the-shelf photonics 
can be used to readout the detector 
so that additional electronics is minimal.  
Scintillator segmented into cells and read out 
with wavelength shifting fiber gives both pulse
 height and position information.  In addition 
scintillator has a fast response and a negligible 
dead time.   Since this type of detector was investigated 
as a candidate for MINOS \cite{border}
very little additional R\&D and engineering is required 
to construct a very large detector from this technology. 

\begin{figure}[htp]  
\begin{center}
{\includegraphics*[width=\textwidth]{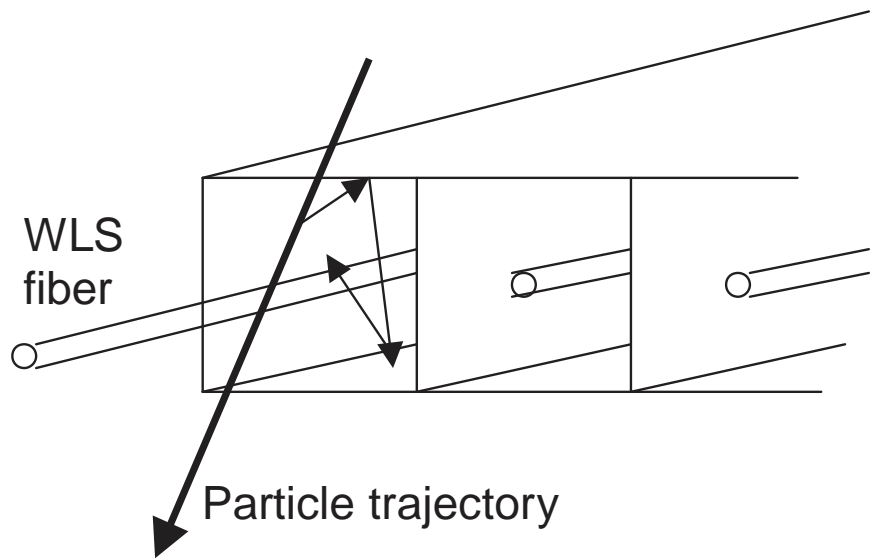}}
\parbox{\hsize} 
{\caption[ Short caption for table of contents ]
{\label{fig:scint1}
Structure of liquid scintillator detector assembled of 15 m long blocks of
PVC extrusions and filled in place.  Extrusions are self supporting in a culvert
covered with a roof supporting several meters of dirt and rock.
}}
\end{center}
\end{figure}

\begin{figure}[htp]  
\begin{center}
{\includegraphics*[width=\textwidth]{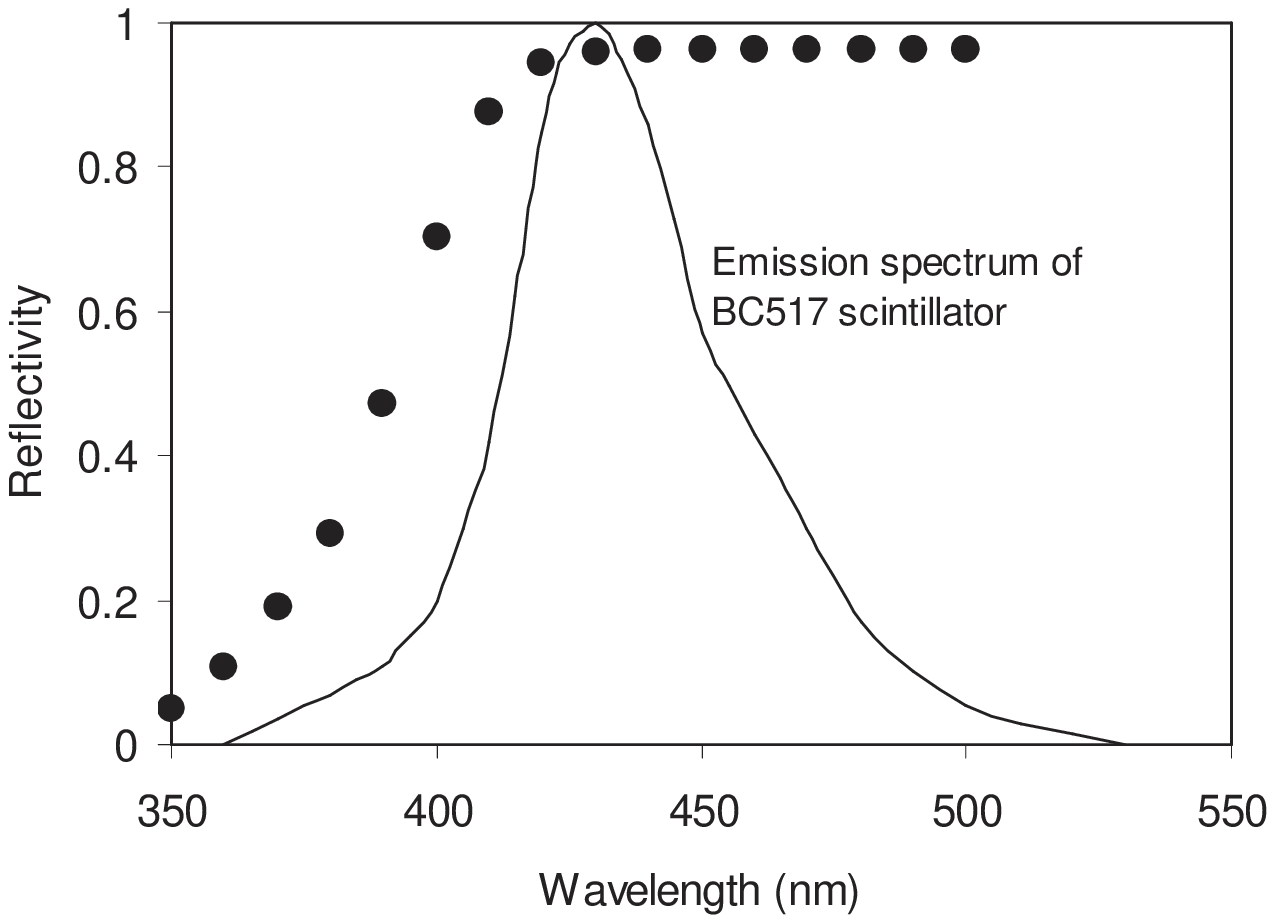}}
{\caption
{\label{fig:scint2}
A charged particle traverses one cell of a multicelled PVC extrusion holding
liquid scintillator.  After reflection by the sides of the extrusion, 
light emitted by the scintillator is collected by a wavelength shifting optical
fiber.
}}
\end{center}
\end{figure}

\underline{\bf Structure}
As a specific example, we present in Figures \ref{fig:scint1} and
\ref{fig:scint2}
a conceptual design based
on a mineral oil based liquid scintillator interspersed
between planes of absorber.  Water based liquid
scintillators do exist and would be less expensive with a
higher light yield.  However, they are more reactive than
mineral oil based scintillator and would require further
study to determine their long-term effects on the fiber.
The liquid scintillator would be contained in modules
consisting of long cells of extruded PVC plastic colored by
titanium dioxide for good reflection.  Light that signals a
charged particle's passage through the scintillator is
captured by a wavelength shifting fiber running the length of
each cell.  Although this is a conceptual design, its
structure has been investigated previously and those
investigations documented in two masters theses/cite{bib:masters}.
The design
we consider here is based on modules with the dimensions of
each cell 3.3 cm x 3.3 cm x 12 m long.  The outer walls are
1 mm thick while the inner webbing of the extrusion is 0.5
mm thick.  Although the cell sizes are chosen here for ease
of calculation and would be optimized by Monte Carlo studies
of neutrino events together with structural studies of the
plastic, they are close to those required for a final
detector and are structurally sound.  For each module, the
cells are sealed at one end by a single plug of PVC glued in
place.  This construction has been tested to be reliable
with no leaks to a pressure of over 6.5 atmospheres.
Alternate planes of scintillator modules would have
perpendicular cells to give an X – Y readout.

\par
\underline{\bf Signal}
\par
\begin{figure}[htp]  
\begin{center}
{\includegraphics*[width=\textwidth]{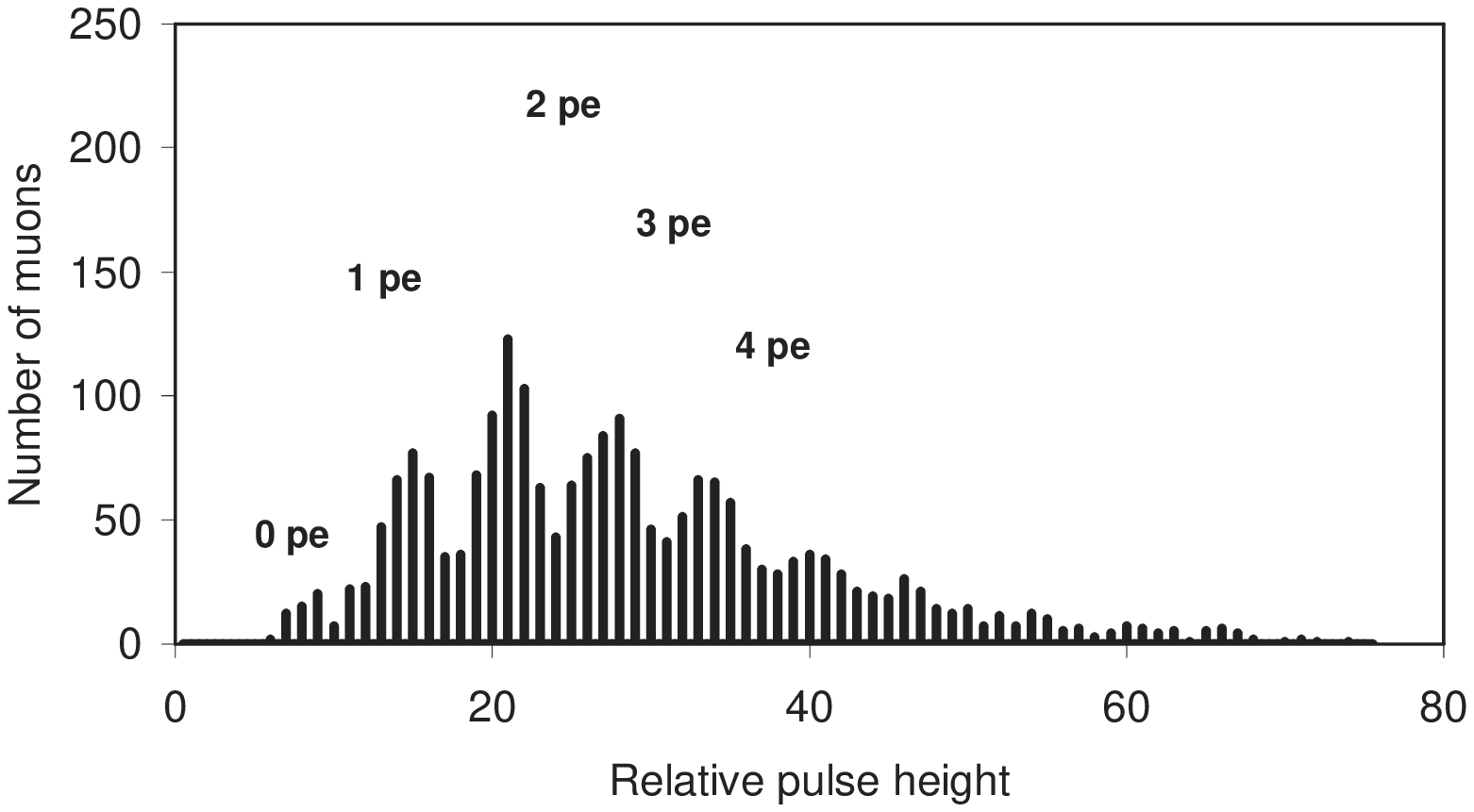}}
{\caption
{\label{fig:scint3}
The measured reflectivity of PCV colored with titanium dioxide and the
light emission spectrum of liquid scintillator.  The reflectivity is 0.965
at 425 nm.
}}
\end{center}
\end{figure}
\begin{figure}[htp]  
\begin{center}
{\includegraphics*[width=\textwidth]{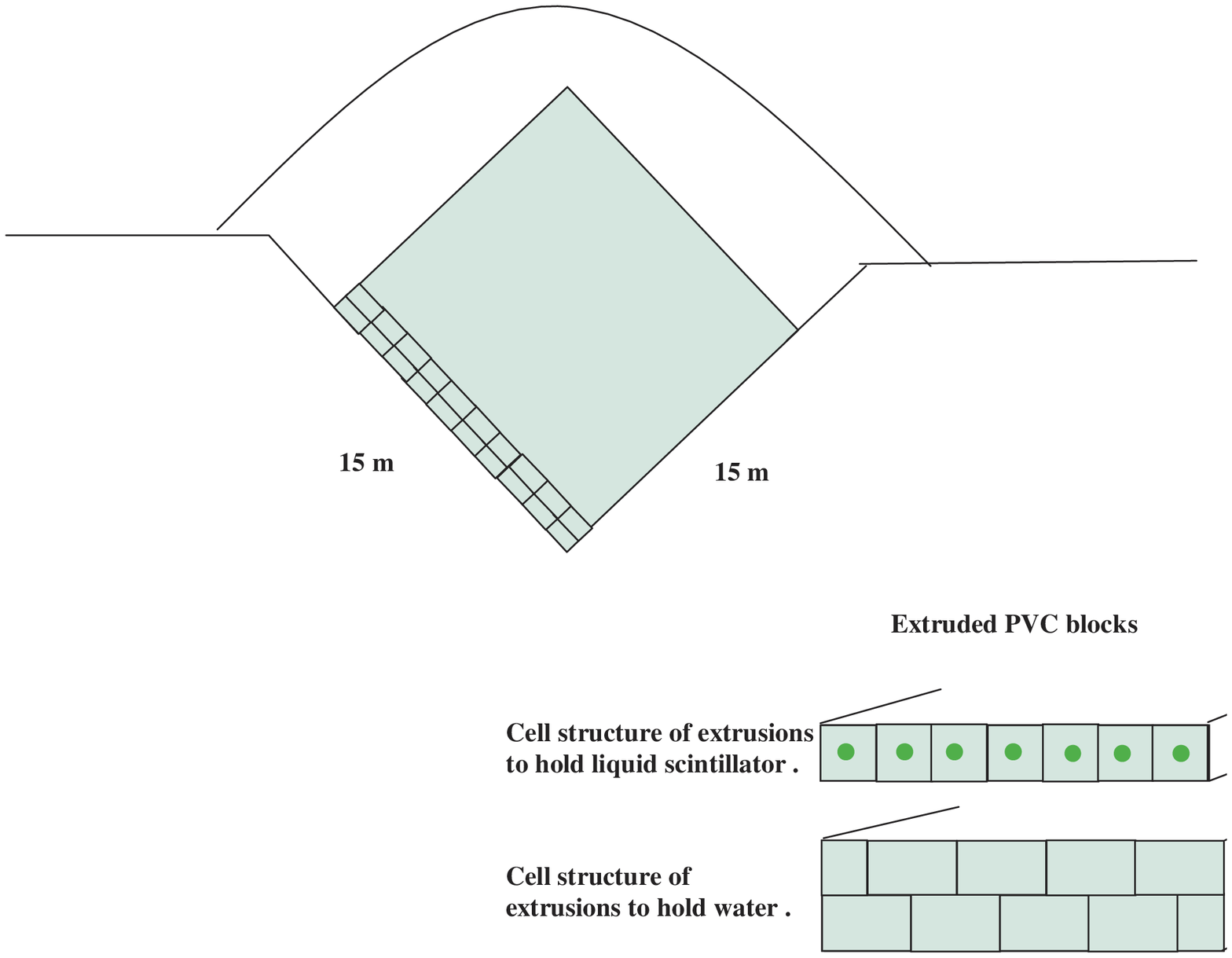}}
{\caption{\label{fig:scint4}
The measured pulse height from the far end of a 7.5 m long extrusion with cross
sectional area 2.1 cm thick x 2.8 cm wide with using minimum ionizing
particles (cosmic ray muons).  This gives an expected yield of
40 photons at the end of a 15 m liquid scintillator (BC517L) PVC extrusion
with cells 3 cm square, optimal reflectivity, and reflective fiber ends.}}
\end{center}
\end{figure}

Light could be collected using a 1 mm diameter wavelength
shifting fiber similar to that used in MINOS.  The light
output of each cell depends on the concentration of fluors,
the thickness of the cell, the width of the cell, the
diameter of the fiber, and the length of the fiber.  
Previous studies have shown that the amount of light
collected by the fiber is not sensitive to the position of
the fiber in the cell.  The end of the fiber will dipped in
white paint and then in epoxy to provide 30\% reflectivity at
the end.  Tests have shown a minimum ionizing particle this
geometry with BC517L scintillator will give 40 photons from
the end of a 12.5 m fiber.  The fibers from module would be
gathered in a manifold to an optical connector similar in
design to that of MINOS.  The design of the PVC manifold
would also include fittings to fill and, if necessary empty
the cells of scintillator.   Chemical activity tests show
there is no measurable effect of BC517L scintillator on
either the fiber or the PVC extrusion over the lifetime of
the experiment.  In Figure \ref{fig:scint3} we show the emission spectrum
of BC517 scintillator, and in Figure \ref{fig:scint4} is
the measured pulse height spectrum from a test of a 7.5 m long
extrusion.


\underline{\bf Support}
The construction of the detector is simplified because the
light plastic extrusions would be stacked to form the
detector before the liquid was added.  Mounting the plastic
extrusions with alternating planes at $45^o$ to the horizontal
and $45^o$ to the vertical would allow the detector to be read
out and filled from the top.  A possible mounting structure
would be a V shaped trench.  The sloping walls could be
supported by earth in a manner similar to culvert wall
construction.  The longitudinal direction the trench would
be sloped at about $1^o$ for ease of construction and to allow
drainage for water seepage from the ground.   The stacked
extrusions would be supported primarily by the sides of the
trench and at the ends of the trench by a bookend structure.
Most of the weight of the detector is supported by
compression on the floor of the trench carried by the entire
length of the extrusion.  In this configuration, all
mechanical stresses are well below the tested strength of
the extruded structure.  We envision that the detector would
be covered with a roof, possibly of Quonset hut design,
which would then be covered by about 3 meters of earth.
This would ensure a stable operating temperature for the
detector and eliminate the soft component of the cosmic ray
flux.
\underline{\bf Readout}
Image intensifiers provide a low cost readout well matched
to the rates of cosmic rays through the detector.  Reading
out fibers into a standard 25 mm image intensifier will
require one image intensifier for every plane of the
detector.  Standard image intensifiers have a quantum
efficiency of better than 10\% in the green and a gain of at
least 105.  An image intensifier would be read out by a
video camera.  The CCD in a modern off-the-shelf video
camera can be gated to have an exposure of as little as 10
microseconds and can be read out at 30~Hz.  For an exposure
gate of 20 microseconds for a detector on the surface, the
occupancy rate of a single cell would then be about 1\%.  The
360 cells of each plane of the detector could be read out by
a standard single image intensifier and video camera.
Standard firewire readout would go into processor for each
camera, the processors would be sparcified and read into a
PC.  If necessary, several PCs would alternate spills to
assure adequate readout time.
\underline{\bf Cost Estimate}
 Now we estimate the costs of constructing the detector
described above.  These costs are based on quotes and
engineering estimates of the liquid scintillation detector
proposed for MINOS and detailed in Reference \cite{border} 
and in the
two engineering theses referred to above.

\begin{table} [h]
\begin{tabular}{|l|l|}  \hline
     Mineral oil:             & \$2.75/gallon\\ \hline
     Fluors:                  & \$2.09/gallon scintillator \\ \hline
     PVC extrusions          &  \$1.5/lb \\ \hline
     Image intensifier 25 mm diameter &  \$2k each \\ \hline
     Video cameras and optics for each image intensifier &  \$1k \\ \hline
    Wavelength shifting fiber 1mm diameter  & \$1.5/m \\ \hline
\end{tabular}
\end{table}

With the type of construction outlined above, we estimate
that a 20 kton liquid scintillator detector using water as
absorber planes could be built for less than \$50M.
Additional cost saving might be achieved by using a water
solvent instead of mineral oil and using higher photon yield
fluors so that the fiber could have a smaller diameter.

\subsection{R\&D Issues for Liquid Scintillator}
Here one can make use of the extensive R\& D that was already 
done for the MINOS experiment, but one oustanding issue would be 
the different requirements due to a more fine-grained transverse
segmentation.  Presumably the change of absorber material from a 
solid (steel) to a liquid (water) would make construction issues 
less complicated, since one could fill the absorber and readout 
extrusions simultaneously. Certainly a test would be required to 
ensure that the extrusions used would not leak, where the extrusions
must be the full size expected to be encountered in the experiment.

\subsection{Streamer Tubes and Particle Board}
\subsubsection{Costs}
\par
In any detector this large, the absorber must be inexpensive. An absorber 
with sufficient strength to support its own (and active) weight 
without additional structure will be the most cost effective. 
Particleboard laminated into blocks with slots for detectors is 
an attractive absorber candidate that may satisfy all of these 
requirements. In particular, particleboard price in large quantities 
is less than \$0.13/lb. 
\par      Minimizing the cost of the active elements and associated 
electronics is the major challenge. The per channel costs of the 
electronics (circuitry, cables, connectors, readout cards, crates, 
power supplies, assembly labor and testing) will be a major cost 
driver. A detector technology capable of readout in strips about 
3 cm wide and 10m long requires approximately 5 $\times 10^5$ electronics
channels. 
If shorter strip lengths must be used, the costs will increase 
dramatically. For example, if readout strips can be 10m long in x, 
but due to mechanical constraints only 2m long in the y, the cost 
multiplier is 5! A geometrical factor such as this will likely dominate 
the final electronics cost. A technology that requires amplifiers 
causing a fractional increase in the cost pales in comparison with 
a doubling of the number of channels. Thus cost considerations will 
likely push the detector solution in the direction of the longest 
possible (10 m) readout strips in both dimensions.
\subsubsection{Readout Geometry}
\par
      The required strip length can be obtained using full-length 
detectors, or by using interconnects to pass signals (and perhaps 
gas) between shorter detectors. Interconnects will be difficult to
 engineer, and for a single detector package performing x and y 
readout over a restricted area, interconnects are required in two 
dimensions. To illustrate, a detector package with dimensions
 2 m $\times$ 5 m, achieves n strips, 10 m long, over a 10m $\times$ 
10m plane 
using 10n electrical interconnects. Mass electrical termination 
may help reduce the cost of these interconnects, but the costs of 
signal concentration and fan out must then be included. 
\par
      In some technologies, no additional costs are incurred in 
constructing separate detector packages for x and y readouts. This 
allows the sampling frequency to be increased by up to factor of 2, 
over detectors that feature two-dimensional readout. Increased 
sampling should improve the interaction energy and vertex z 
resolutions. 
\par
      Based on these observations, to achieve the lowest cost we 
suggest that the following guidelines be used in choosing a detector 
technology. 
\begin{enumerate}
\item To minimize the number of interconnects, single detector 
packages should be constructed with the longest possible (e.g., 10 m) 
readout strips.
\item To retain absorber modularity, simplify assembly and maintenance, 
minimize the number of readout channels, and optimize 
resolutions, detector packages should readout only a single dimension.
\end{enumerate}
\subsubsection{Drift Tube Properties}
\par
      Drift tubes are a good choice for high quality tracking over 
large areas, though often at a high cost. In a long-baseline 
fast-spill neutrino experiment, however, the low (cosmic) rate 
environment and crude position resolution requirements, allow a 
relaxation of most constraints that make drift tube construction 
difficult and costly. 
\par
      For example, the electric field configuration and operating 
gas need not result in a fast saturated drift velocity with a linear 
time-to-distance relationship, nor is a mechanical construction with 
high position accuracy and precision over long distances necessary. 
Because the resolution goals of the neutrino experiment are in mm, 
it is acceptable to have imprecise location of the anode wires in 
the tubes and the tubes with respect to each other. 
\par
      Banks of 8 to 16 tubes, 10m long can be produced using an 
inexpensive thin walled plastic extrusion, stiff enough to be 
inserted into the absorber. A top notch extruding company has 
given us a quote for the 8-tube extrusion, 20 cm wide, of 
\$1.25/linear ft. This converts to a raw materials cost of
 about \$20/sq. meter of active detector. Other detector 
technologies have materials costs 5 - 10 times larger. 
Wire supports will be needed over the 10m length of tube. 
The supports with a slight tension on the tube wall, (or 
perhaps clearance), are attached to the anode wire at three 
(or more) spots along a tube and pulled through the tube during 
the wiring. Tubes of nearly any shape are acceptable and many 
gases can be considered.
\par
      Drift tubes can be run in either the proportional or 
limited streamer mode. In the latter, signals are large enough 
that amplifiers are unnecessary. The low rates, however, imply 
that the proportional mode could use a low cost, large 
integration time (slow), and a high signal to noise ratio 
amplifier. The proportional mode makes analog information (pulse 
height or width) available that might allow crude track counting 
for improved calorimetry or background rejection. With the amplifiers 
described above, signal detection over 10m should not be difficult.

\subsubsection{Integration of Absorber and Drift Tube Extrusions}
\par
\begin{figure}
\epsfxsize=10cm
\epsfysize=6cm
\mbox{\epsfbox{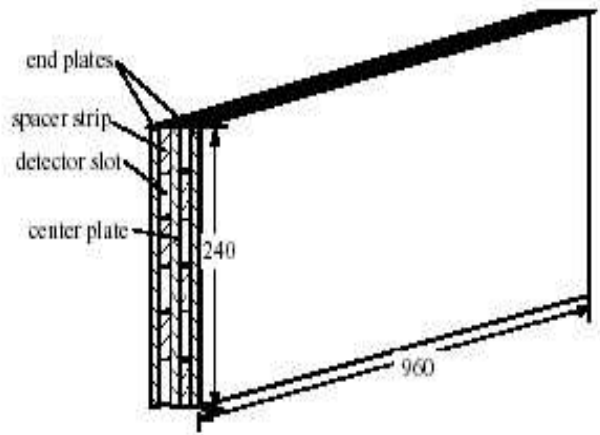}}
\caption{Absorber plates separated by spacer strips. 
Long detector extrusions are inserted in the slots formed 
between spacers. All are shown with equal thickness and not 
to scale. Dimensions are in cm.}
\label{fig:streamer1}
\end{figure}

        Creating a homogeneous interleaving of absorber and active 
elements is a common problem in calorimeter construction. The 
ATLAS Tilecal hadron calorimeter is constructed with alternating 
detector (scintillating tile) slots and spacer (absorber) strips 
attached on either side of a central absorber plate, with spacers 
offset so that detectors in one layer lie over spacers in the next 
layer. Thus in two closely spaced layers one has 100\% coverage 
of the area. A 2.4 m $\times$ 9.6~m absorber module having six long slots 
in which to insert detectors can be constructed in a similar way, 
as illustrated in Figure~\ref{fig:streamer1}.
\par
\begin{figure}
\epsfxsize=12.5cm
\epsfysize=6cm
\mbox{\epsfbox{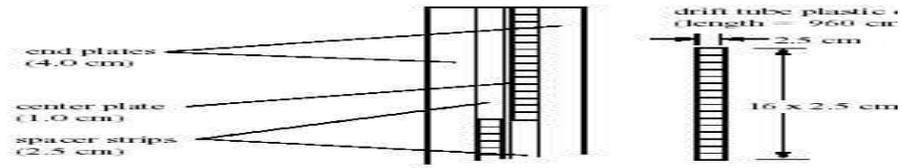}}
\caption{Detail of absorber plates, spacers, and extrusion. Total 
thickness is 14 cm. }
\label{fig:streamer2}
\end{figure}

    A more uniform coverage is obtained with a thin central plate, 
thick end plates, and a spacer matching the extrusion thickness (here 
2.5 cm), as shown in Figure~\ref{fig:streamer2}

\par
\begin{figure}
\epsfxsize=6cm
\epsfysize=6cm
\mbox{\epsfbox{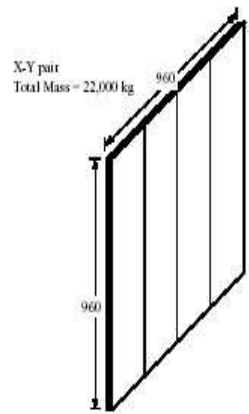}}
\caption{A quadrant of the full detector cross section, with both 
X and Y readout planes. The pair has a total thickness of 28 cm, 
and a mass of 22,000 kg.}
\label{fig:streamer3}
\end{figure}

    A 9.6 m $\times$ 9.6 m plane is formed using four modules (horizontal 
wires) stacked vertically edge to edge. A second plane of 
four modules, turned 90 degrees (vertical wires), is glued (e.g., 
with dowels) to the first plane, thus forming an 
X-Y pair, 28 cm thick, as shown in Figure~\ref{fig:streamer3}. The 
pair is supported on the bottom with vertical wires readout 
from below. Joining 30 of these pairs face to face creates a 
stable platform, 8.4 m long, on which to stack the upper quadrant. 
Assuming sufficient space is available, a drift tube extrusion can 
be removed and replaced at any time during the assembly.

\subsection{R\&D Issues for Streamer Tubes}
\begin{figure}
\epsfxsize=10cm
\epsfysize=6cm
\mbox{\epsfbox{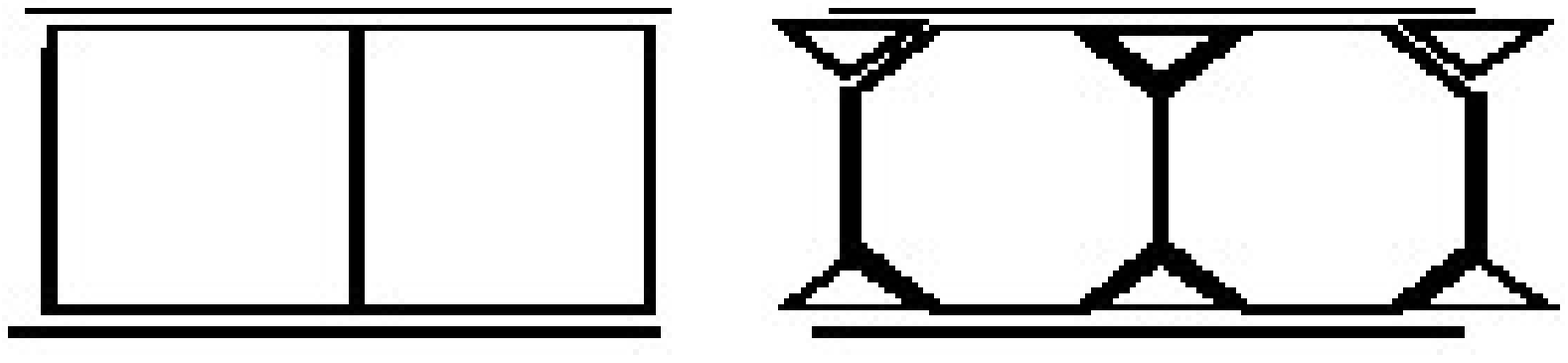}}
\caption{Examples of thin walled extrusions, shaped
 to form drift tubes. }
\label{fig:streamer4}
\end{figure}

    A multi-tube extrusion of carbon loaded plastic (high 
resistivity) looks like a viable method to produce a constant 
cathode potential that will not support a large spark or short circuit 
current. We have determined that an extrusion with a 2.5 cm wide 
cell, and with a wall thickness of about 1 mm (40 - 50 mils) is 
practical, and due to the broad angular spread of secondary tracks 
leads to only a small, $< 1\%$, track inefficiency. Tubes can take any
 shape that has a constant wall thickness, such as the examples with
 a flat exterior wall shown in Figure~\ref{fig:streamer4}. If limited to an
extrusion 
width of 8 (2.5 cm) cells, two extrusions would be joined to form 16
 adjacent drift tube cells. 

\par
    Choosing the ground and high voltage (HV) elements of 
the drift tube is a critical step in the design process. 
Placing the cathode extrusion at negative HV and the wire
 at ground potential allows electronics to be DC coupled 
to the anode wire. The extrusion, however, must then be 
well insulated from any external conductors, like those 
that provide the desirable signal transmission line 
properties. With the cathode extrusion at ground potential 
insulated external conductors are easily attached. The anode
 wire, however, is at positive HV and a blocking capacitor 
is required to de-couple the HV from the electronics.

\par
      Also, there are configurations where the external 
conductor is segmented into strips which are readout 
instead of (or in addition to) the anode wire. Each of 
these configurations has advantages and disadvantages 
that must be evaluated before choosing the appropriate
 one for the neutrino experiment. In the following design
 sketch we will assume that anode wire is at +HV, with the
 resistive extrusion and outer continuous conducting surface
 at (DC) ground potential. 
\par
      The development of a production extrusion die by a 
top-notch company will involve a large up front cost, 
estimated at \$50-100k. A pre-production 
die suitable for R\&D studies might be developed for less, 
but should not be counted on.

\par
\begin{figure}
\epsfxsize=10cm
\epsfysize=6cm
\mbox{\epsfbox{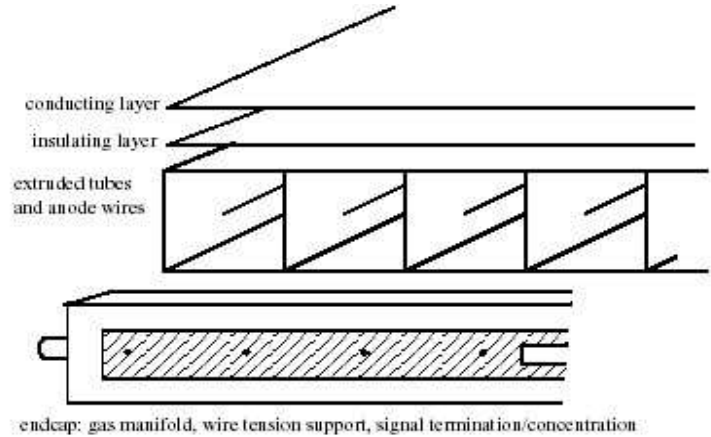}}
\caption{Conceptual design for an extruded drift tube endcap, 
featuring a gas manifold, wire tension support, signal termination
and concentration.}
\label{fig:streamer5}
\end{figure}

Drift tube endcaps must implement the following functions: 
\begin{enumerate}
\item anode wire location, gas seal, and tension support, 
\item  tube gas seal, inlet and outlet gas connections and manifolds, 
\item  HV distribution (far side) and (signal side) capacitor 
de-coupling, 
\item termination, and concentration of the anode signals
\end{enumerate}
Integration of these functions into a reliable, labor and cost 
saving product is a primary task in drift tube detector design. With 
100,000 extrusion ends to cap, one is well into the range in which 
plastic injection molding becomes cost effective. The endcap region 
is shown in Figure~\ref{fig:streamer5}

\par
R\&D will be necessary to minimize the cost of the drift tube 
readout electronics. Signal collection from widely spaced drift tubes 
can be a source of troublesome cross talk or deadly amplifier 
oscillations. Maintaining short and well-shielded transmission lines from 
the anode wire to the amplifier is one way to avoid these 
problems. Placing a differential output amplifier directly
 at the wire termination is another solution. Differential
 outputs allow signals to be safely condensed onto cables 
and transmitted to the (differential input) discriminators.
 Also, it may be desirable to convert large limited streamer 
mode signals to a differential form, before concentrating 
signals onto multiconductor cables. 
\par
      The discriminators are traditionally rack mounted, 
and require a costly cable run of about 400 wires in each
 quadrant layer. Using extrusion mounted discriminators 
and address forming circuitry, only words containing the
 wire address and time (10 MHz experiment clock) of a hit
 need be taken from a chamber. The words can be buffered 
and transmitted on a parallel (e.g., 32 lines) bus for 
about a factor of 10 reduction in cable cost, or in a 
serial fashion (e.g., ethernet) for about a factor of 
100 reduction in cable costs. Hit wire addresses and 
times would then be collected in memory devices located 
beside the detector for retrieval by analysis machines.
\par
      A 10 MHz clock produces hit time measurements in 
100 ns increments that translates, with a slow drift 
time, into a position resolution of a few mm for electron 
tracking. Cosmic rates should be just a few Hz per tube, and 
less than 1 kHz for single 10 m plane, which should allow 
either the parallel or serial readout schemes discussed above.

\subsubsection{R\&D Costs}

\par
      To develop an extruded drift tube technology will take a
 considerable effort involving 1 full-time or 2 half-time 
postdocs, an engineer half-time, a senior technician and, 
and a number of graduate and undergraduate students, 
costing approximately \$250k/year for 1-2 years. Particleboard, 
extrusion die, wire support, and endcap development will cost 
\$100k, mostly in the extrusion die fabrication. Proportional 
(with amplifier) and streamer mode tests, will cost another 
\$15k. Michigan State is interested in pursuing 
this program of R\&D.

\subsection{Resistive Plate Chambers}
Resistive Plate Chambers are an attractive possibility for an active
planar part of a massive neutrino detector because of their
simplicity and inexpensive components.  It may be the least expensive
choice per unit area, but still give excellent timing for
background rejection and adequate efficiency.
\par
\begin{figure}[htp]  
\begin{center}
\includegraphics*[width=5.5in]{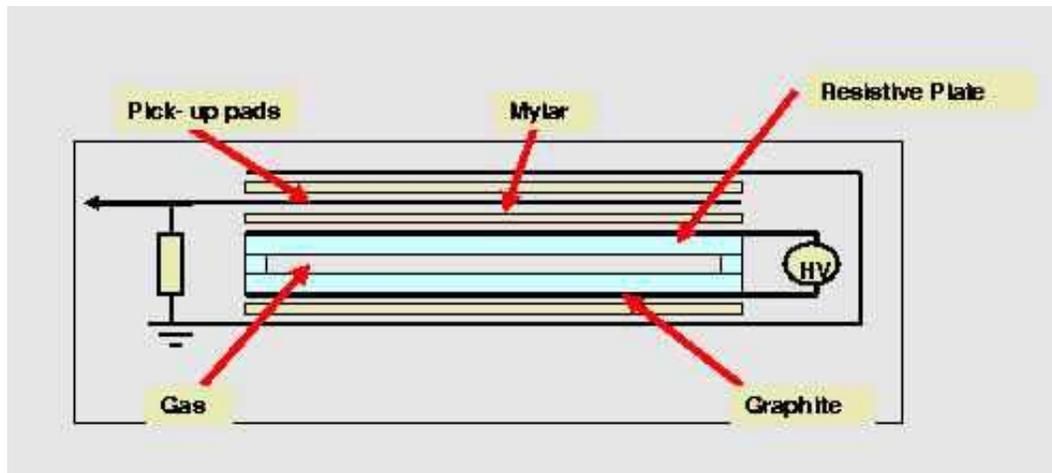}
\parbox{5.5in} 
{\caption[ Short caption for table of contents ]
{\label{fig:rpc} Schematic of 
a resistive plate chamber}}
\end{center}
\end{figure}

 
A sketch of a generic Resistive Plate Chamber is 
shown in Figure \ref{fig:rpc}. Two parallel plates of high 
resistivity, $\rho = 10^{10}$ to $10^{12}$ $\Omega$cm 
generate a uniform, intense 
electric field, about 4kV/mm, in a typically 2 mm wide 
gas gap. The plates are coated, on the external sides, with thin 
graphite layers connected to high voltage and ground, 
respectively. Due to their high surface resistivity of 
about 100 k$\Omega/\square$, these graphite electrodes are transparent to 
the transients of electrical discharges generated in the 
gas. Capacitive signal readout is therefore possible through pads 
which are insulated from the graphite carrying the high 
voltage by a layer of mylar.
\par

The simplicity of the concept of these chambers allows for a 
large variety of design choices. Two types of resistive 
plates have been used for the construction of most 
chambers: glass and bakelite. The advantage of bakelite is 
its somewhat faster recharging capability; however the optimal performance 
requires the application of a coat of linseed oil 
to the inner surface of the plates - a 
somewhat delicate operation. Chambers have been built with one 
single gap, some as wide as 8 mm, or 
multiple and smaller gaps for better timing resolution at 
uncompromised signal efficiency. The thickness of the glass plates, 
typically 2 mm, can be varied; however thinner plates 
will be distorted by the forces resulting from the 
high electric field between the plates. The resistivity of 
the graphite layer can be varied and will affect 
both the rate performance and the amount of cross-talk 
between adjacent readout pads. Finally, the chambers can 
be operated either in the avalanche mode (at a 
lower high voltage setting) or in streamer mode. The 
collected charge in the streamer mode is approximately a 
factor of 50 larger than in the avalanche mode. 
Different gas mixtures have been explored, some with the 
ability of efficiently suppressing streamers.

In general, the physics of RPC's is well understood and 
Monte Carlo programs exist which simulate the various physics 
processes occurring when the chambers are traversed by a 
particle. More details on the current status of research 
related to RPC's and of recent applications of RPC's 
in HEP can be found in the contributions to last 
year's workshop dedicated to these chambers.\cite{rpcwork}

RPC's are slow to recharge: typically recharging times 
of 1 (streamer mode) and 0.1 ms (avalanche mode) have 
been observed. The times depend on several factors, such as 
the resistivity of the graphite layer, the material and 
the resistivity of the plates and the applied high voltage.
Rates are not expected to be an issue for any neutrino detector
operated on the surface, but good timing is required in order
to reduce cosmic ray backgrounds.

\par The electronics for an RPC detector 
used in a neutrino experiment can be 
simple and inexpensive.  The signals produced are large, 
and require no 
additional amplification.  Instead of making 
a pulse height measurement, 
an approach might be to use a simple discriminator as the front end 
electronics.  When a signal is received and 
the associated discriminator 
fires, it would record a timestamp, which 
is essentially the value of a 
counter clocked continuously at a fixed rate.  The timestamp would 
constitute the data read out, 
along with a channel number that identifies 
a physical location.  The data would be read into a trigger farm, which 
would reconstruct the event based using 
algorithms that use the timestamp 
and physical location of the hit.  Accepted events would be written to 
disk, while noise hits would be rejected.  
Counter reset and clock speed 
are parameters based on trigger processing time, 
the number of channels,  
and total data throughput.  The algorithm 
for reconstructing events from 
timestamps is very similar to that 
used in MINOS.  This approach works 
well for detector systems with a low 
event rate and a low spurious noise 
rate.
\subsection{R\&D for Resistive Plate Chambers}

A research program on resistive plate chambers has
already been begun at Argonne to evaluate their potential
use for a calorimeter as part of a linear collider detector.
We propose to join that effort.  With support of about
\$60,000 in the next fiscal year (2002-2003), we could 
support gas tests and electronics conceptual design appropriate
for use of RPC's in an off-axis NuMI detector.  

The two efforts are described as follows:

\par I. We will initiate a detailed R\&D program to evaluate the 
merits of RPC's:
\begin{enumerate}
\item We will complete the evaluation of a 
small number of RPC's which we obtained from other experiments. 
\item	We will construct a small number of test chambers with various 
	glass and gas gap thicknesses
	 resistivity of the layers of ink (distributing the 
high voltage onto the glass)
	geometries of the readout pads.
\item	We will develop a readout system based 
on a one-level discriminator. This system                   
will be used to evaluate the different chamber designs 
and pad geometries.
\item	We will test these chambers in a 
cosmic ray test stand and evaluate their:
	noise characteristics
	signal strength versus applied high voltage 
and for different gas mixtures
	efficiency for the detection of minimum ionizing particles
	cross talk between adjacent channels
	long term stability
\item	Following the completion of the above tests, we 
will design and build a small test section of an 
(electro-magnetic) calorimeter, approximately 25 cm in all three 
dimensions. This test section will 
feature of the order of 10,000 readout channels. The 
electronic readout system will be based on a custom chip. 
The mechanical set-up will be designed such as to allow 
for easy implementation of other active media, as they might 
become available.
\item	We will test this calorimeter in particle beams which 
are available at the major particle physics laboratories, such 
as DESY and CERN. These tests will be important in verifying 
the functionality of the chambers and their electronic readout system.
\item
	Contingent on the successful tests of our small 
(electro-magnetic) calorimeter, we will design and build a test 
section of the hadronic calorimeter, sized $\sim 1 m^3$,
which is sufficient to contain hadronic showers both laterally 
and longitudinally. This calorimeter section will again be subjected 
to extensive tests in particle beams. 
\end{enumerate}
We expect to complete items 1) - 4) in FY 2003, 
items 5) - 6) in FY 2004 and initiate item 7) in FY 2005.

Engineering and technical effort during FY2003
The following engineering and technical activities are planned 
for FY2003:

\begin{enumerate}
\item	Construction of a small number of test chambers with 
different dimensions (glass and gas gap thicknesses.) This 
involves the cutting and gluing of glass.
\item	Development of a technique to apply layers of resistive 
ink (graphite) with different, but uniform thicknesses, leading 
to specific values of the surface resistivity.
\item	Design and production of spacers and rims (out 
of plastic) for the construction of the chambers needed 
for the assembly of the electro-magnetic size calorimeter.
\item	Design and building of a readout system for the 
test chambers and possibly for the electro-magnetic size calorimeter. 
The readout scheme will include only the digital information. 
The large number of channels of both the electromagnetic 
size calorimeter and the test section of the HCAL prevents 
the deployment of an analog readout.
\end{enumerate}

\par II. For Electronics R\&D, we propose 
to use discrete comparators for the front end 
electronics, and programmable logic for 
the timestamp counters.  We would 
build a small system using this approach (several hundred channels) to 
demonstrate the proof of principle.  Later, 
as a second phase, we propose 
to implement the design of the discriminator and the timestamp counter 
inside a custom integrated circuit.  While 
the design of custom circuits 
can be long and difficult, we believe that the circuitry is simple and 
straight-forward, helping to reduce the risk in development and cost.  
This development would be ideal for a national laboratory or 
university with established integrated circuit design groups.

\section{Other R\&D Issues}
\label{sec:other}
\subsection{Operation of a Detector at the Surface}
It needs to be determined whether 
cosmic-ray backgrounds to $\nu_e$ 
interactions will prohibit placing 
the detector at or near the earth's surface. 
To understand the scale 
of the issue, we consider a 20kT plastic/RPC detector 
option, outlined by a 
$20m\times20m\times100m$ box.  The cosmic-ray muon rate 
through the top surface of the detector would be roughly 300,000/sec 
\cite{pdg}.  Assuming the $10\mu sec$ spill length for single turn 
extraction at the Main Injector, and 
the currently foreseen repetition rate of
1.9$sec$, the duty factor of the 
beam will be approximately $5\times10^{-6}$.  
In 5 years of data-taking 
with a 50\% running efficiency, the total beam-on 
live time of the detector 
will be 400 seconds.   In this time, approximately 
$120\times10^{6}$ muons will traverse 
the top surface of the detector.  To 
achieve a cosmic-ray induced background 
on the scale of the approximately 40 
expected intrinsic beam $\nu_e$ events, we therefore 
need a rejection factor 
of approximately $10^{-7}$. 
 
            Although a substantial fraction of cosmic-ray 
interactions
will be easily distinguished from $\nu_e$ events, 
such simple cases may only 
ameliorate the rejection factor 
by an order or two of magnitude.  There are 
wide range of possible interactions 
which must be considered for the remaining 
five for six orders of magnitude.  For 
example, neutrons produced by muons 
which do not cross an active detector 
plane could resemble  $\nu_e$ 
interactions, particularly in a non-proportional 
detector such as the RPC.   
Neutrons 
produced outside the detector, nearly horizontal air showers, muon
decay, and effects not yet 
considered may be important on the scale of the 
rejection factor we need.
 
           Understanding such effects to this 
level is most likely beyond the 
reach of simulations, due to uncertainties on 
physics processes, efficiencies, 
and correlations among efficiencies.  Experience from 
previous (underground) 
experiments  with large cosmic-ray rejection requirements 
provides some 
guidance.   However, we are reluctant to 
use this experience to make a 
definitive statement on the feasibility of 
a surface $\nu_e$ detector due 
to several differences. These include the 
different event energy scale and 
signatures between an off-axis neutrino detector and 
many previous experiments 
(e.g. proton decay), the different detector geometry 
and technology, and the 
different characteristics of the cosmic-ray flux between the surface and
underground.

\par A Fermilab project to study potential cosmic-ray $\nu_e$ backgrounds
has been initiated.  The LoDen 
project has borrowed 20 RPC planes 
($2.2m\times2.7m$) built by Virginia Polytechnic Institute 
and State University 
as spares for the Belle experiment's 
muon system.  Although only a 20T detector will be built with 
these RPCs (with still less fiducial volume), 
the vast difference in mass is 
more than compensated by the ratio of the live 
duty factors.  Within a year of 
study with a test detector, there should be  
new insights into the cosmic-ray 
background issues for a large surface 
detector.


\subsection{Cross-Sections}
\label{cross} 

The measurements of $P(\numu\to\numu)$, $P(\numu\to\nue)$ and
$P(\numubar\to\nuebar)$ in high intensity superbeams at $\dmsq_{\rms
atm}$ will provide our most precise windows into the neutrino
oscillation parameters $\dmsq_{23}$, $\theta_{13}$, $\theta_{23}$ and
$\delta$ until the advent of next generation neutrino sources, such as
muon-based neutrino factories.  Degeneracies and correlations in this
multi-dimensional parameter space~\cite{degenerate1,degenerate2}
affecting the extraction of oscillation parameters from these
measurements will make it especially important to have multiple high
precision measurements at different baselines, energies and neutrino
and anti-neutrino beams in order to determine the neutrino mass
hierarchy and establish CP violation in neutrino oscillations.  The
precision of these measurements rests not only on the statistics which
can be gathered in proposed experiments, but in our abilities to
understand signal and background rates for oscillation processes.
These rates, in turn, depend on a detailed knowledge of neutrino
interaction cross-sections at the low energies proposed for these
experiments.  To illustrate the importance of cross-sections in these
measurements, let us examine three examples.


The measurement of the muon neutrino disappearance in the region of
the peak of the first oscillation maximum is the best method for
determining $\sin^22\theta_{13}$ and $\dmsq_{23}$.  The former
measurement comes from the depth of the oscillation ``dip'' at the
peak and the latter from the location of the peak in energy at fixed
$L$.  Realistic detectors either misidentify other particles as final
state muons or miss some of the final state particles with some finite
probability.  (For example, in water Cerenkov detectors, many
non-quasielastic interactions are misidentified as quasielastic
candidates.)  In general, these misreconstructions feed events from
higher energies into the ``dip''.  The background rates can be
predicted with detailed knowledge of the detector response, the flux
of high energy neutrinos from correctly reconstructed charged-current
events, and knowledge of the cross-sections.

%
%
$P(\numu\to\nue)$ is known to be small at the $\dmsq_{\rms atm}$ from
the CHOOZ non-observation of $\nu_e$ disappearance.  As illustrated in
Figure~\ref{fig:backgds}, there are significant background
from the electron neutrino component of the beam and from
neutral-current misidentification, primarily from single $\pi^0$
production.  To understand the former, particularly at low energy, it
is important to understand the variation of
$\sigma^{CC}_{\nue}/\sigma^{CC}_{\numu}$ with energy.  The latter of
course requires knowledge of the cross-section for the $\pi^0$
production, and particularly the expected $\pi^0$ final state energy
spectrum.

\begin{figure}[tbp]
\epsfxsize=\textwidth\epsfbox{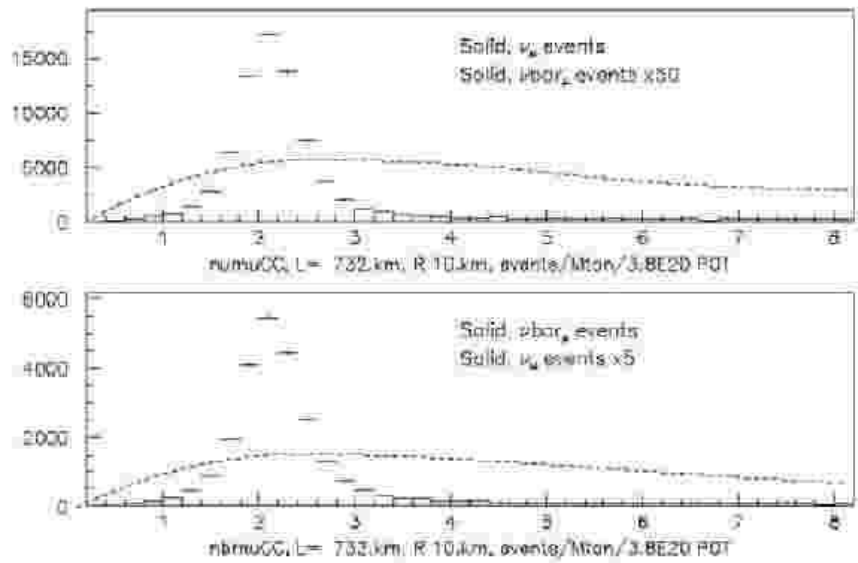}
\caption{$\numu$ and $\numubar$ interaction rates (no oscillations)
for neutrino and antineutrino beams.  Note the difference of a factor
of $10$ in the wrong lepton-number scale.}
\label{fig:wrongsign}
\end{figure}

Finally, comparison of $P(\numu\to\nue)$ and $P(\numubar\to\nuebar)$
requires an excellent understanding of the differences in the two
beams.  A major complication is the difference between the $\numubar$
interactions in the $\numu$ beam, which are a tiny fraction of the
$\numu$ events from the $\numubar$ beam as illustrated in
Figure~\ref{fig:wrongsign}.  To understand this difference, detailed
knowledge of $\sigma_{\nubar}/\sigma_{\nu}$ and its variation of energy 
for the final states of interest will be crucial.

The reconstruction strategies and backgrounds of neutrino experiments
in the $\approx 1$~GeV and $2$--$3$~GeV regions are very
different. Below $1$~GeV, the quasi-elastic cross sections dominate.
Some detector technologies favored in this region, e.g., water
Cerenkov experiments, can only observe the final state muon and are
not sensitive to recoil nucleons.  The energy of the events is
obtained under the assumption that the reaction was quasi-elastic.
Therefore for experiments in the $1$ GeV range, background predictions
require knowledge of how often inelastic scattering events are
misidentified as quasi-elastic events in the detector.  In the
$2$--$3$ GeV region, the inelastic cross section dominates with a
significant contribution from quasi-elastic events.  In this energy
region, the unobserved hadrons are very important.  Even detectors
which are sensitive to all hadrons, e.g., sampling calorimeters, have
a different response to charged and neutral pions as well as mesons
and baryons. The energy calibration and the misidentification of NC
events as CC $\nu_e$ events is very sensitive to the fraction and
fragmentation function of neutral pions in the final state.

\underline{\bf Physics of Low Energy $\nu$ Cross-sections}

At present, the neutrino differential cross sections for CC and NC
events, and the hadronic final states in the $1$--few GeV region are
not well understood. The lack of good data in this region limits the
physics capabilities of any future neutrino oscillations experiment.
The measurements of interest are total and differential cross sections
for charged current and neutral-current interactions with nucleons,
hadronic final states in CC and NC interactions with nucleons, nuclear
effects in the differential cross sections and hadronic final states,
and coherent nuclear processes.  In each case, the current data in the
has large uncertainties and often inconsistencies among experiments.

Theoretical models exist to characterize these
reactions and relate $\nu A$ scattering to well-measured $eA$
processes.  These models don't provide predictions for neutrino
scattering cross-sections and final states, but rather lend
theoretical guidance to interpreting precise low energy data as it
becomes available from measurements such as we describe.
Testing these models with precision low energy neutrino cross-sections
is crucial for future long-baseline experiments, and is also
interesting physics in its own right.

\underline{\bf Quasi-elastic Scattering}

\begin{figure}[t]
\begin{center}
\epsfxsize=0.8\textwidth\epsfbox{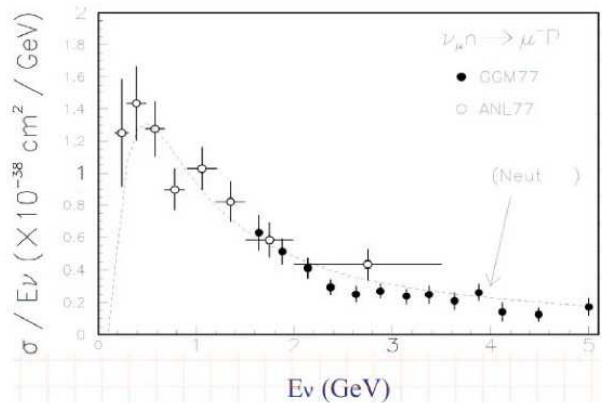}
\end{center}
\caption{Neutrino quasi-elastic cross section data on neutrons.}
\label{fig:figure1.eps}
\end{figure}

The physics of quasi-elastic scattering~\cite{qe} is described in
terms of nucleon weak and electromagnetic form factors.  Some
information on form factors comes from electron-nucleon scattering and
some from measurements with neutrinos. On nuclear targets, the effects
of Fermi motion must be included.  One approach is a Fermi gas
model~\cite{Bodek-Ritchie} which can take phenomenological parameters
from fits to electron scattering data on nuclei and apply them to
neutrino data within the framework of the same model.  For predictions
of far detector cross-sections, this is best done on a target of the
same nuclear composition.  The current data on quasi-elastic neutrino
cross sections \cite{sakuda} are shown in
Figure~\ref{fig:figure1.eps}.  The largest uncertainties for future
experiments come in this case from applying this data corrected to be
a ``free nucleon'' cross-section to nuclear targets.

\underline{\bf Inelastic Scattering}

\begin{figure}[t]
\centerline{\epsfxsize=\textwidth\epsfbox[55 90 530 734]{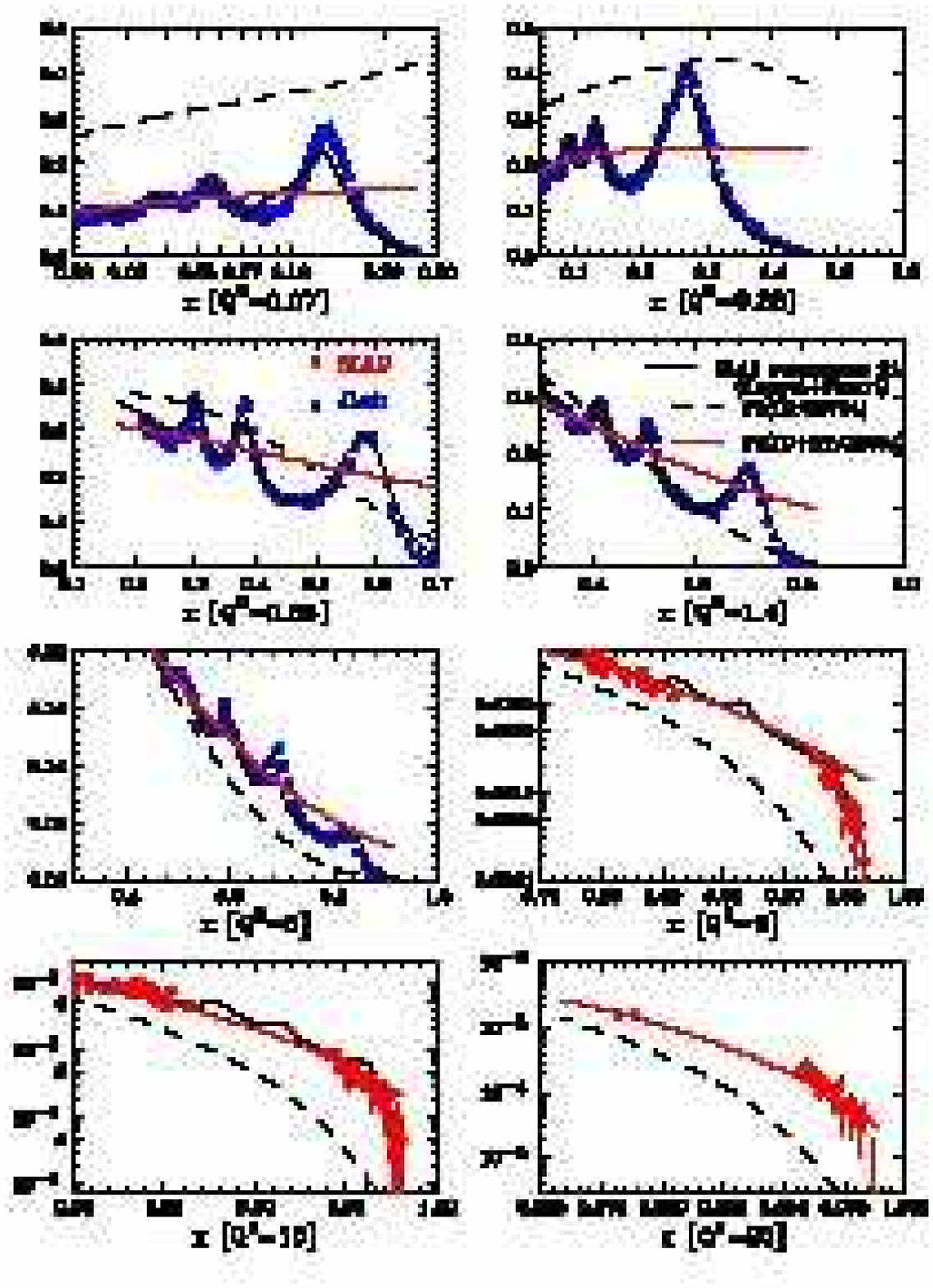}}
\caption{Comparison of a SLAC and JLab low energy
electron scattering
data in the resonance region (or fits to these data)
and the predictions of the GRV94 PDFs with (LO+HT, solid)
and without (LO, dashed) the Bodek-Yang modifications.}
\label{fig:duality_v3.ps}
\end{figure}

In general, all inelastic scattering is described in terms of three
structure functions $xF_1(W,Q^{2})$, $F_2 (W,Q^{2})$ and
$xF_3(W,Q^{2})$.  In the deep-inelastic region (high $W$) and at high
$Q^2$, the relationship between the structure functions measured in
electron and muon scattering and the structure functions in neutrino
and antineutrino scattering are given in terms of Parton Distribution
Functions (PDFs).  In contrast, at low $Q^2$ and in the low $W$
resonance region, a different picture is often used to describe the
data, including vector dominance models, resonance excitation form
factors, production of exclusive channels, etc. In the electron
scattering case, it has been shown experimentally that there is a
duality~\cite{bloom} between the cross section in the resonance
region, and the in the deep inelastic region.  One can relate
cross-sections in these two pictures by either resonance excitation
form factors~\cite{rs} or by treating the resonance as a final state
interaction (as a function of $W$ and $Q^2$) that modulates and
introduces bumps and wiggles into the average cross predicted from the
$F_2(x',Q^{2})$ fits to deep inelastic data~\cite{bodek}.  Models
using PDFs as the basis of a unified description of the inelastic
cross-section can, in principle, be used to predict neutrino and
antineutrino differential cross sections for all energies.  The cross
sections predicted by these modified PDFs can them be compared to
measured low energy neutrino data to determine the deviations from
this application of duality.  It is expected that there will be some
deviations from the expectations of this picture at very low $Q^2$ and
low $W$ because of the axial nature of the $W$ boson, and because
resonances of different isospin are produced at low $W$. In addition,
at very low $Q^2$ nuclear effects in neutrino and electron scattering could
be different due to different final state interactions.

Even if one has a complete description of electron scattering
differential cross sections as well as neutrino charged-current and
neutral-current differential cross sections at all $W$ and $Q^2$, one
still needs to understand the hadronic final states. Hadronic
production as a function of $W$ and $Q^2$ can either be described as
fragmentation of the final state quarks or in terms of decay products
of resonances.  Detailed measurements of final states neutrino
charged-current and neutral-current scattering experiments are
required to test any models relating final states observed in
charged-lepton scattering.  Particularly for neutral current
scattering, these measurements are most easily done in a narrow-band
beam where the initial neutrino energy is known.

\begin{figure}[tbp]
\begin{center}
\epsfxsize=\textwidth\epsfbox{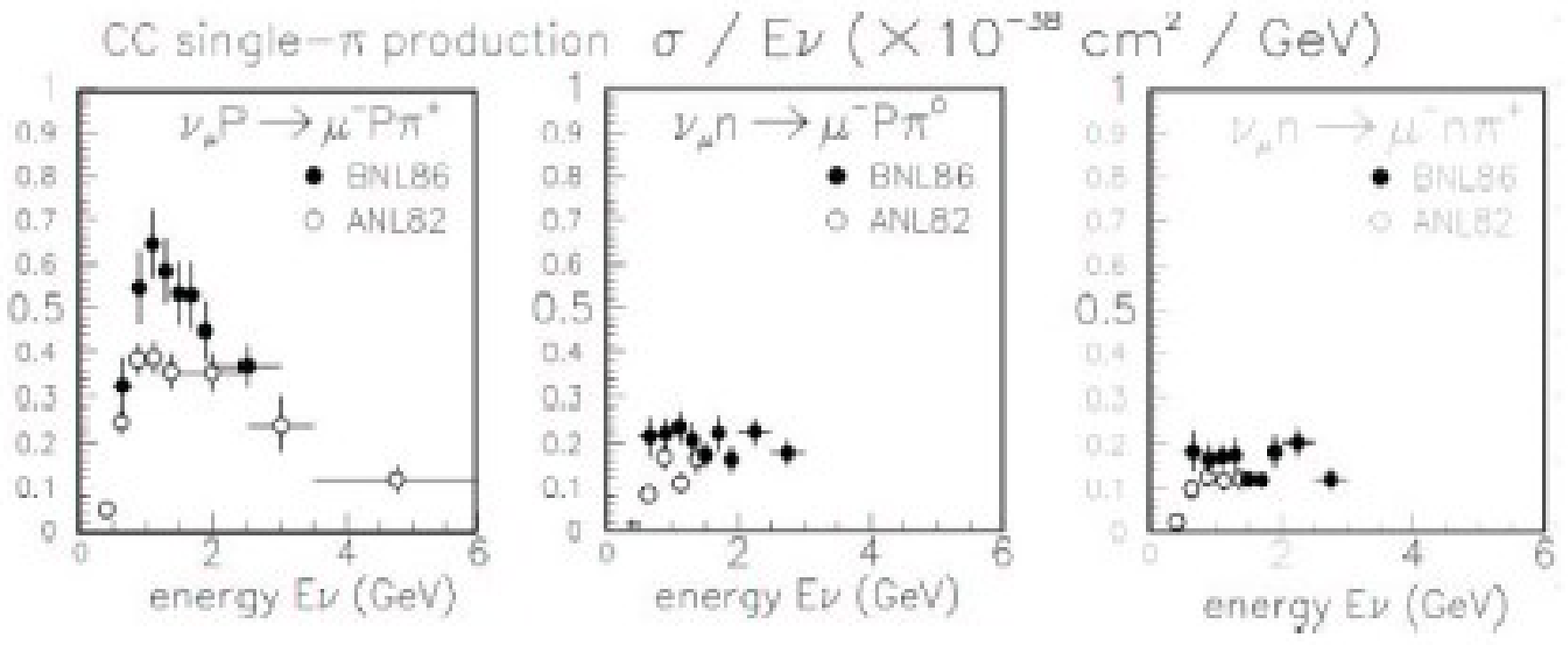}
\end{center}
\caption{ Neutrino charged current single pion production cross section
data. Even in this simplest channel, the errors are large and the
data are not consistent. Note that good measurements
of both the total cross sections
and kinematic distributions of all the final states are needed.}
\label{fig:figure2.eps}
\end{figure}

\begin{figure}[tbp]
     \centerline{\psfig{figure=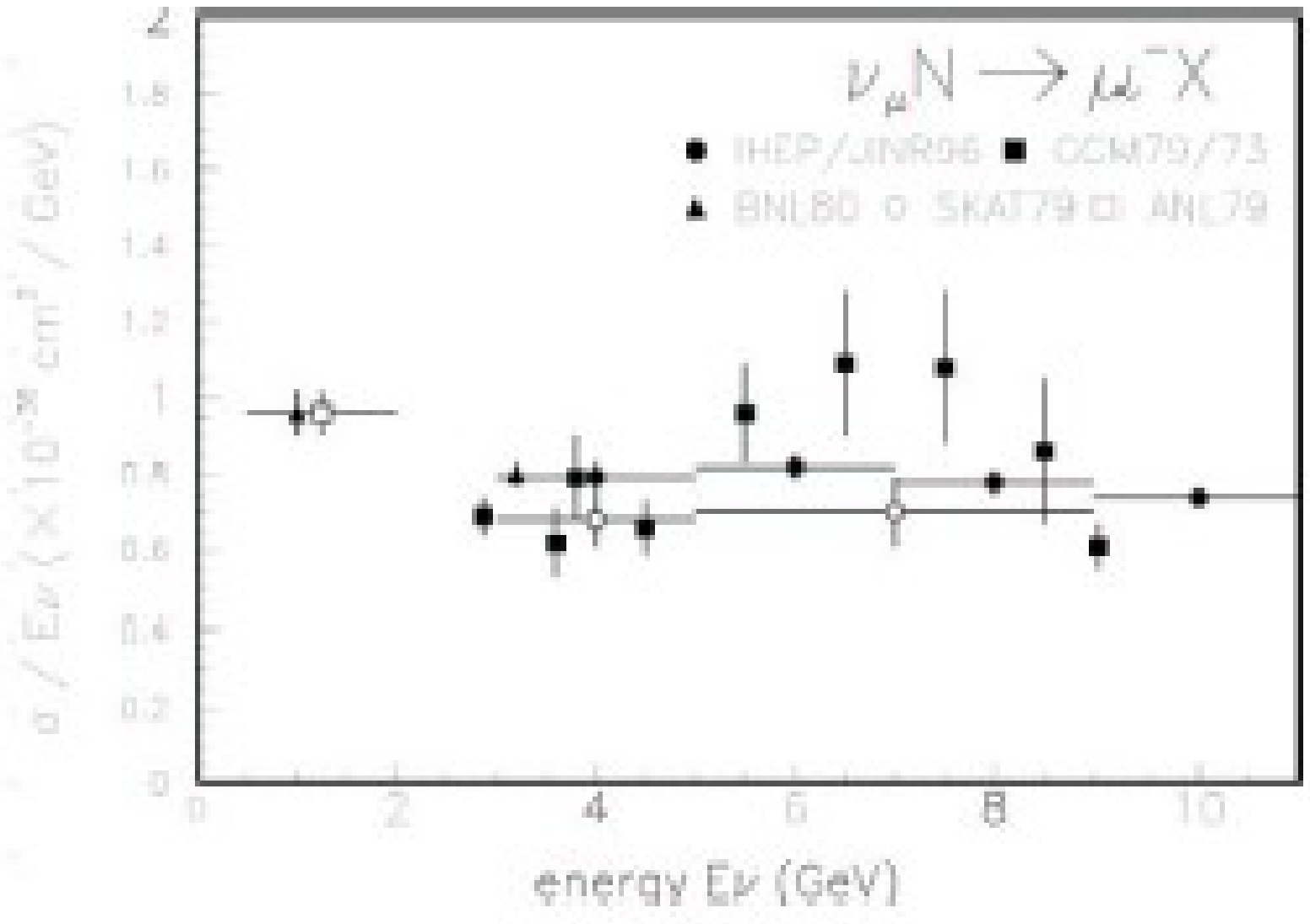,width=\textwidth}}
\caption{ Neutrino total cross section charged-current data (quasi-elastic plus
inelastic.}
\label{fig:figure3.eps}
\end{figure}

The woeful available data on the neutrino cross sections for one
exclusive final state (single pion production) and for the total cross
section (sum of quasi-elastic and inelastic) are shown in
Figures~\ref{fig:figure2.eps} and \ref{fig:figure3.eps}. As can be
seen in the figures, data from different experiments are not in
agreement and the errors are very large in the single pion
channel. Note that in addition to the total cross section for such an
exclusive process, the $W$ and $Q^2$ distributions must also be known
in order for these data to be useful for predicting backgrounds.

\underline{\bf Nuclear Effects}

The nuclear distortion of the inelastic structure functions,
originating from Fermi motion, nuclear energy binding, and shadowing
effects, have been measured in the deep inelastic region for the
structure function $F_2(x,Q^2)$ in electron and muon scattering.
Although not an axiom, particularly at low $Q^2$, most analyses
conventionally assume that for the same value of $x$, these effects
are independent of $Q^2$ or $W$ and are the same for the three
neutrino structure functions $2xF_1(x,Q^2)$, $F_2(x,Q^2)$, and
$xF_3(x,Q^2)$.  To understand the validity of this assumption in
modeling nuclear effects, measurements of $A$-dependent effects with
both electron and neutrino beams~\cite{marteau} at lower energies are
needed.  Furthermore, there is little data on nuclear effects on the
fragmentation functions.  These effects are expected to be significant
and at low energies these may be different in neutrino, antineutrino,
charged-current, neutral-current, and electron scattering, in part
because of final state effects, e.g., neutral pions being absorbed in
the final state nucleus or converted to charged pions.

Coherent particle production processes such as coherent
neutral-current production of neutral pions ($\nu A\to\nu A \pi^0$)
have to be modeled separately since these processes only occur on
nuclei.  The vector dominance models that are used to describe these
processes need to be constrained by direct measurement with neutrino
beams before they can be applied to predict enhancements to
backgrounds to neutrino interactions on nuclear targets.  Studying
such processes on the relevant long-baseline target material would be
possible in only in a high rate neutrino detector.

\underline{\bf Cross-Section Studies in a Near Detector}

\begin{figure}[tbp]
\begin{center}
\epsfxsize=\textwidth\epsfbox{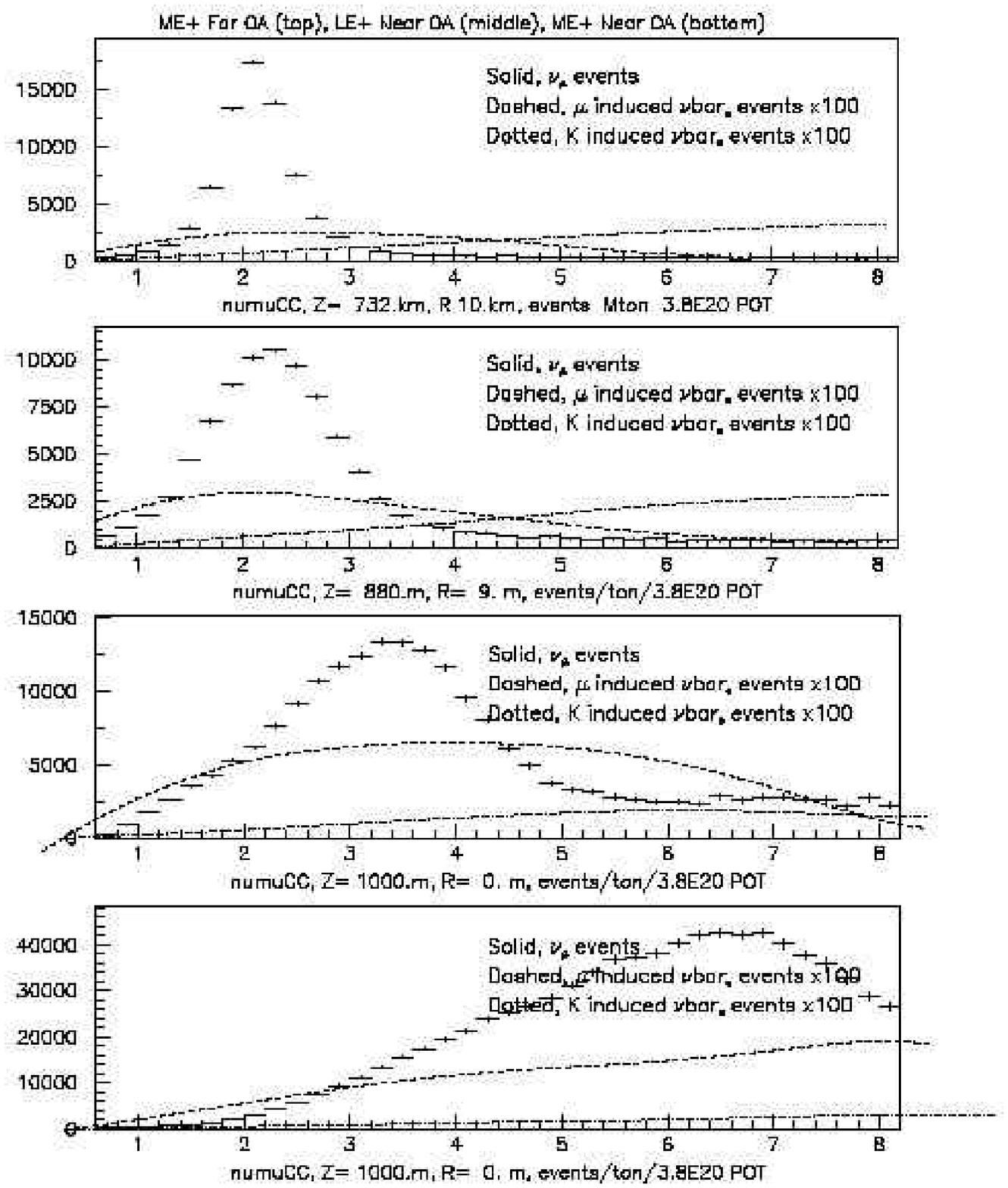}
\end{center}
\caption{Comparison of neutrino interaction rates for the
$0.6^\circ$ ($11$ mr) NUMI
off-axis  long-baseline detector (ME positive configuration, top plot), an off-axis near detector in the (ME positive configuration, second plot),
andand  the
MINOS near detector (third plot, LE positive; fourth plot, ME positive).
$\nu_\mu$ and
electron neutrino [($\nu_e$+$\overline\nu_e$)$\times$ 100]
interaction rates are shown (the contribution from muon decays is
is the dashed line and the contribution from kaon decays is the
dotted line). }
\label{fig:oa-on-2gev}
\end{figure}

Cross-section studies could be performed at high statistics in the
NUMI beam with either on-axis or off-axis near detectors.  As shown in
Figure~\ref{fig:oa-on-2gev}, the advantage of an off-axis beam is
that it provides a relatively monochromatic beam where the neutrino
energies for each event are well-known.  The energy of the neutrinos
depends primarily on the location of the detector, so for an off-axis
detector varying the beam energy would require varying the distance of
the detector from the beamline.  This approach with a nearly
monochromatic beam is particularly valuable for measuring
neutral-current cross-sections where the missing neutrino longitudinal
moment can be known from a beam constraint.  It also allows the direct
comparison of near and far detector rates because of the similar
spectra.  On-axis beams, by contrast, would have higher rates and
allow simultaneous measurement across a broad variety of energies,
include higher neutrino energies than in off-axis case.  Expressions
of interest to perform both kinds of experiments have been submitted
for consideration to the FNAL PAC~\cite{oa-EOI,on-EOI}.

\begin{figure}[tbp]
\centerline{\psfig{figure=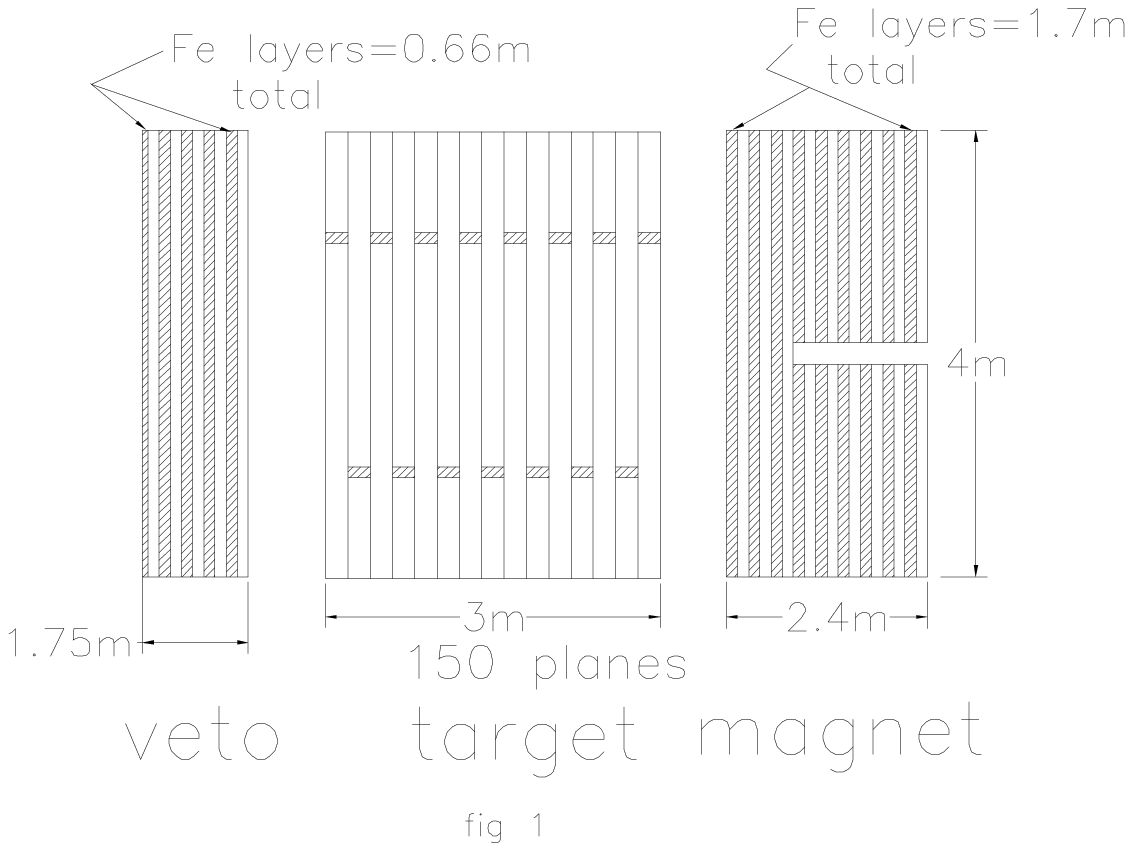,height=.5\textheight}}
\caption{Conceptual layout of the components of a detector suitable
for neutrino cross-section studies on hydrocarbons.  A $4$~m x $4$~m
active target, $3$~m deep, is followed by a sampling calorimeter and
magnetized range detector and preceded by an instrumented upstream
veto.}
\label{fig:protodetector}
\end{figure}

Because high rates can be achieved in small detectors, fully active
detectors are possible, such as a tracking calorimeter constructed
from scintillator strips with wavelength shifting fiber
readout~\cite{oa-EOI, on-EOI}.  A conceptual view of such a detector,
including side and rear sampling calorimeters, a front veto and a muon
ranger is shown in Figure~\ref{fig:protodetector}.  Such a detector
could also support multiple nuclear targets, including, for example,
thin radiators of high $A$ material interspersed in the target, a bulk
or segmented active volume of oxygen-rich material such as
water-miscible scintillator, or even a small test module of a liquid
argon TPC system.  A final possibility for such a detector,
particularly if off-axis, is a component or module to mimic a far
off-axis detector composition as nearly as possible in order to study
detector response to various final states that are perfectly tagged in
the near detector.  This would be particularly appropriate for a
sampling calorimeter, where analyses treating active target material
as absorber could be employed to precisely predict near detector
background and signal efficiencies.

\subsection{Targetry}
The target will remain a critical device in the NuMI
beamline after upgrade of the Fermilab accelerator complex 
with a Proton Driver.
The target must be able to survive a 2MW proton beam with some safety margin.
A good target for a neutrino superbeam facility would satisfy the following
conditions: 
it survives one spill;
a steady state temperature must not exceed a temperature stress limit;
the target lifetime is greater than 6 months;
in the case of multiple choice for a target candidate, the optimal
target is the one that provides the highest $\pi^+$ yield 
in the energy interval of interest.
The first three requirements are quite obvious. 
The last one is driven by the physics case.
An off-axis neutrino beam of a required energy of $\approx$ 3 GeV 
will be formed by secondary pions with energy in the 6--14 GeV interval.
Not all the pions of such energy will contribute
to the neutrino beam. A focusing system acts differently on
pions with different transverse momentum. Moreover, the focusing system itself
would not necessarily be the same for different targets. 
Therefore we use a simplified approach for these studies by
integrating over the $p_T$ distribution of the pion flux. \\
\underline{\bf What survives one spill ?} \\
Several commonly used materials were considered for a target.
The target was simulated with the \textsc{mars14} code~\cite{MARS}.
We assumed that a 120~GeV proton beam
hits a rod target with a length of two interaction lengths.
For the first trial, the beam was assumed to be a Gaussian with 
$\sigma_x$ = $\sigma_y$ = 1~mm (similar to NuMI parameters).
The target radius $R_T$ was optimized for the maximal pion yield
scanning the radius of the target with a step of 0.5~mm starting 
from 2.5~mm (see Table \ref{tab:edFirstTrial}).
In order to achieve a power of 2~MW, the nominal NuMI beam
intensity is rescaled by a factor of 5 that corresponds to
2$\times$10$^{14}$~protons per spill. We also assumed that the Main Injector
repetition period of 1.9~s does not change.

Calculated peak energy deposition densities (ED) on a beam axis are shown 
in Table~\ref{tab:edFirstTrial} for C through Hg targets.
For the beam conditions described above all the materials experience
a significant thermal shock. The stress limits
known for graphite, nickel
and inconel are about 1000~J/g. The limit
for copper is about 600~J/g.
As one can see from the table, none of the solid materials can survive such 
conditions.

\begin{figure}[htb!]
\begin{minipage}[t]{\textwidth}
\begin{center}
\epsfig{figure=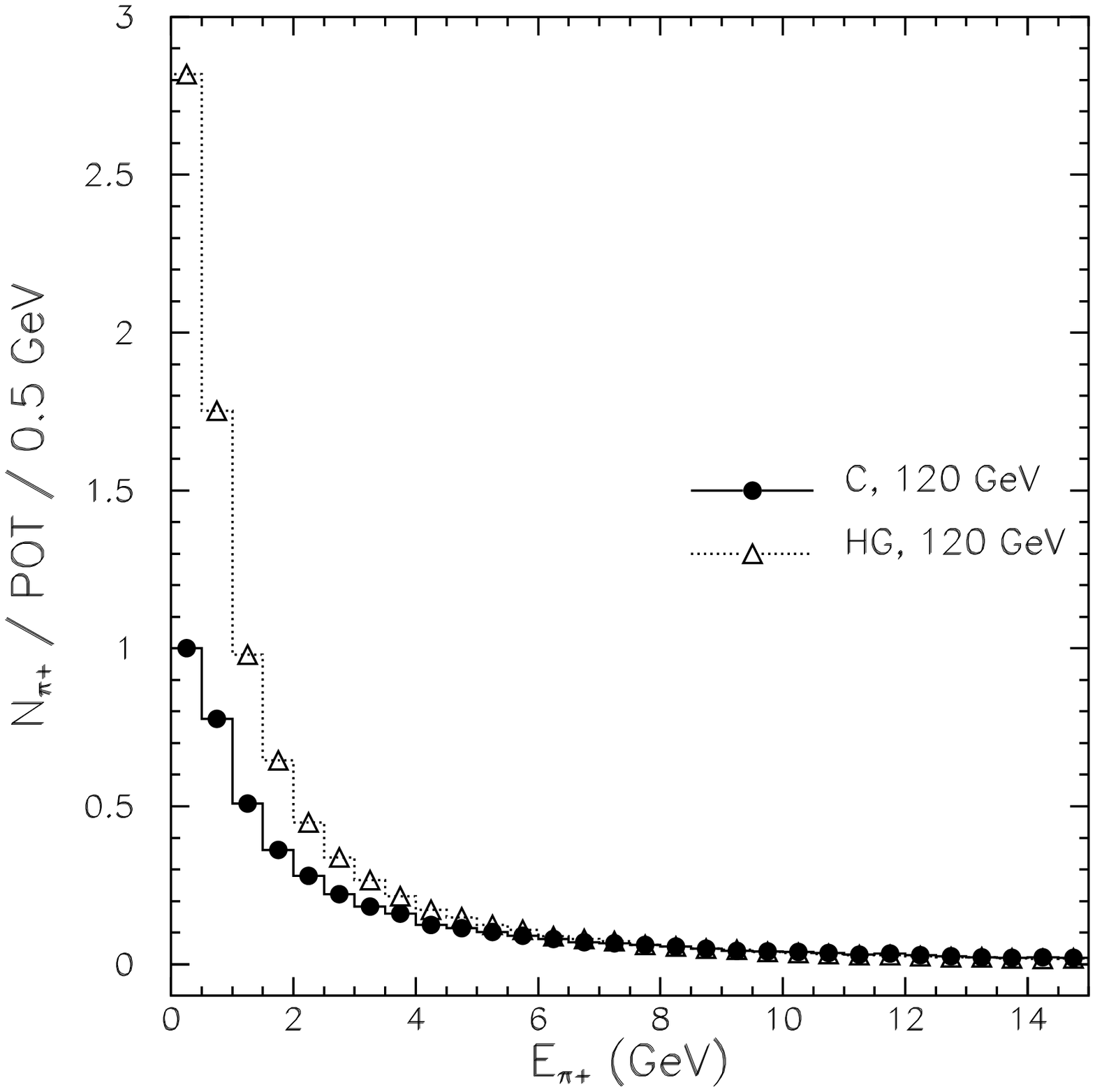, width=.7\textwidth}
\caption{Pion yield versus pion energy. 
The distributions are normalized per the number of protons on target (POT).}
\label{fig:yeildVsE}
\end{center}
\end{minipage}

\begin{minipage}[t]{\textwidth}
\begin{center}
\epsfig{figure=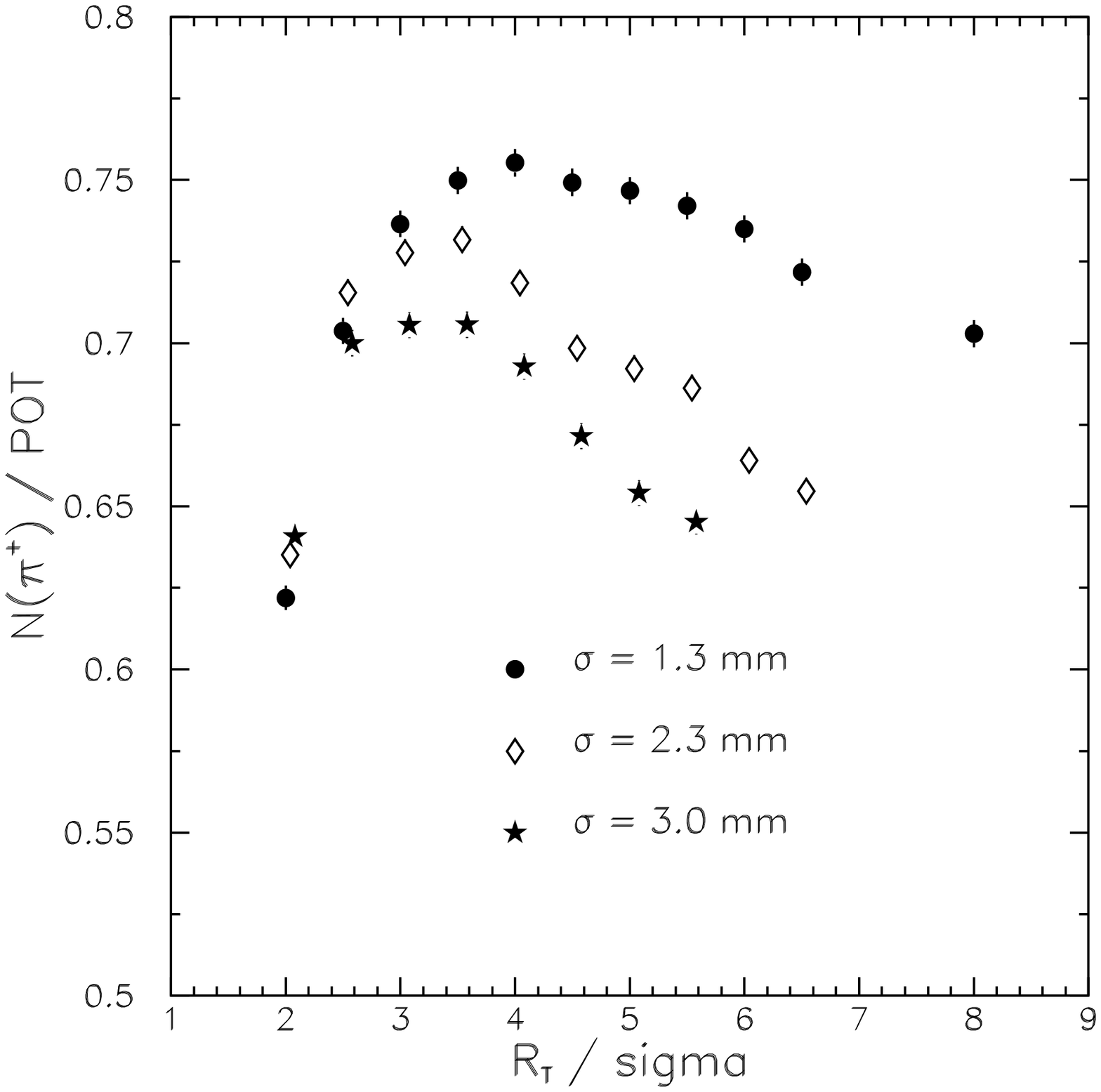, width=.7\textwidth}
\caption{Number of $\pi^+$ coming from a graphite target
in the interval 6--14 GeV against the value of target radius over beam sigma.
Shown are dependencies for different beam RMSs.}
\label{fig:yeildVsSigma}
\end{center}
\end{minipage}
\end{figure}

The pion yield in the defined energy interval
does not vary significantly with the target 
material (Table~\ref{tab:edFirstTrial}).
Figure~\ref{fig:yeildVsE} shows the pion yield versus pion energy
for graphite and mercury. Many more  soft pions 
are produced in a mercury target, but the yield is about
the same in the energy region of interest. 
Pion spectra for the other materials behave similarly.

\begin{figure}[htb!]
\begin{minipage}[t]{\textwidth}
\begin{center}
\epsfig{figure=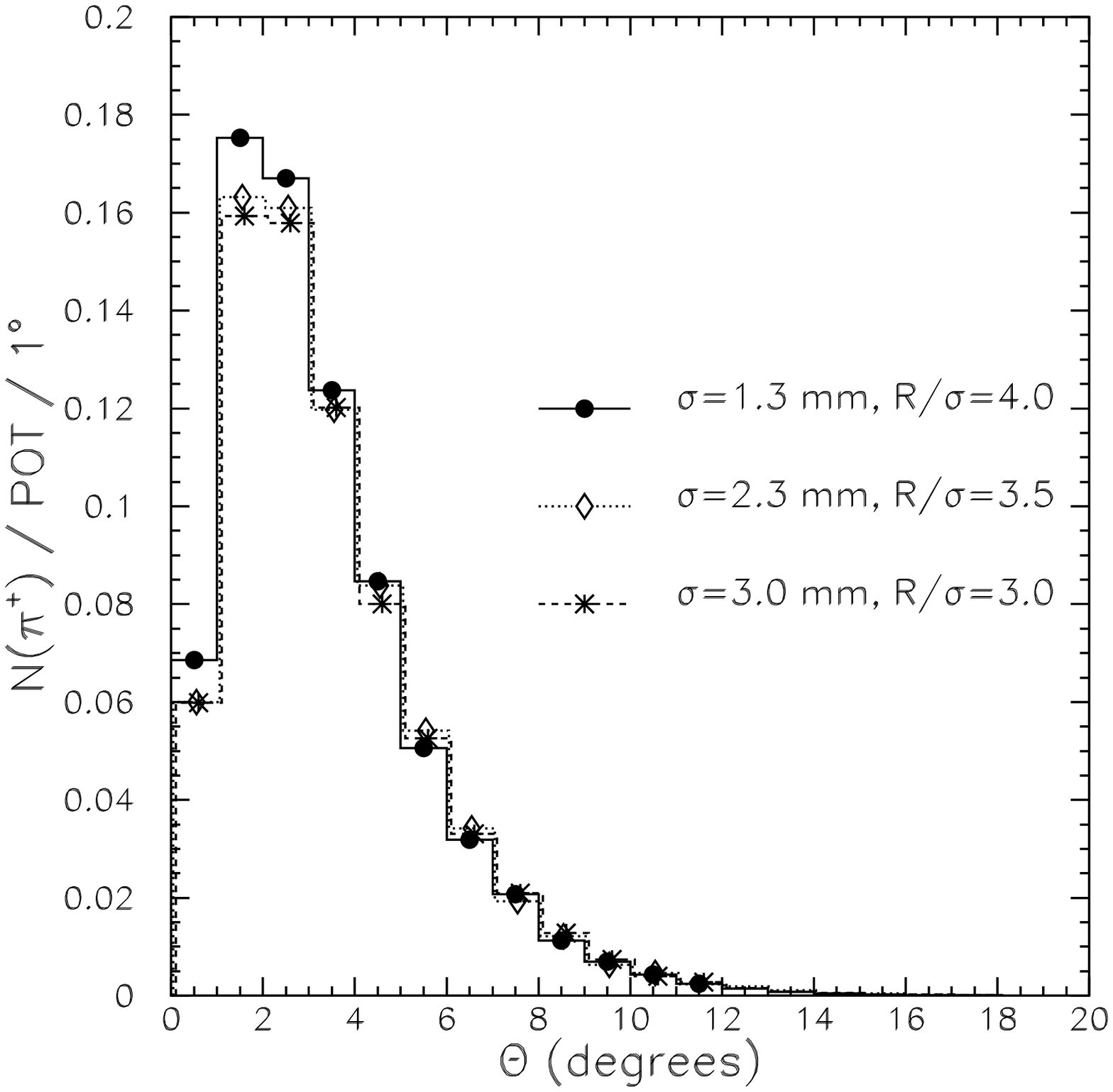, width=.7\textwidth}
\caption{Angular distributions for pions coming off a graphite target.
$\theta$ is an angle between the target axis and pion direction.}
\label{fig:angle}
\end{center}
\end{minipage}
\begin{minipage}[t]{\textwidth}
\begin{center}
\epsfig{figure=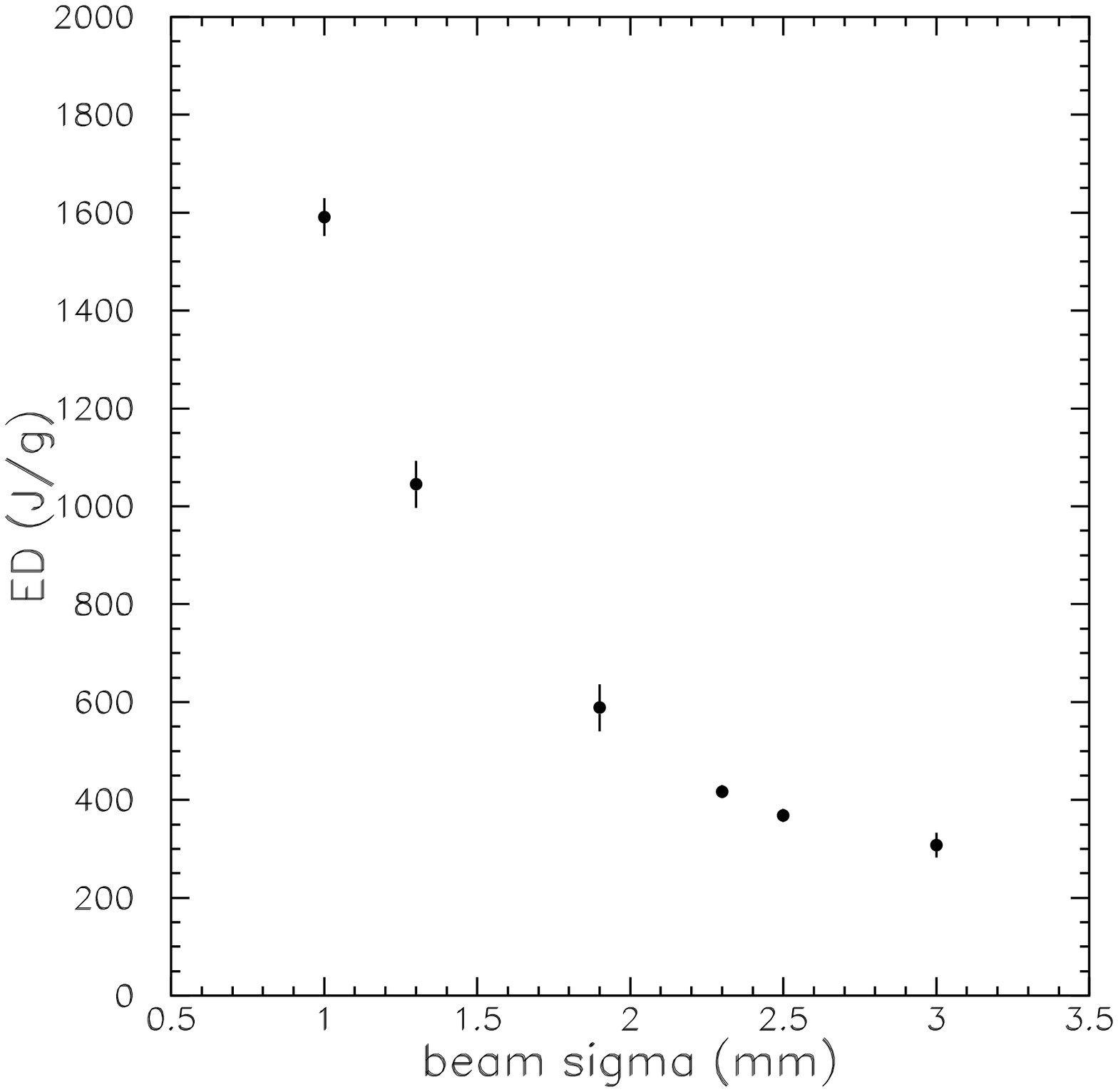, width=.7\textwidth}
\caption{Energy deposition in the hottest cell of a graphite target versus
beam spot size.}
\label{fig:ed}
\end{center}
\end{minipage}
\end{figure}

One obvious solution is to increase the beam transverse spot 
size. In the study below, both the beam spot size $\sigma$ = $\sigma_x$ = $\sigma_y$
and target radius $R_T$ are varied.
For a graphite target the pion yield as a function of
$R_T/\sigma$ is shown in Figure~\ref{fig:yeildVsSigma} 
for different $\sigma$.
As one can see the variation of the maximal pion 
yield for different $\sigma$ 
does not exceed 7\%.
This is due to the fact that pions leave the target from the sides
and not the end, 
that is atan(R$_T/\lambda_I$) $\approx 1^{\circ}$ (Figure~\ref{fig:angle}).
Energy deposition in the hottest cell of the target is acceptable at
$\sigma >$~1.5~mm (Figure~\ref{fig:ed}). For example,
for $\sigma$=3~mm and target radius $R_T$=9~mm the peak energy deposition
is $310$~J/g, well below the limit.
Thus the beam spot size and target radius for graphite can be substantially
increased compared to the NuMI parameters without significant 
loss of yield.

\begin{table}[t]
\begin{center}
\begin{tabular}{c|c|c|c}
\hline
Material  & Peak~ED~(J/g) & $\pi^+$yield (N/POT)  & optimal R$_T$~(mm) \\ \hline
Graphite  & $1581 \pm 18$         & $0.754 \pm 0.007$     &  5.0 \\ \hline
Nickel    & $6520 \pm 251$        & $0.684 \pm 0.006$     &  2.5  \\ \hline
Inconel   & $6011 \pm 259$        & $0.699 \pm 0.006$     &  3.0  \\ \hline
Copper    & $6084 \pm 216$        & $0.690 \pm 0.006$     &  2.5  \\ \hline
Indium    & $5248 \pm 149$        & $0.752 \pm 0.006$     &  3.0  \\ \hline
Mercury   & $10064 \pm 293$       & $0.715 \pm 0.006$     &  3.0  \\ \hline
\end{tabular}
\caption{Density of peak energy deposition and pion yield in the 
6--14 GeV interval for various targets.
Also shown are the optimal target radii $R_T$ for $\sigma$=1~mm.}
\label{tab:edFirstTrial}
\end{center}
\end{table}

The same idea does not work well for other solid dense materials.
An attempt to bring the energy deposition
safely below the limit by increasing $\sigma$ and $R_T$ results in
a substantial reduction of the pion yield.
For example, if one increases the beam $\sigma$ up to 12.5~mm keeping the 
ratio to be optimal $R_T/\sigma=2.5$, 
the pion yield from a
target reduces to
0.43 pions per proton, much smaller than the yield of about 0.71
from an optimal graphite target.
The density of peak energy deposition is 1250~J/g in this case
which is still too high.

\par Graphite is the most convenient material
for an FNAL neutrino superbeam facility. 
Indeed, the dense solid materials do not provide enough of pion flux at the acceptable
beam and target radii.
The use of light materials with large interaction lengths 
such as Li and Na will lead to too long a target, 
making focusing of secondaries difficult.
A mercury jet target seems to be too complex, expensive and hazardous device
and at the same time it does not provide any advantages over a graphite target
for the given experimental conditions.

\underline{\bf Target lifetime}

One of the factors limiting the target lifetime is radiation damage.
The lifetime determined here corresponds to the time when 5~dpa
(displacements per atom) occur in the hottest cell of the target.
The atoms are displaced due to interactions with hadrons
with kinetic energy of $>$0.1~MeV.
The 5~dpa limit corresponds to the integrated hadron flux of 
about 5$\times$10$^{22}$~cm$^{-2}$.
From the \textsc{mars} simulations we have found that 
for the beam $\sigma$=3mm and graphite target radius $R_T$=9mm
the limit of 5~dpa
is reached within 5 years and 8 months assuming 10 months of
operation per year at the full intensity.

\begin{figure}[htb!]
\begin{center}
\epsfig{figure=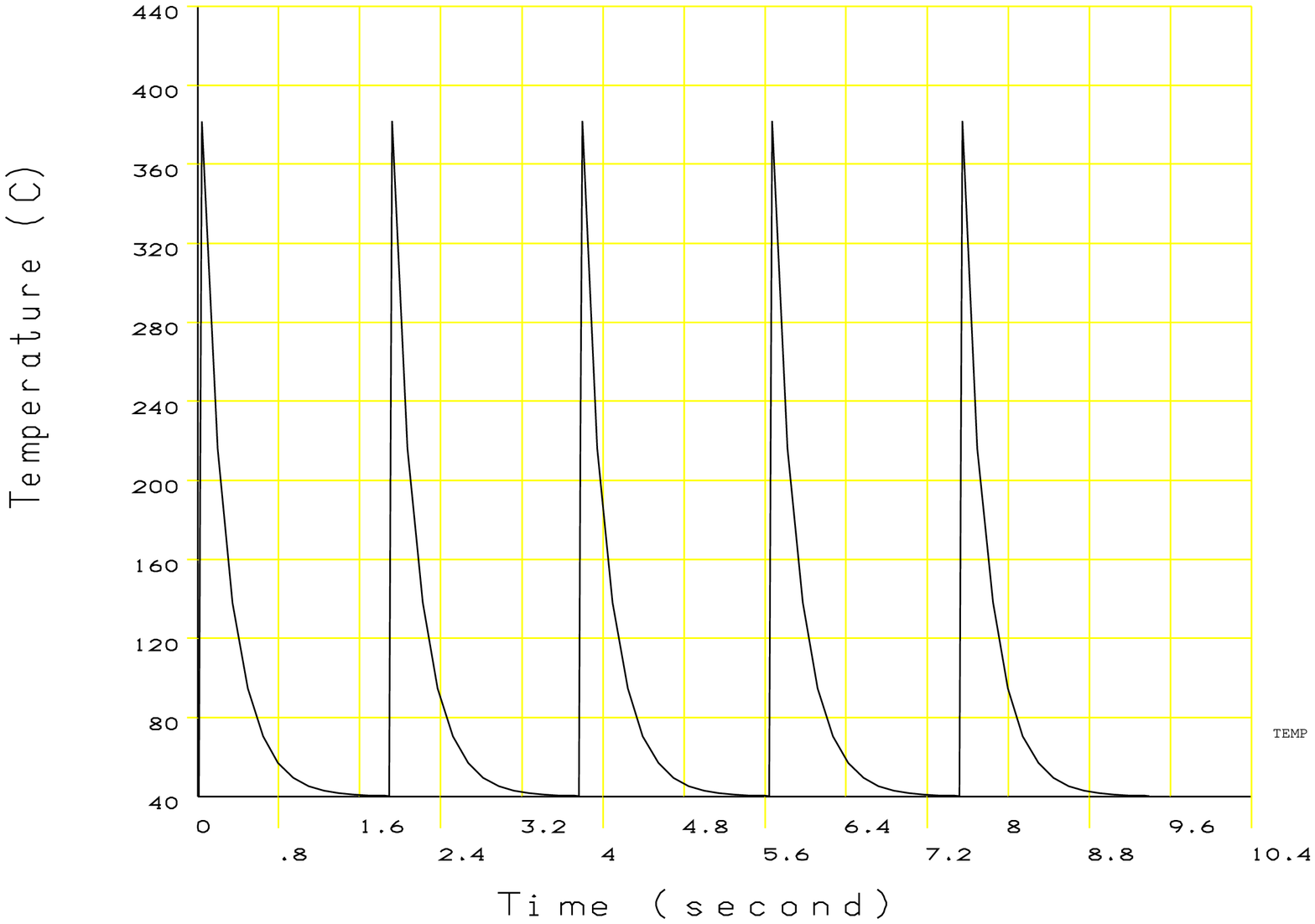, width=\textwidth, angle=-90}
\caption{Temperature evolution in the hottest cell of a graphite target
with 1.9~s repetition rate.}
\label{fig:temp}
\end{center}
\end{figure}

\underline{\bf Temperature buildup}
The temperature evolution in a graphite target has been investigated 
with the \textsc{ansys} code~\cite{ANSYS}. 
An ED distribution in a target with $R_T$=9mm was simulated with \textsc{mars}.
The beam $\sigma$ was chosen to be 3mm. 
The temperature rise for the target 
elements was estimated from the ED distribution and was assumed to be
instanteous. The temperature rise was applied to the target every 1.9 sec.
A simplified cooling system was used in this simulation.
The heat excess was brought away from the target by water with
constant temperature of 40$^{\circ}$~C running around the outer cylindrical
surface of the target.
As Figure~\ref{fig:temp} shows, the temperature in the hottest
cell oscillates between 40$^{\circ}$~C and 380$^{\circ}$~C without any buildup.
The tensile stress obtained from the simulation is about 20~MPa 
at the tensile strength for graphite of about 90~MPa.



A graphite target for the FNAL neutrino superbeam facility
satisfies all
requirements: it survives one spill;
the lifetime due to radiation damage is acceptable;
there is no temperature buildup in target; and the 
pion yield is quite high.

\subsection{NuMI Optics for High Intensity Beams}
\noindent
Design of the NuMI beamline optics for Run II Main Injector beam parameters 
is now well established. At $4*10^{13}$ ppp and a 40$\pi$ 
(95~$\%$, normalized) beam, specifications call for a $\sigma$=1~mm 
round beam on the target. With the order-of-magnitude higher intensities 
projected for the era when a new Proton Driver supplants the current 
8~GeV Booster, spot size must be greatly increased to avoid destruction 
of the target. However, increasing the beam size to $\sigma$=3~mm to 
accommodate these higher intensities is far beyond the tuning range
of the final-focus quadrupole configuration in the baseline NuMI design --
requiring as it does a nine-fold growth in $\beta^*$ on the target.

Creating the desired target beam parameters involves re-locating some, and
re-powering all, of the final 6 quadrupoles in the line. (Details of the
baseline NuMI lattice are described in \cite{JohnJonston}). 
In the high-intensity configuration the lattice is completely unchanged 
from quads Q101 through Q115. Quadrupoles Q116~$\to$~Q118 are altered 
both in gradients and locations. Gradients of the last 3 final-focus 
quads change (the polarity of Q119 also changes), but they remain in 
their current locations to preserve the bend center of the final
vertical dipole string. The final-focus is tuned to $\beta^*$=172.8~m 
at the target, with dispersion Dx~=~Dy~=~0 -- giving $\sigma$=3.00~mm 
for a 40$\pi$ beam.

The B2 dipoles comprising the final vertical bend have a 2"(H)~$\times$~4"(V) 
aperture. In the horizontal plane, despite the much large final $\beta^*$ 
objective, the beam is actually smaller through this restricted 
aperture than in the baseline design, with $\beta_{max}$(x)=78~m 
here compared to 99~m previously. Vertically, the beam grows 
significantly,
with $\beta_{max}$(y) reaching 300~m. For a 40$\pi$ beam this
translates into $\sigma_y$=4.0~mm and, again, no aperture troubles are readily
apparent.

\begin{figure}
\begin{center}
  \includegraphics[height=.32\textheight]{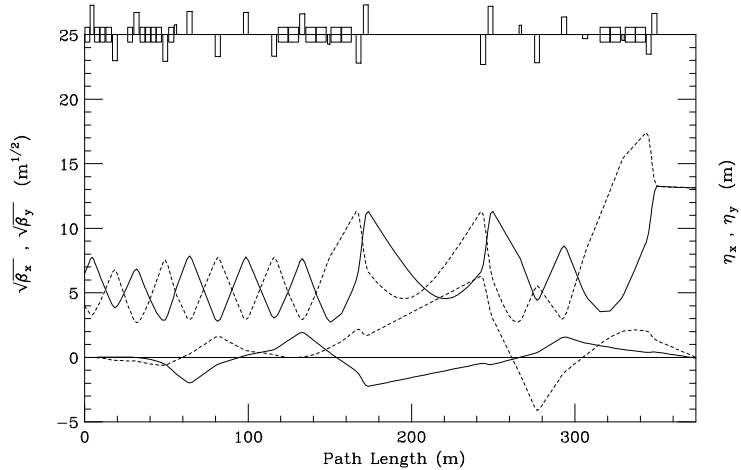}
  \caption{A new lattice.}
  \label{fig:newlattice}
\end{center}
\end{figure}
\subsection {Proton Intensity}
   In order to maximize the number of neutrino events in any off axis
detector, a crucial issue is the proton intensity delivered to the neutrino
beam target. An interesting question in the design of  
future neutrino experiments
 will be whether more neutrino events can be realized
by investing money into more detector mass or into
the ability to accelerate more 
protons. The best experiments in the world
will no doubt result from investment in 
both. An additional incentive for
investment in the ability to accelerate protons is that this can provide
improved experimental capabilities for other experiments 
at the same time
as increasing the number of neutrino event.
 Of
course, in the NuMI beamline the number of events in the MINOS detector
will benefit directly from increasing the protons on target. However,
other experiments can also benefit from higher proton intensities, although
in general there will be only partial overlap in the specific accelerator
improvement projects which benefits each experiment the most.

  There are three main paths to increasing the 
proton intensity:
\begin{enumerate}
\item {\bf Invest in the existing accelerator complex to increase the
number of protons which can be accelerated:} For NuMI, this specifically
means investment in the Booster and Main Injector to increase the number
of protons which can be handled per acceleration cycle and to reduce the
cycle time so more total protons are delivered per unit time.
This is likely the only means of increasing the proton intensity within 
the next five years. A recent study \cite {NUMIPIWG} has identified 
a number of upgrades possible in the Booster and Main Injector which 
can yield significant increases in the proton intensity for NuMI. 
Total proton beam
power from 0.6-0.8 MW should  be possible for an investment in 
the range of \$50M over five years (a factor of at least 4 increase from
the capability with no new investment).
\item {\bf Build new accelerators to increase the intensity:} 
The proton driver is a new accelerator to replace
the Booster as an injector to the Main Injector. Both a new synchrotron and
a LINAC have been studied \cite {synchrotron,LINAC}.
Improvements
in the Main Injector for handling high beam intensity 
would be essential.  A
faster cycle time in the Main Injector 
presents an attractive means of yet higher protons on target
than just the new
proton driver. With this approach, total proton beam power greater than 2 MW
should be possible with total cost in the range \$200M-\$350M. 
A series of neutrino experiments over many years could form 
the core of the justification for such a machine \cite {PDphysics}.
\item {\bf Reconfigure the existing complex to be used in new ways that will
maximize protons for NuMI:} An example here is to use a slightly reconfigured
recycler ring as a stacking system at 8 GeV to accumulate protons which
are then injected into the Main Injector \cite {Marriner}.
This can increase the total
number of protons per cycle and decrease the MI cycle time since stacking
can occur while the MI is ramping from a previous injection. Of course,
this kind of scheme will only be possible once collider operation had 
finished. In this case, Main Injector improvements for both proton intensity
and cycle time are required.
Total proton beam energy in the 
range 0.8-1.0 MW should be possible at a total
cost in the range of \$70M. This approach could be taken as a ``next step''
following the various improvements in the existing complex (\$50M in item 1
above) for an incremental investment of about \$25M.
\end{enumerate}

Any of these paths to increasing proton intensity will require significant
manpower resources beyond those currently available 
within Fermilab Beams Division 
Some of  that manpower investment must come from groups interested in the neutrino
experiments (MINOS, off axis,...) for these projects to move forward. An
ongoing NuMI Proton Intensity Working group is in formation and work in
this direction should be considered with comparable importance as R\&D for
detector construction. Many examples of possible upgrade projects which can
be undertaken in the near term are presented in reference \cite {NUMIPIWG}. 
\section{ Summary}

\par 
We are at an important stage of the field of neutrino oscillations.  
Between the currently running generation of experiments and the next 
we will be going from confirmation of oscillations to precision measurements
of the atmospheric parameters.  At the same time, we as a field are trying 
to determine not only how to see evidence for the last undiscovered 
mixing angle $\theta_{13}$, but how to get ultimately to precise 
measurements of $\nu_\mu \to \nu_e$ probabilities.  Although there are 
many suggestions for how to get to precision, certainly the experiments 
with the highest reach for seeing a non-zero $\nu_\mu \to \nu_e$ probability 
for a given proton beam power and detector mass are those which use a 
very narrow band neutrino beam to minimize backgrounds.  The off-axis 
technique, suggested originally by Brookhaven and adapted by the JHF to 
SuperK proposal, is a powerful one to achieve such a narrow band beam.  

\par 
Given the intense neurino beamline that is currently being built at 
Fermilab, and the long distances the resulting neutrinos will travel, 
there is an enormous opportunity not only for seeing $\nu_\mu \to \nu_e$ 
transitions, but to get to the underlying physics:  determining the 
mass hierarchy and ultimately measuring CP violation.  Because of the 
off-axis technique and the lack of a far detector location at present, 
there is a wide range of energies and baselines that can be chosen.  The 
narrowest neutrino beam produced by the NUMI beamline is at about 2~GeV 
and emerges about 14~mrad from the beamline axis, 
but ultimately the most 
precise measurements of CP violation or the mass hierarchy may come from 
placing a detector elsewhere off-axis, from neutrino energies from 0.6 to 
as much as 3~GeV.  
In this document we have therefore 
focused on detectors which are suitable for measurements at 2~GeV, 
but where relevant we have tried to comment on their 
appropriateness at other energies in that range.  

\par
Over its century history, the field of particle physics has developed
expertise in a large number of techniques for detecting ionizing
radiation.  The preferred technology
would likely be the least expensive one as a function of physics reach,
but direct cost comparisons are difficult for at least two
reasons:  1.)  The capabilities of each system are different and it
is not always clear how to compare the value of an additional
capability, and  2.) choosing between certain designs with comparable 
capabilities would require a 
level of detail in the cost estimate which is not currently 
available.  Therefore, this document must at best recommend the 
steps that need to be taken to be able to ultimately chose a detector 
technology (or technologies) for the NuMI off-axis beam.  

\par Here we summarize some of the salient features of each
technology considered:

Conclusions about Water Cerenkov:  
\begin{itemize} 
\item Much expertise in the field with large detector performance 
\item 20 kton fiducial mass proof of principle exists. 
\item Chain reaction of phototube implosions now understood.
\item Costs driven almost entirely by phototubes.
\item operation at the surface not obvious but perhaps possible (K2K). 
\item Could be promising for high angle lowest energy (sub-GeV) beams, but 
\item Monte Carlo studies show 
$\nu_e$ identification at 2GeV compromised due to inability of detector 
to discriminate between high energy neutral current $\pi^0$ production, and 
charged current $\nu_e$ interactions.  
\item R\&D efforts being pursued elsewhere already for sJHF to HK, which 
include developing cheaper and more robust photodetectors.  This won't 
change the background rejection capabilities, however.  
\item since individual particle energy resolution is not a limiting 
factor, the 
AQUARICH technology is not likely to have very different conclusions 
than regular water cerenkov devices.  
\end{itemize} 

Conclusions about Liquid Argon TPC's:  
\begin{itemize} 
\item Very detailed pattern recognition capabilities, especially for
electron identification.
\item Monte Carlo Studies show this to be the most efficienct detector for 
keeping signal and rejecting background. 
\item Cosmic ray studies in Pavia show that backgrounds at the ground level 
are manageable assuming acceptable data handling capabilities. 
\item Economies of scale and experience of Liquid Natural Gas industry promising for a large (phase III) single-volume detector. 
\item Favorable scaling for large size.
\item Need to verify that particle identification works as well as predicted
in simulations--this could be a promising phase III detector, but we strongly 
recommend placing a prototype detector in a neutrino beam which could prove the 
performance in the first few radiation lengths of a neutrino interaction.  
\end{itemize} 

Conclusions about Fine-Grained Calorimetry: 
\begin{itemize} 
\item Monte Carlo studies show this detector has adequate background 
discrimination and energy resolution, and the processes that generate 
the signals are well-understood (thresholds well below those for 
water cerenkov, for example, and there's a long history in the field of 
sampling calorimetry).  

\item Low Z absorber would provide the maximum amount of mass per 
readout plane, but low density induces large separations between consecutive
readout planes.  Backgrounds induced by operation at the surface must be verified.  

\item Different readout technologies have different risks associated with them: 
\begin{itemize} 
\item RPC's:  possibly the cheapest readount per $m^2$, but operational difficulties
have been encountered in the past. 
\item Streamer Tubes:  are likely to be the next cheapest readout.
\item Liquid or Solid Scintillator is the easiest to operate, no tricky gas or 
high voltage systems to build.  
\begin{itemize} 
\item Depending on light collection technique, the integration time could 
be quite long, implying bigger cosmic ray problems. 
\item Minimum R\&D, can use much of what was learned while designing MINOS. 
\item Gains in recent past to reduce fabrication costs for solid scintillator. 
\item Liquid scintillator would be easy to install in situ.  
\end{itemize} 
\end{itemize} 

\item Different absorber ideas have different risks associated with them: 
\begin{itemize} 
\item Is the cost of containing the water for a water-absorber detector prohibitively high? 
\item Would particle board warp too much to be acceptable for housing detector elements? 
\item Can any solid low z material provide enough mechanical support for readout?  
\end{itemize} 

\item Finally, before one embarks on a full-scale construction of any fine-grained  
calorimeter, one should certainly produce a prototype, where 
at least one dimension of the prototype would be the size of a single 
module.  
\end{itemize} 

There are a few issues which must be addressed regardless of detector 
technology:  for example, what is the the optimal segmentation that is required
to get an acceptable neutral current rejection factor?  Also, does the detector 
technology respond as predicted to charged particle beams?  

\par For phase II, we specifically recommend focused R\&D on fine-grained 
calorimetry:  this technique appears to have the smallest amount of risk 
associated with it, and although there are several options for absorber and 
readout technology, the outstanding issues are largely engineering ones, 
and can be addressed relatively quickly.  

\par For both phases, we will need to improve our understanding of 
neutrino interactions in the NuMI Off-axis energy regime.  
In phase II this is critical to get to the 
best precision on measuring the $\nu_\mu$ 
disappearance probability, and in phase III this will be essential to 
optimize the design of what is likely to be a $>100$M\$ detector.  We 
therefore recommend that as early as phase I that there be a program 
established to study neutrino interactions in a location 
underground at the NuMI beamline facility.

\par For phase III, large water Cerenkov detectors or liquid
argon offer scaling advantages.  In addition to sensitivity for
$\theta_{13}$, if placed underground,
such detectors would be sensitive to proton decay
and other topics of underground physics.  Since the 
time scale for phase III R\&D will take longer, it is important that 
this effort start now.  We recommend building a small prototype to test
in (but slightly off the axis of) the NuMI beamline, somewhere in the 
near detector hall.  

\par Finally, the most sensible path to the physics
is not simply to 
improve the far detector's size and/or performance.  
Investments in both the proton source (as early as phase I) and the 
beamline itself (phase III) will improve the experiment's sensitivity dramatically, 
and in a more economical way than by simply increasing the detector size.  

\par The writers of this report look forward to joining
such R\&D programs and collaborations which are forming to pursue
future neutrino initiatives.


\end{document}